\documentclass[twocolumn]{aastex631}
\usepackage{lmodern}
\usepackage{natbib}
\usepackage{graphicx}	
\usepackage{amsmath}	
\usepackage{amssymb}	
\usepackage{float}
\usepackage{url}

\newcommand{\planck}{{\it Planck\ }}
\newcommand{\ba}{\begin{eqnarray}}
\newcommand{\ea}{\end{eqnarray}}
\newcommand{\n}      {{\bf{n}}}

\newcommand  \beq    {\begin{equation}}

\newcommand  \eeq    {\end{equation}}
\newcommand  \gtsim  {\lower.5ex\hbox{$\; \buildrel > \over \sim \;$}} 
\newcommand  \ltsim  {\lower.5ex\hbox{$\; \buildrel < \over \sim \;$}}

\newcommand{\LCDM} {$\Lambda$CDM}

\newcommand{\ee}{\end{equation}}

\begin{document}
\label{firstpage}



\title[SO-Planck power spectra]{ The Simons Observatory: a new open-source power spectrum pipeline \\ applied to the Planck legacy data}

\author{Zack Li}
\affiliation{Department of Astrophysical Sciences,  Princeton University, Princeton, NJ 08544, USA}
\affiliation{Canadian Institute for Theoretical Astrophysics, 60 St. George Street, University of Toronto, Toronto, ON, M5S 3H8, Canada}
\author{Thibaut Louis}
\affiliation{Universit\'e Paris-Saclay, CNRS/IN2P3, IJCLab, 91405 Orsay, France}
\author{Erminia Calabrese}
\affiliation{School of Physics and Astronomy, Cardiff University, The Parade, Cardiff, Wales CF24 3AA, UK}
\author{Hidde Jense}
\affiliation{School of Physics and Astronomy, Cardiff University, The Parade, Cardiff, Wales CF24 3AA, UK}
\author{David Alonso}
\affiliation{Department of Physics, University of Oxford, Denys Wilkinson Building, Keble Road, Oxford OX1 3RH, United Kingdom}

\author{J. Richard Bond}
\affiliation{Canadian Institute for Theoretical Astrophysics, 60 St. George Street, University of Toronto, Toronto, ON, M5S 3H8, Canada}

\author{Steve K. Choi}
\affiliation{Department of Physics, Cornell University, Ithaca, NY 14853, USA}
\affiliation{Department of Astronomy, Cornell University, Ithaca, NY 14853, USA}

\author{Jo Dunkley}
\affiliation{Department of Astrophysical Sciences,  Princeton University, Princeton, NJ 08544, USA}
\affiliation{Joseph Henry Laboratories of Physics, Jadwin Hall, Princeton University, Princeton, NJ, USA 08544}

\author{Giulio Fabbian}
\affiliation{Center for Computational Astrophysics, Flatiron Institute, 162 5th Ave, New York, NY 10010, USA}
\affiliation{School of Physics and Astronomy, Cardiff University, The Parade, Cardiff, Wales CF24 3AA, UK}

\author{Xavier Garrido}
\affiliation{Universit\'e Paris-Saclay, CNRS/IN2P3, IJCLab, 91405 Orsay, France}

\author{Andrew H. Jaffe}
\affiliation{Blackett Laboratory, Prince Consort Road, Imperial College, London SW7 2AZ UK}

\author{Mathew S. Madhavacheril}
\affiliation{Perimeter Institute for Theoretical Physics, 31 Caroline Street N, Waterloo ON N2L 2Y5 Canada}
\affiliation{Department of Physics and Astronomy, University of Southern California, Los Angeles, CA, 90007, USA}

\author{P. Daniel Meerburg}
\affiliation{Van Swinderen Institute for Particle Physics and Gravity, University of Groningen, Nijenborgh 4, 9747 AG Groningen, The Netherlands}

\author{Umberto Natale}
\affiliation{School of Physics and Astronomy, Cardiff University, The Parade, Cardiff, Wales CF24 3AA, UK}

\author{Frank J. Qu}
\affiliation{DAMTP, Centre for Mathematical Sciences, Wilberforce Road, Cambridge CB3 0WA, UK}

\begin{abstract}
We present a reproduction 
of the \planck 2018 angular power spectra at $\ell > 30$, and associated covariance matrices, for intensity and polarization maps at 100, 143 and 217 GHz. This uses a new, publicly available, pipeline that is part of the \texttt{PSpipe} package.
As a test case we use the same input maps, ancillary products, and analysis choices as in the \planck 2018 analysis, and find that we can reproduce the spectra to 0.1$\sigma$ precision, and the covariance matrices to 10\%. We show that cosmological parameters estimated from our re-derived products agree with the public \planck products to 0.1$\sigma$, providing an independent cross-check of the \planck team's analysis. 
Going forward, the publicly-available code can be easily adapted to use alternative input maps, data selections and analysis choices, for future optimal analysis of \planck data with new ground-based Cosmic Microwave Background data. 
\end{abstract}

\section{Introduction}
\label{sec:intro}

The Cosmic Microwave Background (CMB) has helped build a precisely-measured Standard Model of Cosmology.
The current state-of-the-art is dominated by legacy data from the \planck  satellite mission, which obtained cosmological constraints from analysis of the CMB anisotropies in temperature and polarization. 
Results from the final collaboration data release were presented in \cite{planck_like2018} and \cite{Planck:2018} (hereafter, PL20 and PC20) for the power spectra and likelihood, and derived cosmology. A description of a re-analysis using the \planck team's alternative {\texttt{CamSpec}} code was also presented in \cite{efstathiou/gratton:2019}. We expect to match and then surpass the statistical power of \planck from the ground in the coming decade, through ongoing and future ground-based surveys of the CMB, such as from the Atacama Cosmology Telescope \citep[ACT,][]{thornton/etal:2016}, the South Pole Telescope \citep[SPT,][]{benson/etal:2014}, the upcoming Simons Observatory \citep[SO,][]{so_forecast:2019}, and CMB-S4 \citep{cmbs4:2019} at the end of the decade. 

\planck will provide a critical part of new cosmological constraints coming from ground-based surveys at the South Pole and Chile, for at least the next decade. \planck measured the temperature power spectrum to the cosmic variance limit up to multipoles of $\ell \sim 1800$ with excellent control of systematic effects, and such measurements at low to moderate scales in temperature are challenging to make robustly from the ground due to the atmosphere and ground pick-up. Polarization measurements 
from \planck were also made across the full sky, containing cosmological information currently inaccessible from the ground. 
Until the planned LiteBIRD satellite mission delivers science results near the end of this decade \citep{litebird:2020}, the \planck  legacy dataset will be a part of all cosmological constraints from the CMB at large scales in both temperature and polarization. The scales measured by \planck anchor cosmology in important ways, particularly for constraining the primordial power spectrum parameters $n_s$ and $A_s$, and the optical depth to reionization, $\tau_{\text{reio}}$, as well as  
properties such 
as extra light relic particles, 
isocurvature fluctuations, and dark matter interactions with baryons. 

The set of power spectra and the associated covariance matrix is the core product for many analyses of the CMB. Until now, joint analysis of \planck with high-resolution experiments like ACT \citep{aiola/etal:2020,choi/etal:2020} and SPT \citep{dutcher/etal:2021} has usually been done by combining power spectrum likelihoods, avoiding calculating the covariance between overlapping maps from the different experiments. 
With the sky coverage of ground-based surveys now increasing considerably \citep{2016SPIE.9910E..14D, naess/etal:2020}, future state-of-the-art cosmological model constraints 
will require estimating new cross-spectra and covariance matrices for the \planck maps in combination with the new datasets. One of the goals of this paper is to take a step towards these future data combinations by demonstrating a new, publicly-available, pipeline for estimating the \planck spectra and covariances. The modern cosmological model derived from the \planck data also exhibits a number of possible tensions \citep{planck_like2018}, 
either internally between different subsets of the data, or with external data (for parameters including the Hubble constant, $H_0$, the amplitude of structure formation, $\sigma_8$, and the lensing amplitude, $A_L$) that could be hints of new components in the model. Further scrutiny of the \planck maps with an independent pipeline therefore serves as an important check for these glimpses of possible discoveries.

In this paper, we provide a detailed description of the construction of cross-spectra, covariance matrices, and the resulting multi-frequency likelihood obtained from analysis of the high-$\ell$ \planck 2018 data release (PR3), made using the power spectrum pipeline developed for the Simons Observatory. While this paper focuses on reproducing and testing the approach implemented by the \planck collaboration in PL20, the new machinery is flexible for modified use, for example for computing statistics over different sky regions and in cross-correlation with new ground-based data. 
We chose to implement the analysis choices in PL20 in this paper, to provide a concrete comparison point from which to make future changes. The \planck analysis is complicated, and an open-source pipeline will allow different groups to explore optimum analysis and map choices, for example following those made for the \texttt{CamSpec} likelihood in \cite{efstathiou/gratton:2019}, or for the use of new \texttt{NPIPE} maps in \cite{planck_npipe:2020}. We anticipate making our own new analysis choices for the combination of \planck\ with Simons Observatory data in future work. The pipeline for this particular reproduction is released as part of the Simons Observatory power spectrum package, \texttt{PSPipe}\footnote{\url{https://github.com/simonsobs/PSpipe/blob/master/project/Planck_cov}}. We present detailed documentation to accompany this code\footnote{\url{https://simonsobs.github.io/planck-pr3-web/}}. 


The paper is organized as follows. In Sec.~\ref{sec:products} we describe the data products which enter into the analysis. In Sec.~\ref{sec:spectra} we present the power spectrum methods, results, and comparisons to the public \planck data products. In Sec.~\ref{sec:covmat} we describe the methods used to estimate analytic covariance matrices. In Sec.~\ref{sec:likelihood} we summarize the modeling of the spectra and in Sec.~\ref{sec:params} we show cosmological parameters estimated using our power spectrum products. We conclude in Sec.~\ref{sec:conclude}.




\begin{figure*}
\centering
\includegraphics[width=0.95\textwidth]{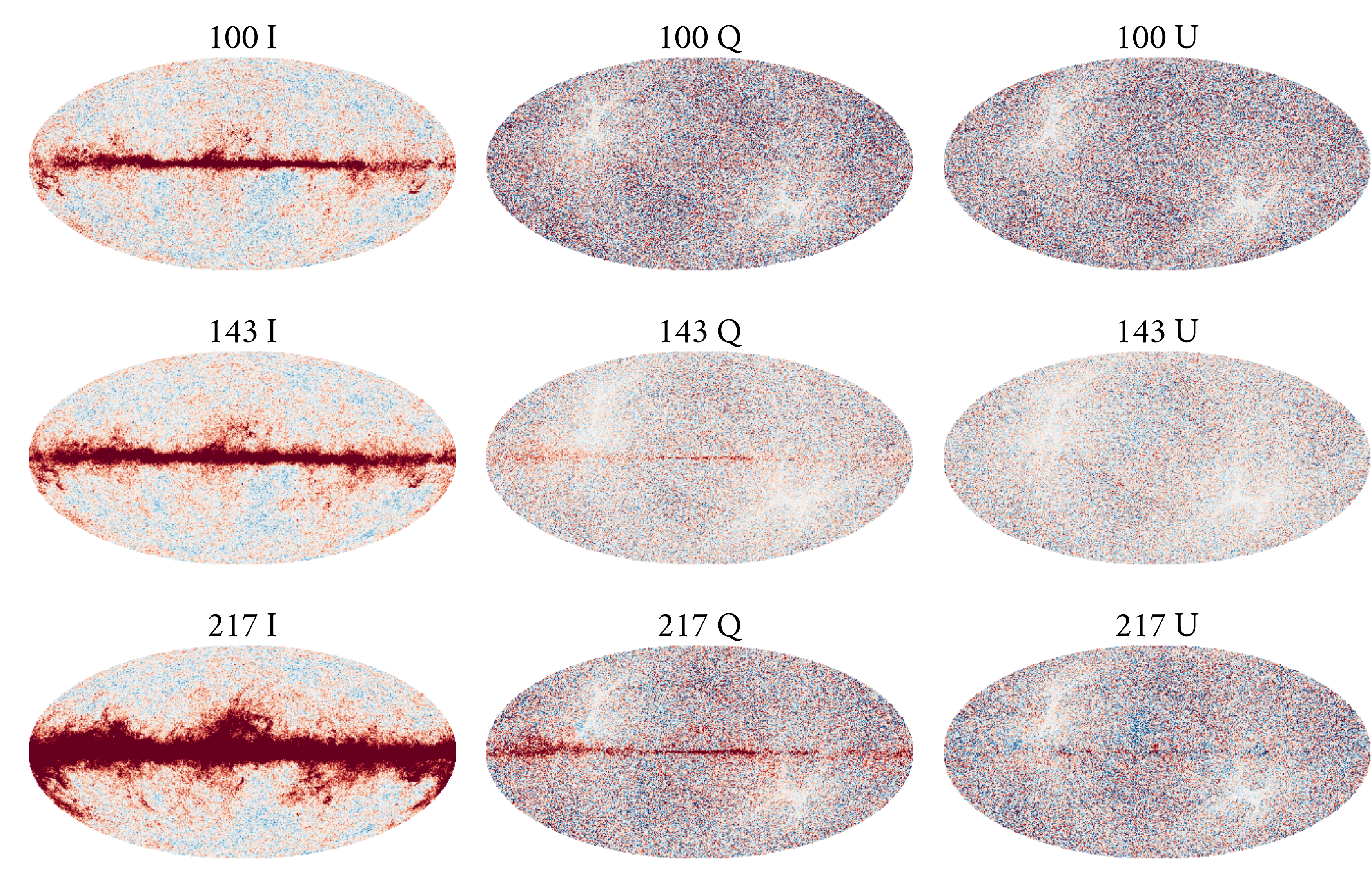}
\caption{\planck 2018 maps from the first half-mission at 100, 143 and 217 GHz \citep{Planck:2018}, with $I$, $Q$, and $U$ Stokes vectors, as produced and presented in \citet{Planck:2018} and used here. The second half-mission maps look similar, and we use the two half-mission map splits to compute spectra, as done in PL20. Maps are Mollweide projections with a color scale range of $\pm 500$ $\mu$K for Stokes I and $\pm 200$ $\mu$K in Stokes Q/U. }
\vspace{1em}
\label{fig:maps}
\end{figure*}

\section{Data products}
\label{sec:products}

We use the maps and ancillary data products as described in PL20, and summarized below. Most of the products require some manipulation in order to closely reproduce the \planck PR3 analysis, and we summarize these modifications here.

\subsection{Maps}

The primary inputs for this analysis are the publicly available \planck HFI half-mission maps from the 2018 data release. We expect to provide an analogous analysis of \texttt{NPIPE} maps \citep{planck_npipe:2020} in future work. We follow previous \planck cosmology analyses and focus on the 100, 143, and 217 GHz half-mission frequency maps, shown in Figure~\ref{fig:maps}. All of these maps are in HEALPix pixelization with $N_{\text{side}} = 2048$, and have missing pixels marked with the HEALPix \texttt{UNSEEN} pixel value. As in the PL20 analysis, we restrict our analysis to computing power spectra between different half-mission splits, to avoid both noise bias and time-dependent systematic effects. We use the maps in $I$, $Q$, $U$ Stokes vectors, as well as the pixel variance $II$, $QQ$, and $UU$. We do not use the $QU$ intra-pixel covariances.
We use these maps to estimate the spectra as well as the noise in the construction of the covariance matrix, described in Sec.~\ref{sec:covmat}. This differs from what is done in PL20, in which `half-ring half-mission maps', that were internally available to the \planck team, are used for the noise estimation.

{\it Polarization calibration:} The calibration of the \planck polarization maps relative to the temperature maps is uncertain, as they lack the bright dipole used to calibrate the temperature maps. The relative calibration is limited by the accuracy with which their polarization efficiencies could be determined either on the ground or in flight. 
To best match the spectra in the public \texttt{Plik} likelihood we determined that calibration factors of 0.9995 for 100 GHz, 0.999 for 143 GHz, and 0.999 for 217 GHz should be multiplied with the publicly released maps.
We note that the uncertainty in polarization calibration ($\sim 1$\%) dwarfs these ad-hoc calibration factors. 

\subsection{Masks}
\label{subsec:mask}

\begin{figure}
\centering
\includegraphics[width=0.5\textwidth]{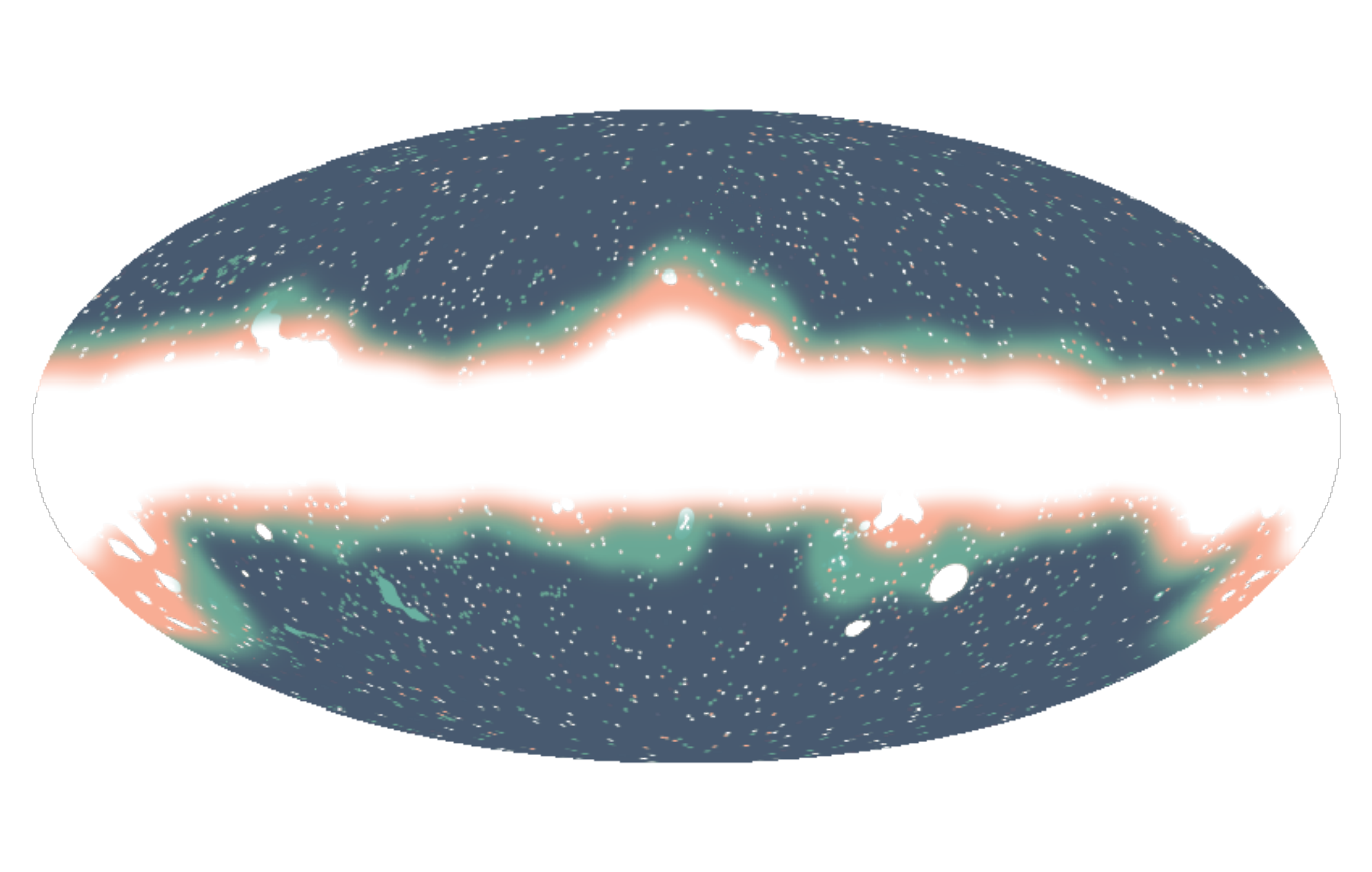}
\caption{The \planck masks which combine Galactic and point source masking. We show the 100 GHz (orange), 143 GHz (green), and 217 GHz (dark blue) masks in temperature overlaid. The corresponding polarization mask for each frequency does not have the point source holes. Point sources detected with $S/N > 5$ are masked, with holes of radius $\sigma = FWHM / \sqrt{\ln 8}$ of the effective Gaussian beam at that frequency \citep{planckspectra2015}. The colors correspond to 66\%, 57\% and 47\%
retained sky fraction at each frequency respectively, corresponding to the average of the square of the weight function. For each frequency, the parts of the mask that are colored refer to areas which are not masked at that frequency.}
\label{fig:mask}
\end{figure}

As in PL20, we use the publicly released \texttt{plik} sky masks shown in Figure~\ref{fig:mask} to avoid bright contaminated regions caused by the Galaxy and by extragalactic point sources. These comprise the temperature and polarization masks at 100, 143 and 217 GHz since both the point source and Galactic masks are different at each frequency. Polarized point source emission is negligible at \planck noise levels, so the polarization masks do not include point source holes. The Galactic masks cover 66\%, 57\%, and 47\% of the sky at 100, 143, and 217 GHz \citep{planck_like2015,planck_like2018} and are apodized with a 4.71 degree FWHM ($\sigma = 2^{\circ}$) Gaussian window function, whereas the point source mask is apodized with a 30 arcminute FWHM. 
The masks used for estimating the \planck power spectrum do not weight for noise or hits. 

These masks also include a sharp mask on the `missing pixels' in the maps by setting the mask values at the locations of the missing pixels to zero\footnote{There are $2,528$ missing pixels for 100 GHz hm1, $50,917$ pixels for 100 GHz hm2, $6,531$ pixels for 143 GHz hm1, $64,307$ pixels for 143 GHz hm2, $13,358$ pixels for 217 GHz hm1, and $115,439$ pixels for 217 GHz hm2.}. This represents a small fraction of the total pixels at $N_{\text{side}}=2048$ (corresponding to 50,331,648 pixels), with the most impacted frequency map (217 GHz hm2) missing 0.2\% of the pixels. Although the mask is not band-limited, simulations have shown that the estimated pseudo-$C_{\ell}$ spectra are not significantly biased \cite{planckspectra2015}.

The PR3 maps also feature a few thousand pixels at moderate Galactic latitude with extraordinarily high estimated pixel variance, $\sigma_p$, particularly in polarization. These are understood to arise from pixels with very few hits, which sometimes led to the inversion of a nearly-degenerate matrix during map processing. These are most noticeable at 100~GHz, where these pixels can have $\sim 10^6 \times $ more variance than the typical pixel. In this analysis we apply an additional mask which imposes a sharp pixel cut on pixels with $QQ$ and $UU$ variance above $10^6$ $\mu \mathrm{K}_{\mathrm{CMB}}^2$. These pixels do not have an effect on spectra estimation, regardless of mask -- however, their huge estimated variance would contaminate the covariance estimation, where the noise-like term involves a power spectrum of the noise variance map.
We use the union of these missing pixels masks across the $I$, $Q$, and $U$ channels for all as the missing pixel mask for all channels.

\begin{figure}
\centering
\includegraphics[width=0.5\textwidth]{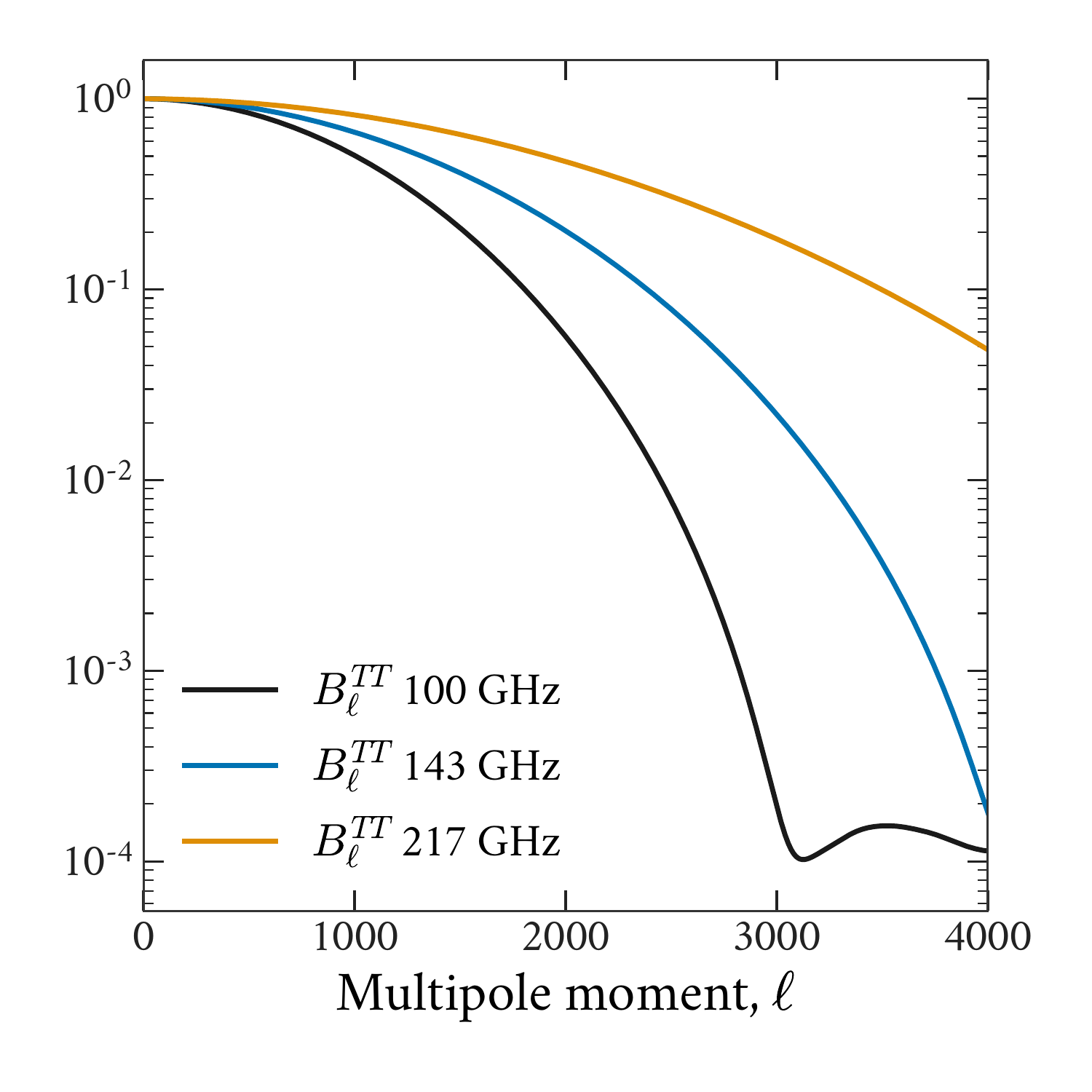}
\caption{The temperature beam window $B_{\ell}^{TT} = (W_{\ell}^{TT,TT})^{1/2}$  for three half-mission-1 frequency maps, as released in the Planck Legacy Archive \texttt{RIMO R3.01}. Polarization beams differ by a few percent at the smallest scales that \planck can measure ($\ell \sim 2500$).}
\label{fig:beam}
\end{figure}

\subsection{Beams}
The beam window function relates the signal angular power spectrum $C_{\text{map}}(\ell)$ of a map to the true underlying sky angular power spectrum $C_{\text{sky}}(\ell)$. Consider spectrum channels $A$, $B$ $\in \{ T, E, B \}$. We use the beams generated from the \texttt{QuickPol} formalism \citep{Hivon:2017}, which computes a beam matrix $W_{AB, A^\prime B^\prime}(\ell)$ , such that
\begin{equation}
    C_{\text{map}}^{AB}(\ell) = \sum_{A^\prime B^\prime} W_{AB, A^\prime B^\prime}(\ell) \, w^2_{\text{pix}}(\ell) \, C_{\text{sky}}^{A^\prime B^\prime}(\ell).
\end{equation}
Here $W_{AB, A^\prime B^\prime}(\ell)$ is the instrumental beam transfer function for the contribution of spectrum $A^\prime B^\prime$ to spectrum $AB$. The binning into pixels also induces $w_{\text{pix}}(\ell)$, the pixel window. For example, $W_{EE, TT}(\ell)$ is the mixing of the true $TT$ spectrum into the observed $EE$ spectrum.
The \planck beam is azimuthally asymmetric, so the average of the scanning beams for a given pixel depends on the location of the pixel and the scanning strategy. The choice of mask therefore defines an effective, azimuthally and spatially averaged, beam. Since we use the likelihood mask following the PL20 analysis, these beams were already computed and released in the \texttt{RIMO R3.01} for multipoles up to 4000, and are shown in Figure~\ref{fig:beam}. We smoothly extend the beam in a power law for multipoles from 4000 up to the maximum multipole $\sim 3n_{\text{side}}$ induced by the Nyquist frequency. We also multiply this beam correction by the corresponding temperature or polarization HEALPix pixel window function for $N_{\text{side}}=2048$. These beams are sufficient for this re-analysis of the \planck maps, but future cross-correlations with ground-based surveys will require recomputing the \planck beam for each new mask using \texttt{QuickPol} \citep{quickpol}. We also separately provide a \texttt{QuickPol} implementation in the software package developed for this analysis, \texttt{PowerSpectra.jl}.

\subsection{Binning}
We first compute unbinned, mask-deconvolved spectra in $2 \leq \ell \leq 2508$. To pass these new inputs to the \planck likelihood for cosmological analysis and also make comparisons with the \planck spectra, we then bin these spectra with the same scheme as PL20. We use an $\ell (\ell+1)$ weighting for $C_{\ell}$ inside each bin, over bins of $\Delta \ell = 5$ for $30 \leq \ell \leq 99$, $\Delta \ell = 9$ for $100 \leq \ell \leq 1503$, $\Delta \ell = 17$ for $1504 \leq \ell \leq 2013$, and $\Delta \ell = 33$ for $2014 \leq \ell \leq 2508$. 
Typically the mode-coupling matrix is binned before it is inverted. The binning operation regularizes the matrix, making it more tractable to invert, particularly for ground-based experiments with partial sky coverage. However, \planck is a full-sky experiment and the mode-coupling matrix of these masks is well-conditioned. Inverting before binning allows the \planck bandpower window function to be a top-hat with a $\ell (\ell+1)$ weighting within the bins.

\section{Power spectra}
\label{sec:spectra}

The analysis in this paper uses \texttt{PSPipe}\footnote{\url{https://github.com/simonsobs/PSpipe}}, a general purpose public pipeline for computation and analysis of CMB power spectra, and in this paper we expand it to deliver the full \planck\ high-$\ell$ power spectrum analysis pipeline. 

In the nominal analysis shown in this paper we use a \texttt{PSPipe} pipeline implementation in the Julia language, \texttt{PowerSpectra.jl}, as this high-level language allowed for the easy construction of the approximate covariance matrix computed in this work. In particular, the summations involved in these expressions map easily to for-loops that are inefficient in Python. We also perform a number of cross-checks of the power spectra computed with the other two codes in \texttt{PSPipe}: \texttt{NaMaster}  \citep{alonso/etal:2019} and \texttt{pspy}  \citep{louis/etal:2020}. The three codes within the pipeline have been tested against each other, and agree to numerical precision. 
The multiple backends within \texttt{PSPipe} accelerates the analysis process, by providing independent checks and reference implementations, complementing existing testing suites for the individual codes. 

As part of \texttt{PSPipe}, we provide a full suite of scripts which reproduce every step of the power spectrum analysis presented in the paper, from automatic download of the data to cosmological parameter estimation. We expect this to be an important resource to be used both directly and as a reference implementation, for future large-sky-area surveys in the millimeter band.




\subsection{Power spectrum estimation}\label{sec:specs}

\begin{figure*}
\includegraphics[width=\textwidth]{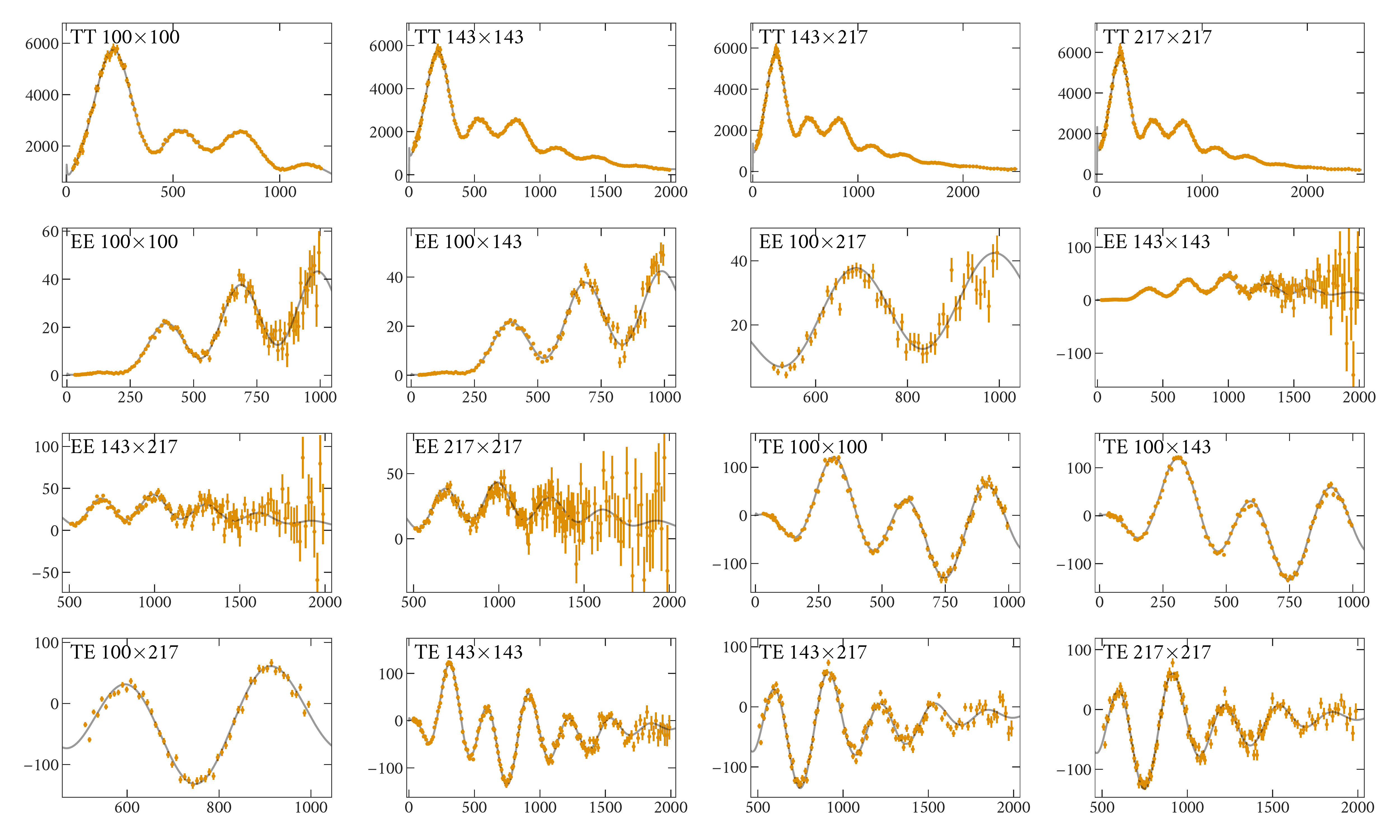}
\caption{Re-estimated \planck spectra and errors at 100, 143 and 217 GHz, and the cross-frequency spectra, in terms of $D_{\ell} = \ell (\ell+1) C_{\ell} / 2\pi$ in units of $\mu$K$^2$. We show just the angular range and suite of cross-spectra used in the likelihood, following the multipole cuts in PL20. We show the PR3 best-fit theory in the solid black line.}
\vspace{1em}
\label{fig:spectra}
\end{figure*}

We compute the power spectra at 100, 143 and 217 GHz, using the products described in Sec.~\ref{sec:products}. 
We begin by producing the masked maps, multiplying the half-mission maps with the masks described in Sec.~\ref{subsec:mask}. 

{\it Monopole and Dipole Subtraction:} We then estimate and subtract from each frequency map a monopole and dipole, using the unmasked pixels. The monopole, in particular, is extremely large in comparison to the scalar fluctuations even at large scales. The mask couples the monopole to other scales, and we encounter issues in accurately treating this monopole mode coupling, as the monopole is so much larger relative to the power on other scales. Small errors in estimating the mode-coupling matrix elements related to the monopole are multiplied by the large monopole, leading to significant bias on the resulting spectra. The power spectrum analysis itself does not use information from the monopole and dipole modes. 
We fit both the monopole and dipole simultaneously using the \texttt{healpy} package, and conservatively restrict this fit to map pixels where the corresponding mask pixel has value 1, excluding regions where the mask is 0 or apodized to a value less than 1. We confirmed that this choice does not substantially affect the determination of the monopole and dipole, compared to simply using the mask. 
Fitting with the mask avoids bias in the monopole and dipole from the point sources, the Galaxy, and missing pixels. 

{\it Mode-Coupling:} We then compute the mode-coupled spectrum, $\tilde{C}_{\ell}$, between pairs of fields,
\begin{equation}
    \tilde{C}^{AB \it{ij}}_{\ell} = \frac{1}{2\ell+1} \sum_m \mathsf{m}^{i,A}_{\ell m} \mathsf{m}^{*j,B}_{\ell m},
\end{equation}
where $\mathsf{m}^{i,A}_{\ell m}$ are the spherical harmonic coefficients of the masked map $i$ of channel $A \in \{T, E\}$. The effect of the mask is then decoupled, by solving for the decoupled spectrum $\hat{C}_{\ell}$ following
\begin{equation}
    \tilde{C}^{AB \it{ij}}_{\ell} = \sum_{\ell'} M^{AB \it{ij}}_{\ell \ell'} \hat{C}^{AB \it{ij}}_{\ell'},
\end{equation}
where $M^{AB \it{ij}}_{\ell \ell'}$
is the mode-coupling matrix between fields $i$ and $j$ for the cross-spectrum 
$AB \in \{TT, TE, EE, BB\}$ \citep{hivon2001,alonso/etal:2019}. A field $i$ refers to some combination of frequency and data split. Note that although we compute the $BB$ pseudo-spectrum, we only use it in this analysis in order to remove its contribution to the $EE$ power spectrum through mode coupling. This is important primarily for the noise power spectrum, where $EE$ and $BB$ contributions are comparable.
The PL20 analysis used \texttt{PolSpice} for this calculation \citep{chon/etal:2004}, which performs this same operation using correlation functions. 

We compute mode-decoupled spectra for every cross spectrum $C_{\ell}(\mathsf{m}^{i,A},\mathsf{m}^{j,B})$ for frequency-split combinations with frequencies $\in \{100, \,143,\, 217\}$ and splits $ \in \{\mathrm{hm1},\, \mathrm{hm2}\}$. Cross-spectra within the same half-mission are not used for signal calculations. 
We combine TE and ET spectra with a flat average, i.e. $(C_{\ell}^{TE} + C_{\ell}^{ET})/2$.
To estimate the data cross-spectra used in the likelihood, we then correct for the beam and pixel window by applying their inverse to the decoupled spectra.
Note that this differs from PL20, which used inverse-variance weighting in order to combine spectra.

\begin{figure}
\centering
\includegraphics[width=0.45\textwidth]{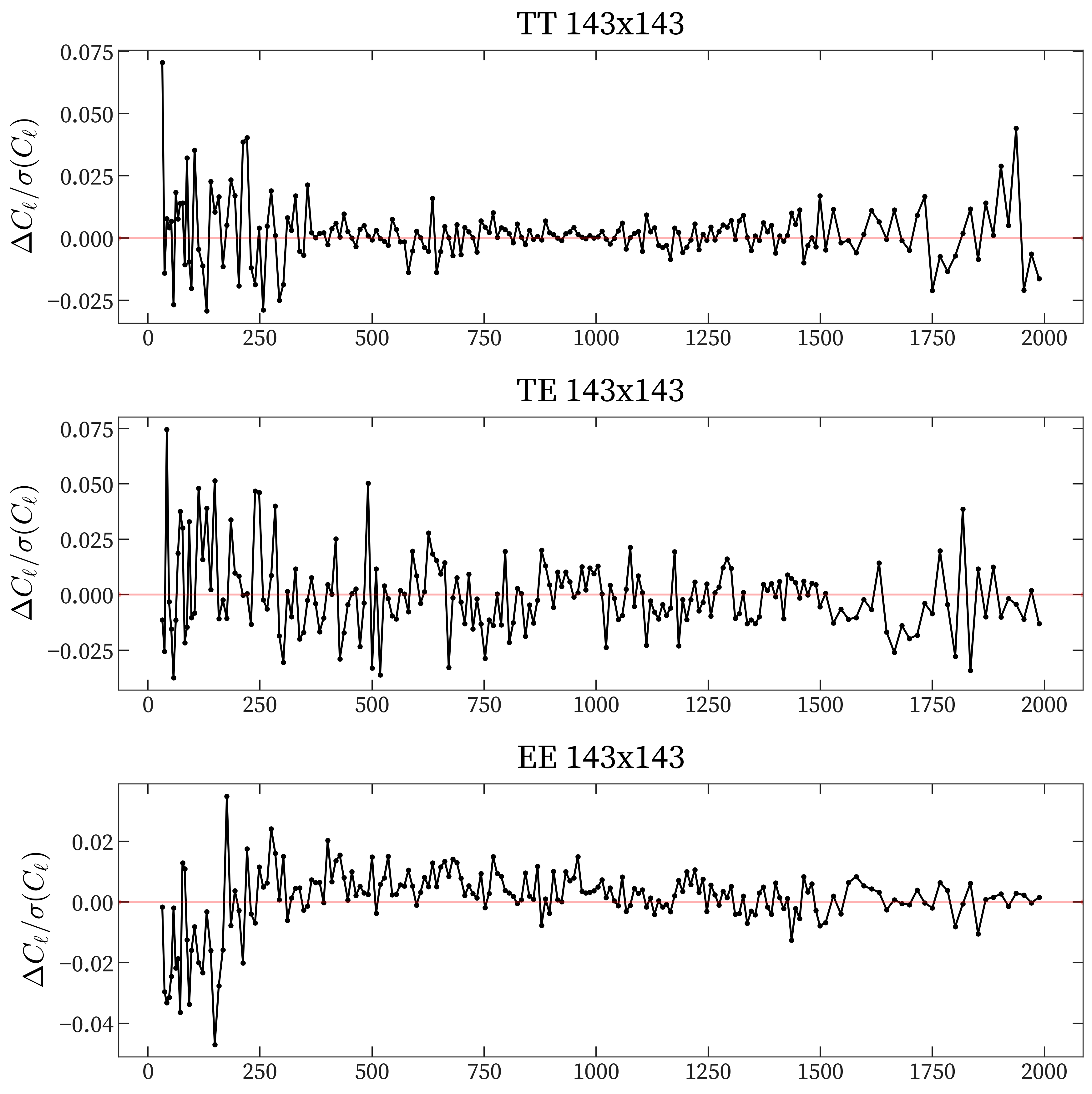}
\caption{Residuals between our re-estimated power spectra and the public \planck PL20 spectra at 143~GHz, shown as a fraction of a standard deviation. The agreement is within 0.08$\sigma$ over the full angular range; the $EE$ agrees to within 0.05$\sigma$, with the largest deviation at large angular scales. The residuals for the full suite of spectra are in Figure \ref{fig:all_resid} in the Appendix.}
\label{fig:143resid}
\end{figure}

\subsection{Comparison to PL20 spectra} 

The spectra at each of the three frequencies, for $TT$, $TE$ and $EE$, are shown in Figure \ref{fig:spectra}. 
For spectra which are combined, such as $TT$ from different frequencies and half-missions or $TE$ and $ET$, are combined together using uniform weighting. The residuals compared to the PL20 spectra for 143~GHz $TT$, $TE$ and $EE$ are in Figure~\ref{fig:143resid}, computed here using the errors from the PL20 products. The residuals compared to the full suite of spectra are shown in the Appendix in Figure~\ref{fig:all_resid}, including cross-frequencies. We find general agreement to generally at the 0.05$\sigma$ level, with a notable deviation in $EE$ at large scales up to $0.3\sigma$, and $TE$ at small scales to $0.1 \sigma$. Considering the same angular ranges as in PL20, the full set of binned spectra can be concatenated to form a single data vector of total length 2290. The ordering of this vector is shown in Table \ref{table:multipoleranges}, together with the multipole ranges.

\begin{table}
\hspace{1em}
\label{table:multipoleranges}
\begin{tabular}{ccccc}
\hline\hline
$AB$ & $f_1$ & $f_2$ & $\ell_{\mathrm{min}}$ & $\ell_{\mathrm{max}}$ \\ \hline 
    TT& 100& 100& 30& 1197 \\
    TT& 143& 143& 30& 1996 \\
    TT& 143& 217& 30& 2508 \\
    TT& 217& 217& 30& 2508 \\
    \hline
    EE& 100& 100& 30& 999 \\
    EE& 100& 143& 30& 999 \\
    EE& 100& 217& 505& 999 \\
    EE& 143& 143& 30& 1996 \\ 
    EE& 143& 217& 505& 1996 \\
    EE& 217& 217& 505& 1996 \\
    \hline
    TE& 100& 100& 30& 999 \\
    TE& 100& 143& 30& 999 \\
    TE& 100& 217& 505& 999 \\ 
    TE& 143& 143& 30& 1996\\ 
    TE& 143& 217& 505& 1996\\ 
    TE& 217& 217& 505& 1996 \\
\hline\hline\\
\end{tabular}
\caption{Multipole ranges for the data vector  for each cross-spectrum between frequency $f_1$ and $f_2$, using the same choices as in PL20. The $TE$ is an average of $TE$ and $ET$ for cross-frequency spectra. All cross-spectra that arises from different half-missions are included in an average, for any combination of frequencies and channel $AB \in \{TT, TE, EE\}$. The ordering of the rows matches the ordering in the data vector used in the likelihood.}  
\end{table}








\section{Covariance matrix}
\label{sec:covmat}

We now review the method for hybrid estimation of covariance matrices, building on \citet{hansen2002, hinshaw2003, efstathiou2004, brown/etal:2005,challinorchon2005, efstathiou:2006,planck_like2013, planck_like2015}, used in PL20 to produce the \textit{Planck} 2018 likelihood. We describe the details of our implementation, and note where this differs from PL20 due to the available products, and where the approach differs from the analytic covariance method used in \texttt{NaMaster} \citep{garcia-garcia:2019} and \texttt{pspy}. 

Following PL20, the model describing the time-ordered data is 
\begin{equation}
   d_t = (\mathbf{P}_{tp} \mathbf{B}_t \mathbf{w}_p) \, s_p + n_t,
\end{equation}
for data $d_t$, pointing matrix $\mathbf{P}_{tp}$, beam operator $\mathbf{B}_{tp}$, pixel window operator $\mathbf{w}_p$, true underlying signal map $s_p$, and noise $n_t$ \citep{hivon2001}. Note that the noise is not affected by any of the beam-related operators.
The likelihood assumes Gaussian variance about a fiducial signal power spectrum 
defined by 
\begin{equation}
    C_{\ell}^{AB, \it{ij}} \equiv W^{AB, \it{ij}}_{\ell} \, p_{\ell}^2 \left( C_{\ell}^{\mathrm{AB,CMB}} + C_{\ell}^{\mathrm{AB,FG}}(f_i, f_j) \right),
    \label{eqn:signal}
\end{equation}
where $f_k$ is the frequency of the field $k$, $W^{AB, \it{ij}}_{\ell}$ is the beam window function, and $p_\ell$ the pixel window function. In units of $\mu\mathrm{K}_{\mathrm{CMB}}$, the CMB power spectrum stays constant between frequencies, but the foreground signal varies. As in PL20, the fiducial CMB spectrum is given by the \planck best-fitting $\Lambda$CDM model, with foreground model given in Sec.~\ref{sec:likelihood}.

The covariance matrix for the suite of cross-spectra is estimated for the combination of signal and noise on the masked sky. It combines the fiducial power spectra and projector functions (the product of the two masks) that account for the mode-coupling from the mask and non-uniform noise, as described in \citet{planckspectra2015} (PL15). We include the expression here for the covariance matrix between a $TT$ cross-spectrum from maps $i,j$ and a $TT$ cross-spectrum from maps $p,q$, to demonstrate the overall structure of the calculation: 

\begin{equation}
\scalebox{0.9}{%
$\begin{split}
    \mathrm{C}&\mathrm{ov}(\hat{C}_{\ell}^{TT \, i,j}, \hat{C}_{\ell}^{TT \, p,q}) \\
        & \approx \sqrt{ C_{\ell}^{TT \, i,p}  C_{\ell^{\prime}}^{TT \, i,p} C_{\ell}^{TT \, j,q}  C_{\ell^\prime}^{TT \, j,q} }  \, \Xi^{\emptyset \emptyset,\emptyset\emptyset}_{TT} \left[ (i,p)^{TT}, (j, q)^{TT}\right]_{\ell \ell^{\prime}} \\
        &+ \sqrt{ C_{\ell}^{TT \, i,q}  C_{\ell^{\prime}}^{TT \, i,q} C_{\ell}^{TT \, j,p}  C_{\ell^\prime}^{TT \, j,p} }  \, \Xi^{\emptyset \emptyset,\emptyset\emptyset}_{TT} \left[ (i,q)^{TT}, (j, p)^{TT}\right]_{\ell \ell^{\prime}} \\
        &+ \sqrt{ C_{\ell}^{TT \, i,p}  C_{\ell^{\prime}}^{TT \, i,p}  } \, \Xi^{\emptyset\emptyset,TT}_{TT} \left[ (i,p)^{TT}, (j, q)^{TT}\right]_{\ell \ell^{\prime}} \\
        &+ \sqrt{ C_{\ell}^{TT \, j,q}  C_{\ell^{\prime}}^{TT \, j,q}  } \,  \Xi^{\emptyset\emptyset, TT}_{TT} \left[ (j,q)^{TT}, (i,p)^{TT}\right]_{\ell \ell^{\prime}} \\  
        &+ \sqrt{ C_{\ell}^{TT \, i,q}  C_{\ell^{\prime}}^{TT \, i,q}  } \,  \Xi^{\emptyset\emptyset,TT}_{TT} \left[ (i,q)^{TT}, (j, p)^{TT}\right]_{\ell \ell^{\prime}} \\
        &+ \sqrt{ C_{\ell}^{TT \, j,p}  C_{\ell^{\prime}}^{TT \, j,p}  } \,  \Xi^{\emptyset\emptyset, TT}_{TT} \left[ (j,p)^{TT}, (i,q)^{TT}\right]_{\ell \ell^{\prime}} \\  
        &+ \Xi^{TT,TT}_{TT} \left[ (i,p)^{TT}, (j, q)^{TT}\right]_{\ell \ell^{\prime}} + \Xi^{TT,TT}_{TT} \left[ (i,q)^{TT}, (j, p)^{TT}\right]_{\ell \ell^{\prime}}
\end{split}$%
}
\label{eq:covTTTT}
\end{equation}

The $\Xi^{X,Y}_{AB}$ terms are projector functions describing mode-coupling from the mask for channel $AB \in \{TT, TE, EE\}$, and $X,Y$ refer to the kind of noise-variance weighting applied to the mask.
$X=\emptyset\emptyset$ refers to the mask alone, $X = TT$ refers to weighting by the $II$ variance, and $PP$ refers to weighting by the $QQ$ and $UU$ variance. The suite of projector functions are given in the Appendix, and follow those in Appendix C1 of PL16.  


The noise power spectrum is an important input for the covariance matrix. If the noise was diagonal in pixel space, leading to a white noise power spectrum, then Equation~\ref{eq:covTTTT} would be a good approximation for the spectrum covariance. However, we see from Figure \ref{fig:noise_ratio} that this is not the case (we describe the estimation of these noise terms in the next section), with many spectrum channels deviating significantly from the white noise prediction for the noise power spectrum. Following PL16, PL20, we define the ratio of the measured noise power spectrum with the predicted power spectrum from the measured white noise level,
\begin{equation}
    \mathcal{R}^{i,X}_{\ell, \ell^\prime} = \sqrt{\frac{ N_{\ell,\mathrm{data}}^{i,X} N_{\ell^\prime,\mathrm{data}}^{i,X}  }{ N_{\ell,\mathrm{white}}^{i,X}  N_{\ell^\prime,\mathrm{white}}^{i,X} }}.
\end{equation}
In this case $N_{\ell,\mathrm{data}}$ refers to the effective white noise level presented in Table~\ref{tab:effectivewhitenoiselevel}, constant with $\ell$.
A factor of $\mathcal{R}^{i,X}_{\ell} \mathcal{R}^{i,Y}_{\ell}$ is multiplied with each $\Xi^{X,Y}_{\ell, \ell^\prime}$ for each factor of $X,Y \in \{TT, PP\}$, where $N_{\ell}^{PP} = N_{\ell}^{EE}$. For example, Equation \ref{eq:covTTTT} transforms to
\begin{equation}
\small
\label{eq:covTTTTnonwhite}
\scalebox{0.9}{%
$\begin{split}
    \mathrm{V}&\mathrm{ar}(\hat{C}_{\ell}^{TT \, i,j}, \hat{C}_{\ell}^{TT \, p,q})_{\mathrm{nonwhite}} \\
        & \approx \sqrt{ C_{\ell}^{TT \, i,p}  C_{\ell^{\prime}}^{TT \, i,p} C_{\ell}^{TT \, j,q}  C_{\ell^\prime}^{TT \, j,q} }  \, \Xi^{\emptyset \emptyset,\emptyset\emptyset}_{TT} \left[ (i,p)^{TT}, (j, q)^{TT}\right]_{\ell \ell^{\prime}} \\
        &+ \sqrt{ C_{\ell}^{TT \, i,q}  C_{\ell^{\prime}}^{TT \, i,q} C_{\ell}^{TT \, j,p}  C_{\ell^\prime}^{TT \, j,p} }  \, \Xi^{\emptyset \emptyset,\emptyset\emptyset}_{TT} \left[ (i,q)^{TT}, (j, p)^{TT}\right]_{\ell \ell^{\prime}} \\
        &+ \sqrt{ C_{\ell}^{TT \, i,p}  C_{\ell^{\prime}}^{TT \, i,p}  } \, \Xi^{\emptyset\emptyset,TT}_{TT} \left[ (i,p)^{TT}, (j, q)^{TT}\right]_{\ell \ell^{\prime}}  \mathcal{R}_{\ell, \ell^\prime}^{j,TT}
         \mathcal{R}_{\ell, \ell^\prime}^{q,TT} \\
        &+ \sqrt{ C_{\ell}^{TT \, j,q}  C_{\ell^{\prime}}^{TT \, j,q}  } \,  \Xi^{\emptyset\emptyset, TT}_{TT} \left[ (j,q)^{TT}, (i,p)^{TT}\right]_{\ell \ell^{\prime}}  \mathcal{R}_{\ell, \ell^\prime}^{i,TT}
         \mathcal{R}_{\ell, \ell^\prime}^{p,TT} \\  
        &+ \sqrt{ C_{\ell}^{TT \, i,q}  C_{\ell^{\prime}}^{TT \, i,q}  } \,  \Xi^{\emptyset\emptyset,TT}_{TT} \left[ (i,q)^{TT}, (j, p)^{TT}\right]_{\ell \ell^{\prime}}  \mathcal{R}_{\ell, \ell^\prime}^{j,TT}
         \mathcal{R}_{\ell, \ell^\prime}^{p,TT} \\
        &+ \sqrt{ C_{\ell}^{TT \, j,p}  C_{\ell^{\prime}}^{TT \, j,p}  } \,  \Xi^{\emptyset\emptyset, TT}_{TT} \left[ (j,p)^{TT}, (i,q)^{TT}\right]_{\ell \ell^{\prime}}  \mathcal{R}_{\ell, \ell^\prime}^{i,TT}
         \mathcal{R}_{\ell, \ell^\prime}^{q,TT} \\  
        &+ \Xi^{TT,TT}_{TT} \left[ (i,p)^{TT}, (j, q)^{TT}\right]_{\ell \ell^{\prime}}  \mathcal{R}_{\ell, \ell^\prime}^{i,TT}
         \mathcal{R}_{\ell, \ell^\prime}^{p,TT}  \mathcal{R}_{\ell, \ell^\prime}^{j,TT}
         \mathcal{R}_{\ell, \ell^\prime}^{q,TT} \\
        &+ \Xi^{TT,TT}_{TT} \left[ (i,q)^{TT}, (j, p)^{TT}\right]_{\ell \ell^{\prime}}  \mathcal{R}_{\ell, \ell^\prime}^{i,TT}
         \mathcal{R}_{\ell, \ell^\prime}^{q,TT} \mathcal{R}_{\ell, \ell^\prime}^{j,TT}
         \mathcal{R}_{\ell, \ell^\prime}^{p,TT}.
\end{split}$%
}
\end{equation}
The full suite of the blocks for $TE$, $EE$, and the covariances between these three sets of spectra is given in PL16 and  repeated in the Appendix for completeness. 

This covariance matrix expression is not quite the same as the Gaussian, uniform-noise approximations typically used by \texttt{NaMaster} and \texttt{pspy}. 
There one defines the total power spectrum 
\begin{equation}
{P}_{\ell}^{i,A} \equiv C_{\ell}^{i,A} + N_{\ell}^{i,A},
\end{equation}
with covariance, for the $TT$ block for example, given by
\begin{equation}
\small
\label{eq:covTTTTuniform}
\begin{split}
    \mathrm{V}&\mathrm{ar}(\hat{C}_{\ell}^{TT \, i,j}, \hat{C}_{\ell}^{TT \, p,q})_{\mathrm{uniform}} \\
        & \approx \sqrt{ P_{\ell}^{TT \, i,p}  P_{\ell^{\prime}}^{TT \, i,p} P_{\ell}^{TT \, j,q}  P_{\ell^\prime}^{TT \, j,q} }  \, \Xi^{\emptyset \emptyset,\emptyset\emptyset}_{TT} \left[ (i,p)^{TT}, (j, q)^{TT}\right]_{\ell \ell^{\prime}} \\
        &+ \sqrt{ P_{\ell}^{TT \, i,q}  P_{\ell^{\prime}}^{TT \, i,q} P_{\ell}^{TT \, j,p}  P_{\ell^\prime}^{TT \, j,p} }  \, \Xi^{\emptyset \emptyset,\emptyset\emptyset}_{TT} \left[ (i,q)^{TT}, (j, p)^{TT}\right]_{\ell \ell^{\prime}}. 
\end{split}
\end{equation}
We note that \texttt{NaMaster} uses the arithmetic and not geometric mean for these covariances.
In the limit of uniform noise, the projector functions $\Xi^{TTTT}_{TT}$ are proportional to $N_{\ell}^2 \Xi^{\emptyset \emptyset,\emptyset\emptyset}_{TT}$ and Equation \ref{eq:covTTTTnonwhite} is equivalent to Equation \ref{eq:covTTTTuniform}. The \planck maps are sufficiently non-uniform in noise that approximating them as uniform, as implemented in \texttt{NaMaster} and \texttt{pspy}, can cause order 10\% inaccuracies in the covariance matrix. While this may not be enough to substantially affect cosmological parameters, mis-estimating the covariance matrices to such an extent can have consequences for null tests. As a rough rule of thumb, \cite{efstathiou/gratton:2019} requires that for a data vector of length $N$, the noise estimates need to satisfy 
\begin{equation}
    \frac{\Delta \sigma^2}{\sigma^2} \lessapprox \sqrt{\frac{2}{N}},
\end{equation}
for accurate $\chi^2$, which for \planck requires $\sim 1\%$ estimates of the covariance matrices. We found that failing to treat the noise anisotropy leads to covariance matrix differences on the order of $10\%$.

\subsection{Noise Model}


\begin{figure}
\centering
\includegraphics[width=0.45\textwidth]{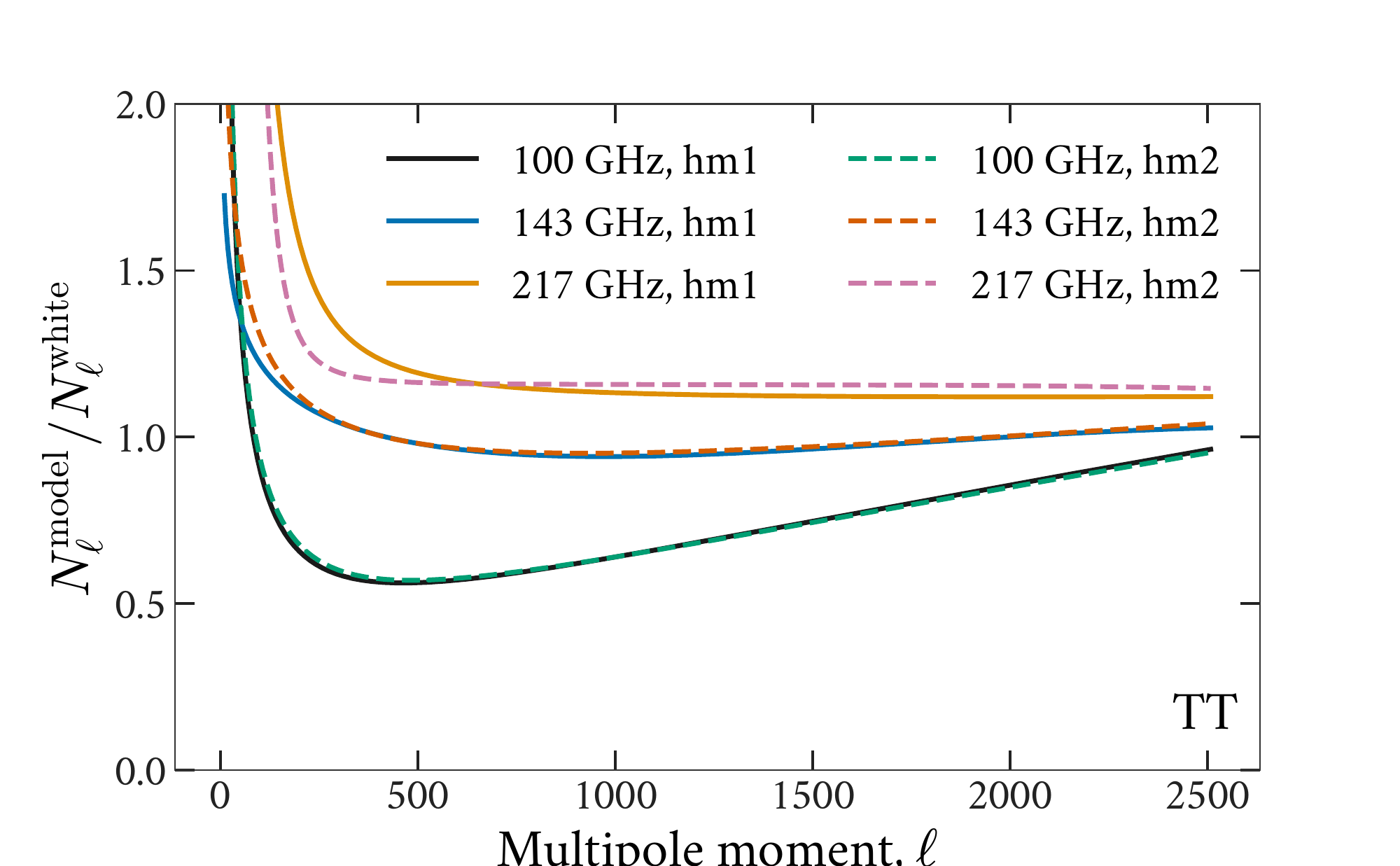}
\includegraphics[width=0.45\textwidth]{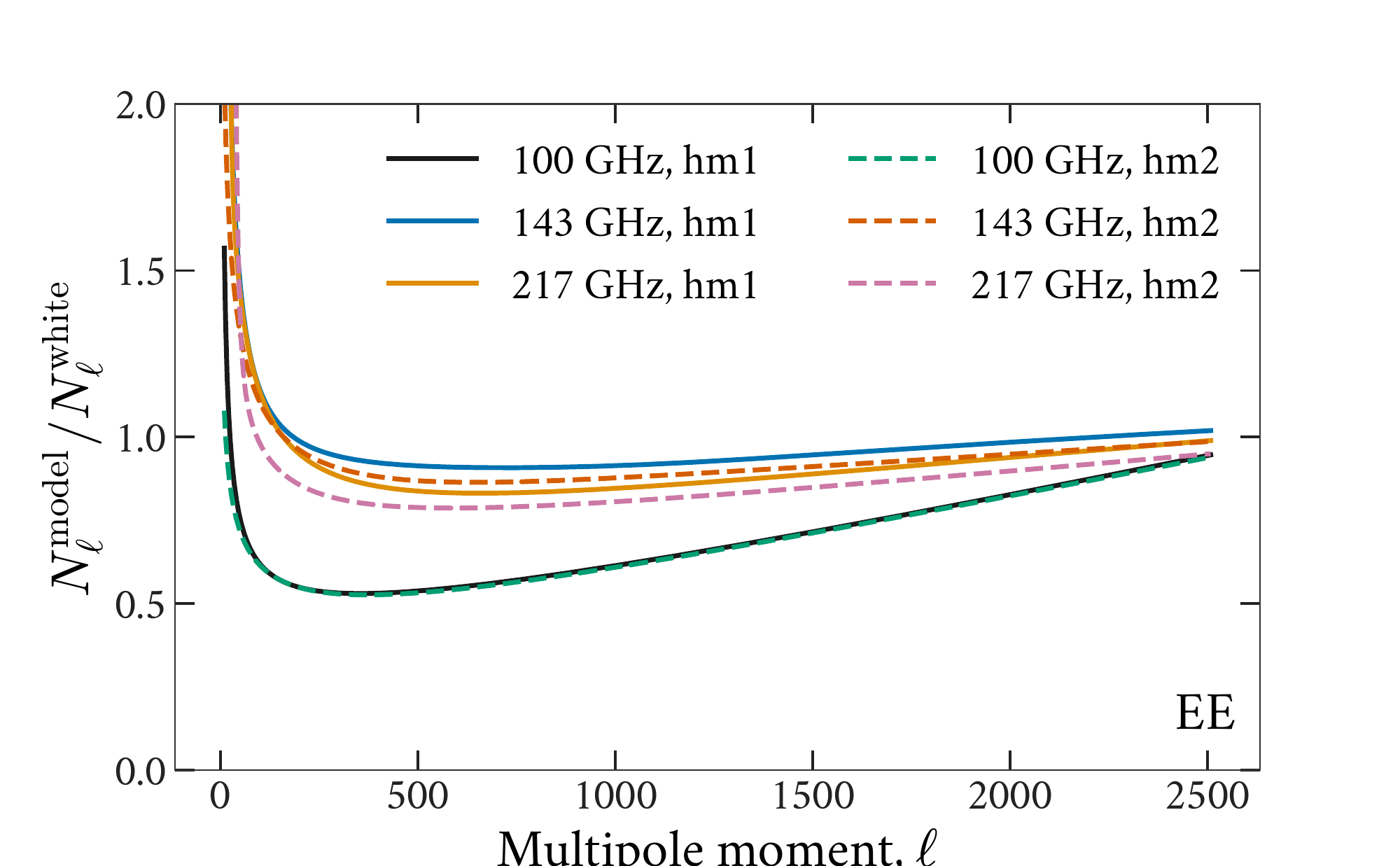}
\caption{The ratio between the white noise and measured noise model in temperature (top) and polarization (bottom), estimated by differencing the auto and cross-spectra of the half-mission maps. This is a key quantity in the covariance matrix calculation. The noise model parameters estimated for each 100, 143 and 217~GHz half-mission map are reported in Table \ref{table:noise}. The white noise levels $N_{\ell,\mathrm{data}}$ are presented in Table~\ref{tab:effectivewhitenoiselevel}, and are constant with $\ell$. The temperature noise spectra (top panel) are poorly determined for $\ell \ltsim 500$. However, the covariance matrices are not sensitive to the noise power spectrum at these signal-dominated scales.}
\label{fig:noise_ratio}
\end{figure}

\begin{figure}
\centering
\includegraphics[width=0.5\textwidth]{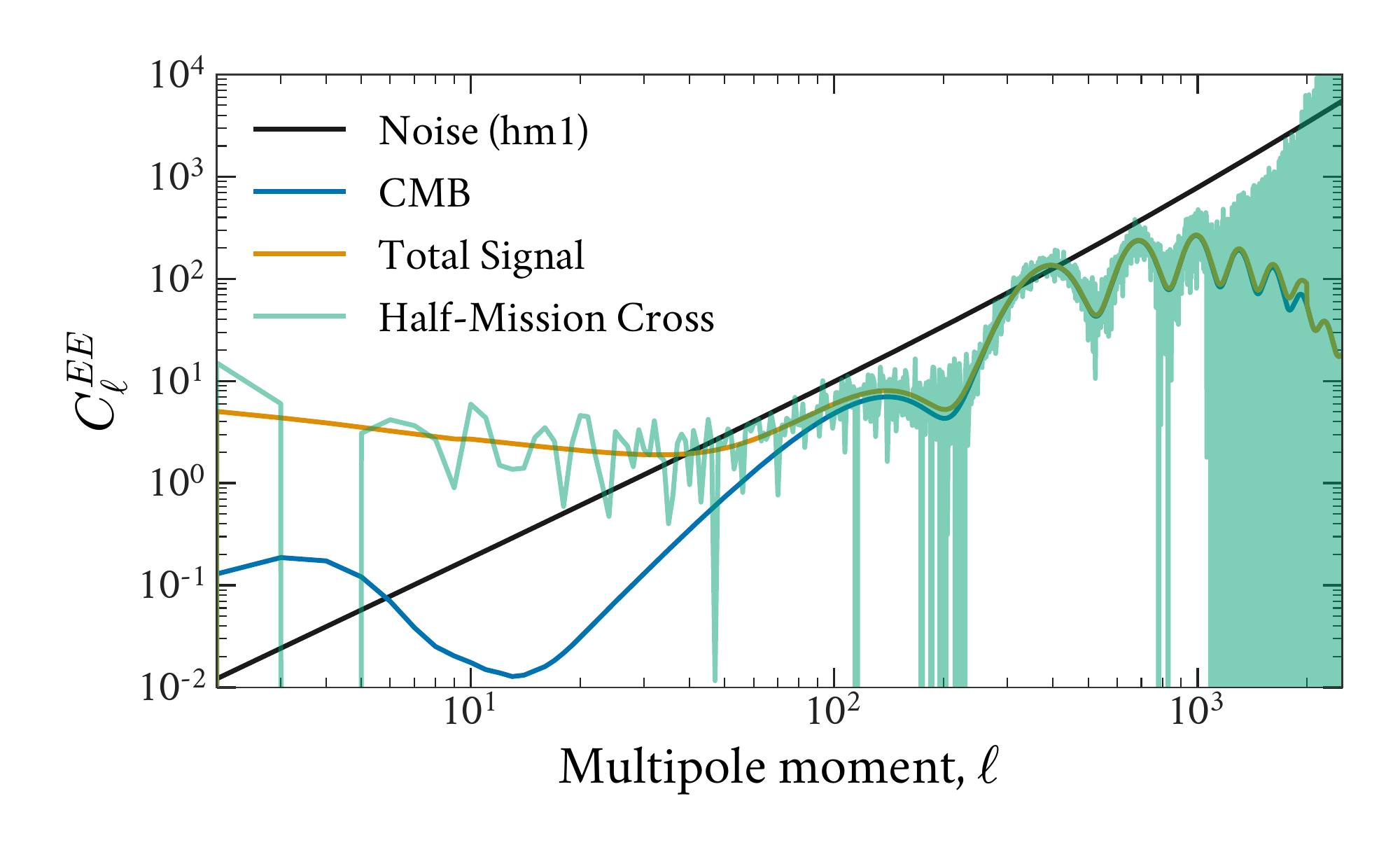}
\caption{ The $EE$ noise model (black) for half-mission $\rm hm1$ at 143~GHz, compared to the total signal (the sum of theory and foreground components) and the estimated half-mission cross-spectrum.}
\label{fig:ee_noise}
\end{figure}

\begin{table*}
    \resizebox{0.9\textwidth}{!}{
\begin{tabular}{ccccccccc}
 \hline \hline
freq & split & $A$ & $B$ & $\alpha$ & $\beta$ & $\ell_c$ & $\gamma$ & $\delta$ \\ \hline 
100 & 1 & $0.000849$ & $0.705$ & $0.117$ & $0.654$ & $5 \cdot 10^{5}$ & $8.51 \cdot 10^{-11}$ & $7.82$\\
100 & 2 & $0.000849$ & $0.705$ & $0.11$ & $0.654$ & $5 \cdot 10^{5}$ & $8.51 \cdot 10^{-11}$ & $7.82$\\
143 & 1 & $0.000255$ & $0.15$ & $1.75 \cdot 10^{-5}$ & $1.96$ & $8.23 \cdot 10^{3}$ & $2.13$ & $8$\\
143 & 2 & $0.00024$ & $0.289$ & $0.00848$ & $2.13 \cdot 10^{-11}$ & $5.49 \cdot 10^{5}$ & $-0.286$ & $2.52$\\
217 & 1 & $0.000734$ & $1.91$ & $0.000454$ & $0$ & $1 \cdot 10^{4}$ & $2.12$ & $-0.135$\\
217 & 2 & $0.000579$ & $3.41$ & $0.000434$ & $0$ & $7.3 \cdot 10^{3}$ & $6$ & $6$\\
\hline%
100 & 1 & $0.00103$ & $0.0985$ & $6.29 \cdot 10^{-5}$ & $-0.805$ & $309$ & $1.46$ & $-1.28$\\
100 & 2 & $0.00138$ & $0.268$ & $0.000703$ & $0.93$ & $5 \cdot 10^{5}$ & $-7.73 \cdot 10^{-8}$ & $-7.86 \cdot 10^{-8}$\\
143 & 1 & $0.000407$ & $0.877$ & $0.000823$ & $0.0725$ & $1.16 \cdot 10^{3}$ & $6$ & $-0.0188$\\
143 & 2 & $0.000891$ & $0.313$ & $0.000357$ & $0.486$ & $1.14 \cdot 10^{3}$ & $-6$ & $0.0132$\\
217 & 1 & $1.24 \cdot 10^{-8}$ & $8$ & $0.000852$ & $0.541$ & $813$ & $-0.799$ & $-1.25$\\
217 & 2 & $4.26 \cdot 10^{-6}$ & $6.38$ & $7.27 \cdot 10^{-9}$ & $2.38$ & $3.43 \cdot 10^{4}$ & $-0.365$ & $-8$\\
\hline\hline
\end{tabular}
}
    \vspace{0.5em}
    \caption{Noise model parameters for the $TT$ (top) and $EE$ (bottom) power spectra, for each frequency and half-mission map. We emphasize that the estimation of \planck noise spectra is a challenging problem, particularly at large scales, and this noise model we present here is primarily an intermediary product used for this power spectrum analysis. They are representative of, but do not fully characterize, the \planck noise properties. These model parameters provide a smooth fit to the estimated noise spectra (see Appendix Figure \ref{fig:TT_noise_models} and \ref{fig:EE_noise_models}). This full table is available in the \href{https://github.com/simonsobs/PSpipe/blob/planckcov/project/Planck_cov/input/noisecoeffs.dat}{PSpipe repository}.}
    \label{tab:noise_params}
\end{table*}


We require sufficiently accurate estimation of the $TT$, $TE$ and $EE$ noise power spectra to compute the contribution of noise to the covariance matrices. 
Here we use the model described in \citet{planck_like2013, efstathiou/gratton:2019} for $N_{\ell}$, assuming the \planck noise power spectrum follows a smooth power-law behavior at small and large scales:
\begin{equation}
    N_{\ell}^{\mathrm{model}} = A \, \left( \frac{100}{\ell} \right)^{\alpha} + B \, \frac{(\ell/1000)^{\beta}}{(1+(\ell / \ell_c)^{\gamma})^{\delta}}.
    \label{eqn:noise}
\end{equation}
This differs from the \texttt{plik} model used in PL20, which uses a more general polynomial expression. The \planck PL20 analysis relied on jointly fitting the large-scale term for both half-mission maps at a given frequency by using the half-mission half-ring difference maps, which can be used to estimate noise power spectra for each half-mission map. This two-stage fitting process makes it easier to use the general expression used in 
PL20. 

These half-mission half-ring difference maps are not publicly available, so we use the half-mission maps instead, together with this simpler model. We find that our analysis, which uses only the publicly available products, still reproduces the intermediate products and cosmology to high fidelity.

We compute the noise power spectrum for channel $X \in \{ TT, EE \}$ as the difference between the mask-deconvolved auto and cross-spectra,
\begin{equation}
    \begin{split}
        N_{\ell}^{X, \mathrm{hm1} \times \mathrm{hm1}} &=  C_{\ell}^{X, \mathrm{hm1} \times \mathrm{hm1}} -  C_{\ell}^{X, \mathrm{hm1} \times \mathrm{hm2}}, \\
        N_{\ell}^{X, \mathrm{hm2} \times \mathrm{hm2}} &=  C_{\ell}^{X, \mathrm{hm2} \times \mathrm{hm2}} -  C_{\ell}^{X, \mathrm{hm1} \times \mathrm{hm2}}.
    \end{split}
\end{equation}
We note that the noise in our convention is not modified by the beam. Since the half-mission maps have very slightly different beams  (their difference is $\sim 10^{-4}$ the value of the beam transfer function), we compute these half-mission differences by correcting for the beam within the auto- and cross-spectra, differencing the spectra, and then re-applying the beam appropriately for that half-mission.
We also check that we reproduce the qualitative differences reported in PL20 between noise power spectra estimated from either half-mission or half-ring maps. 

In order to estimate the parameters of the noise model in Equation \ref{eqn:noise}, we optimize a Gaussian likelihood with a diagonal covariance matrix that simply weights modes as $N_{\ell}/\sqrt{2\ell+1}$ when obtaining the noise model fit. This likelihood is then optimized through an adaptive differential evolution optimizer \citep{wang2014} implemented in \texttt{BlackBoxOptim.jl} \citep{bbopt}. 
We report the noise model parameters for each of the \planck half-mission frequency maps in Table \ref{table:noise}, and plot the resulting noise curve ratios in Figure~\ref{fig:noise_ratio}. 
The total noise for $EE$ at 143~GHz is shown in Figure~\ref{fig:ee_noise}, where we also show various contributions to the signal spectrum in comparison to the noise.
We show the estimated noise spectra and our smooth fits to those spectra in Figure~\ref{fig:TT_noise_models} and Figure~\ref{fig:EE_noise_models}.

The \planck noise power spectra are challenging to determine, even in polarization. The \planck polarization maps are individually noise-dominated at most scales, allowing for a more accurate determination of the noise power spectrum. In contrast, the temperature maps contain substantial signal, so the estimation of the noise power spectrum is more challenging. There is therefore uncertainty in the large-scale noise power spectra shown in Figure~\ref{fig:noise_ratio}, that is difficult to characterize using the available public data products from the \planck collaboration. We note however that, at these scales, the noise contribution to the covariance matrix is subdominant and thus does not affect cosmological constraints; these noise curves are therefore a sufficient intermediate product for the power spectrum estimation. We show this behavior in Figure~\ref{fig:signal_to_noise}, where we plot the signal-to-noise ratios from some of the \planck frequency maps.

\begin{table}
\label{table:noise}
\hspace{-1em}
\begin{tabular}{cccc}
\hline\hline
freq & split & TT [$10^{-3}\,\mu$K$^2$] & EE [$10^{-3}\,\mu$K$^2$] \\ \hline 
100 & 1 & 1.06583 & 2.47488\\
100 & 2 & 1.01859 & 2.38146\\
143 & 1 & 0.208448 & 0.965957\\
143 & 2 & 0.194149 & 0.893841\\
217 & 1 & 0.409548 & 2.16995\\
217 & 2 & 0.374725 & 1.91914\\
\hline\hline
\label{tab:effectivewhitenoiselevel}
\end{tabular}
\caption{The amplitude of the white noise power spectrum $N_{\ell}$ estimated for the \planck frequency maps. We include more significant figures than necessary to aid in reproducibility.} 

\end{table}




\subsection{Point source treatment} \label{subsec:pstreat}
\begin{figure}
\centering
\includegraphics[width=0.5\textwidth]{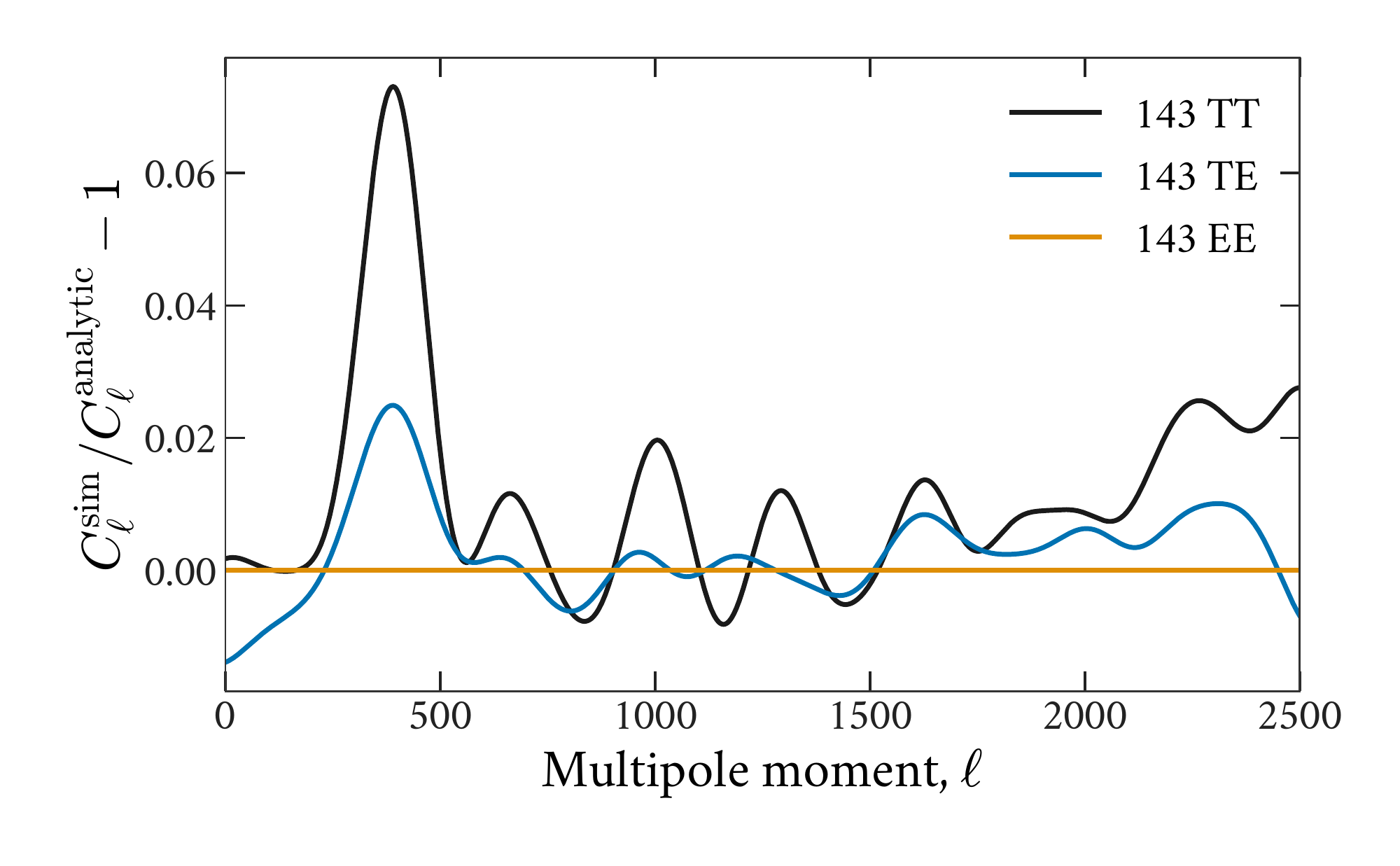}
\caption{The ratio of the covariance estimated from signal-only simulations compared to the analytic estimate. This ratio is non-unity due to the presence of sharply-masked point sources in the simulations. This correction ratio is applied to the fiducial power spectrum used in the data covariance matrix.}
\label{fig:ps_correct}
\end{figure}

Point sources violate the approximation in the covariance matrix that the spectra of the masks are declining rapidly in power towards small scales. The apodization around each point source required to satisfy this assumption would result in a substantial reduction in signal. To accommodate the low apodization, we instead follow PL20 by computing signal-only simulations of each cross-spectrum used in the analysis. Using only the fiducial signal power spectrum $C_{\ell}$, we obtain a correction ratio by dividing the predicted analytic covariance diagonal with that from the simulations. This correction ratio is applied to the fiducial power spectrum in the covariance estimation. This is the same approach as in PL20, but we generate our own simulations for this purpose. 

In Figure~\ref{fig:ps_correct} we show the correction ratio for each multipole estimated using this method at 143~GHz, based on a smooth interpolation of the ratio estimates from 1000 simulations using a Gaussian Process with an exponential kernel. PL20 used a spline fit, but the two different non-parametric models should produce nearly identical results. We distribute these correction ratios and code to generate these correction factors within \texttt{PSPipe}. 







\begin{figure}
\centering
\includegraphics[width=0.5\textwidth]{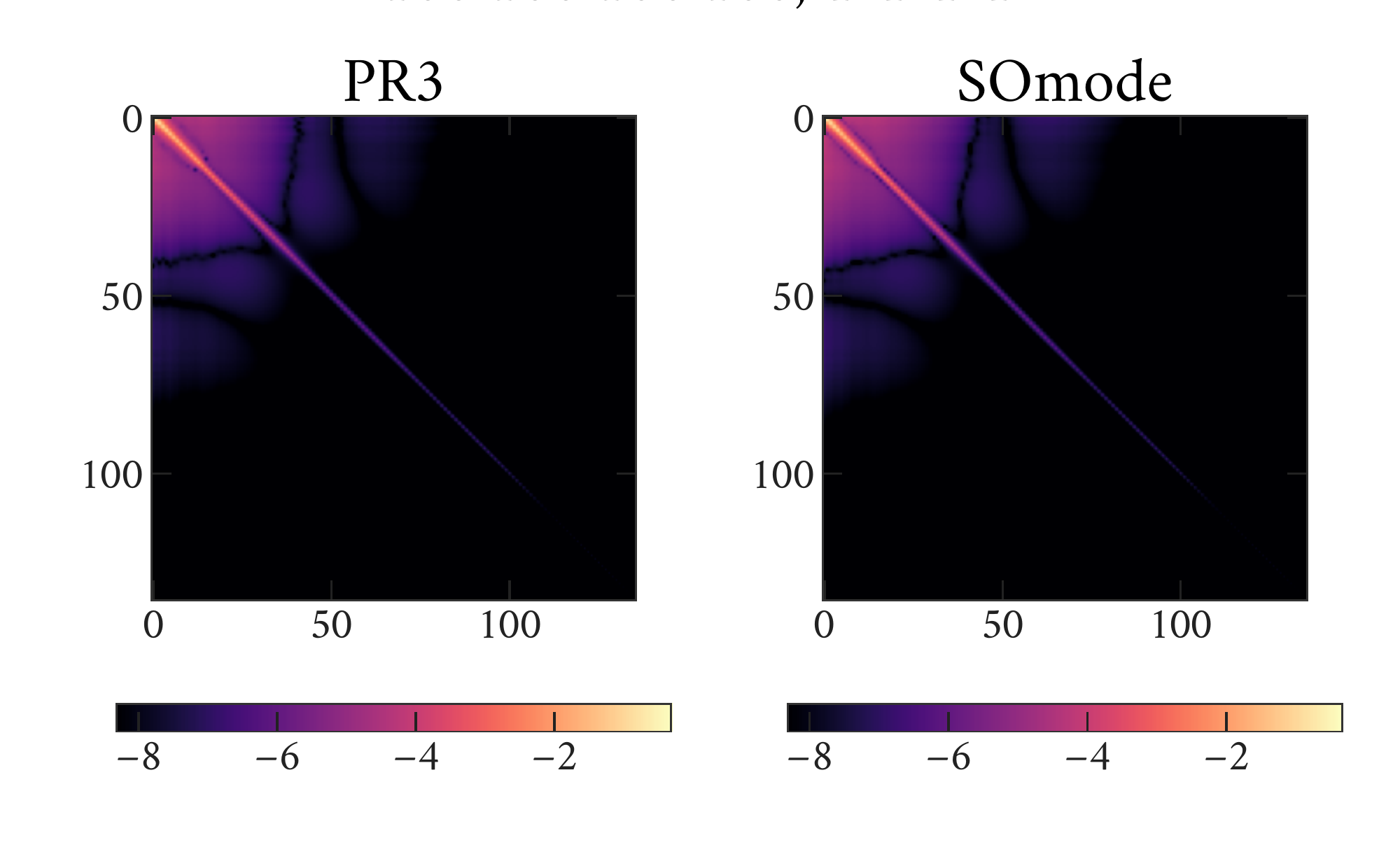}
\includegraphics[width=0.5\textwidth]{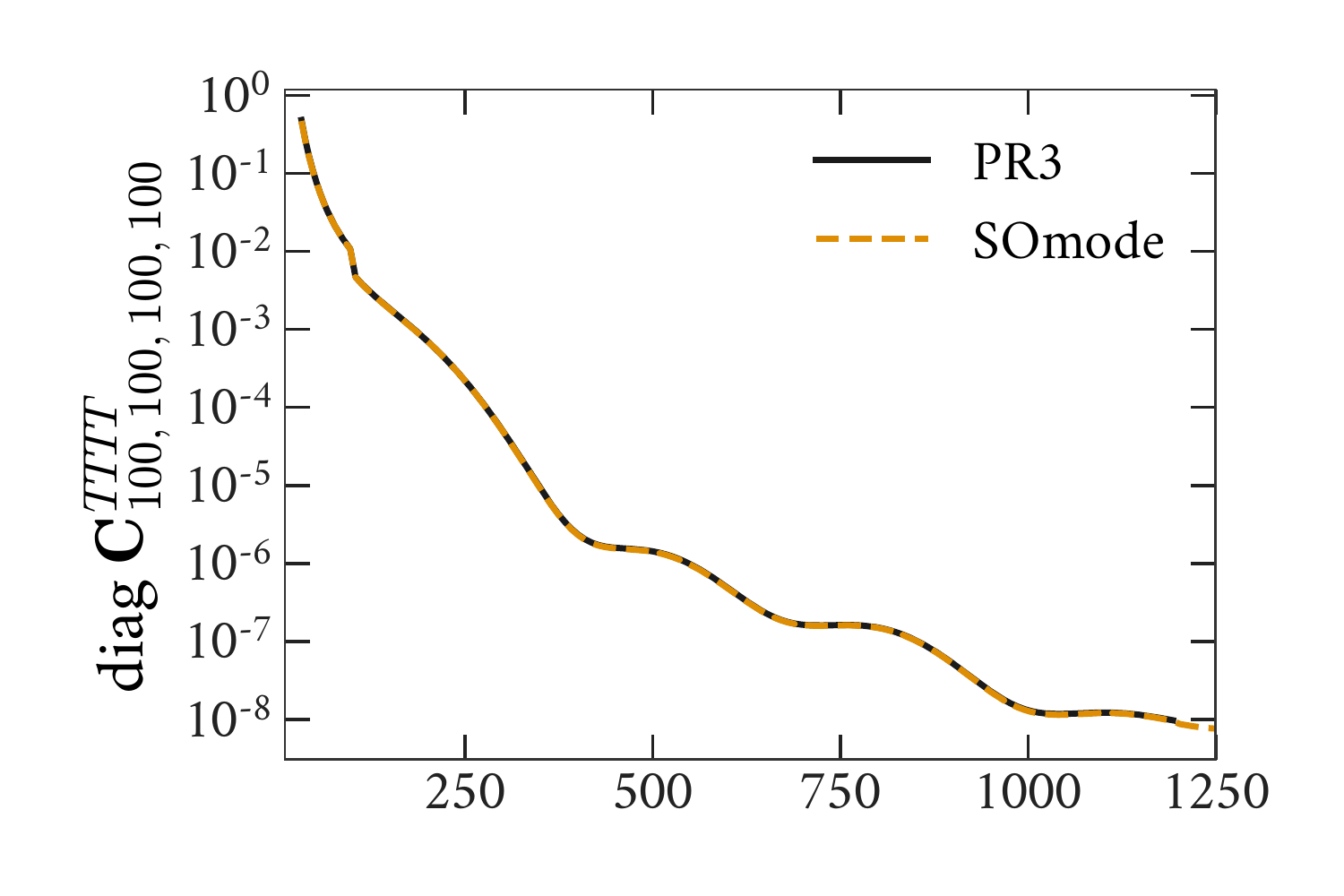}
\includegraphics[width=0.5\textwidth]{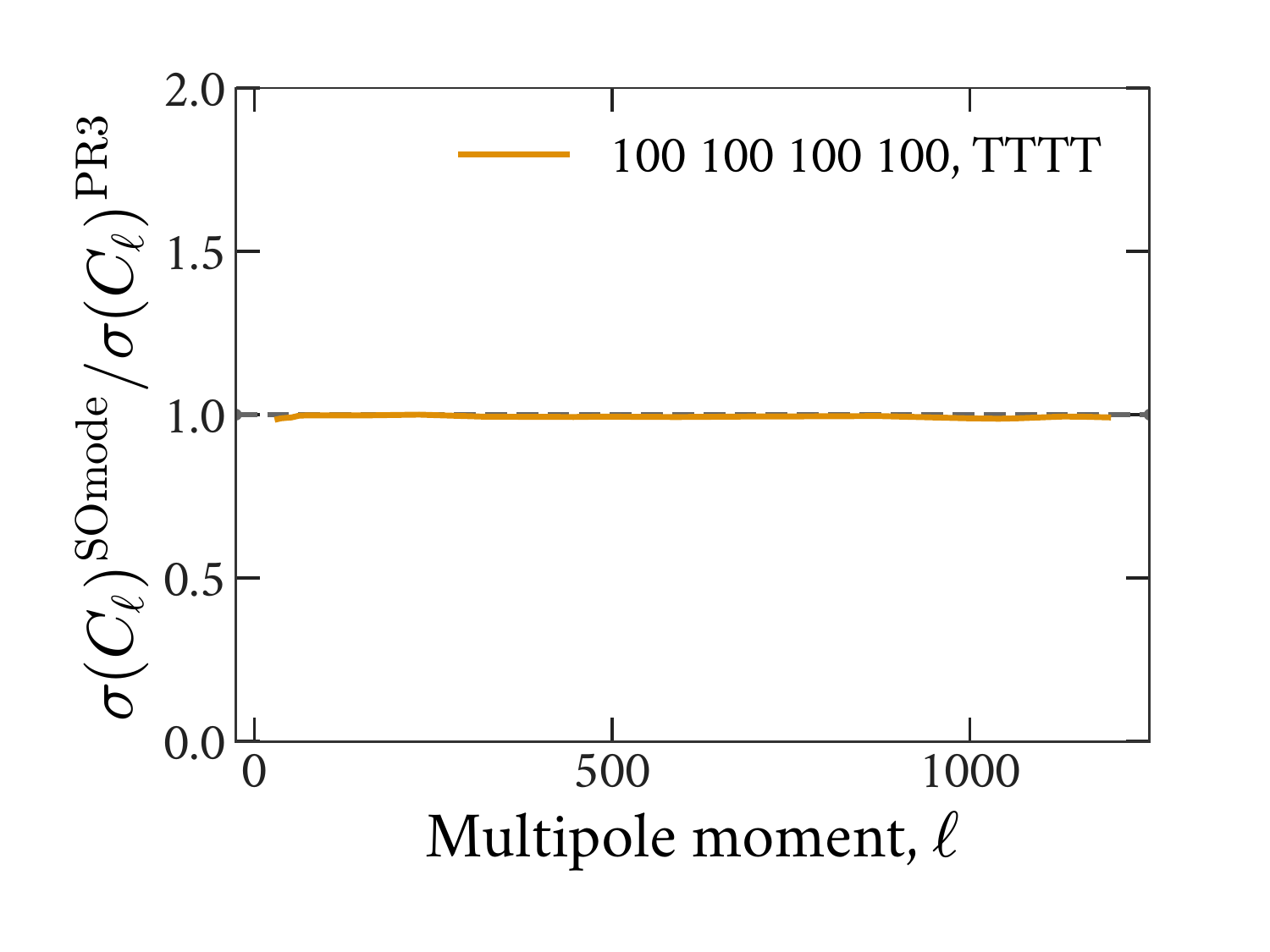}
\caption{A comparison of the temperature covariance sub-blocks for cross-spectra between the half-missions. We note the excellent agreement in both the diagonal and off-diagonal structure. In the heatmaps above, we show the logarithm of the absolute value of the covariance matrix elements. }
\label{fig:tt_sub}
\end{figure}

\begin{figure}[h]
\centering
\includegraphics[width=0.5\textwidth]{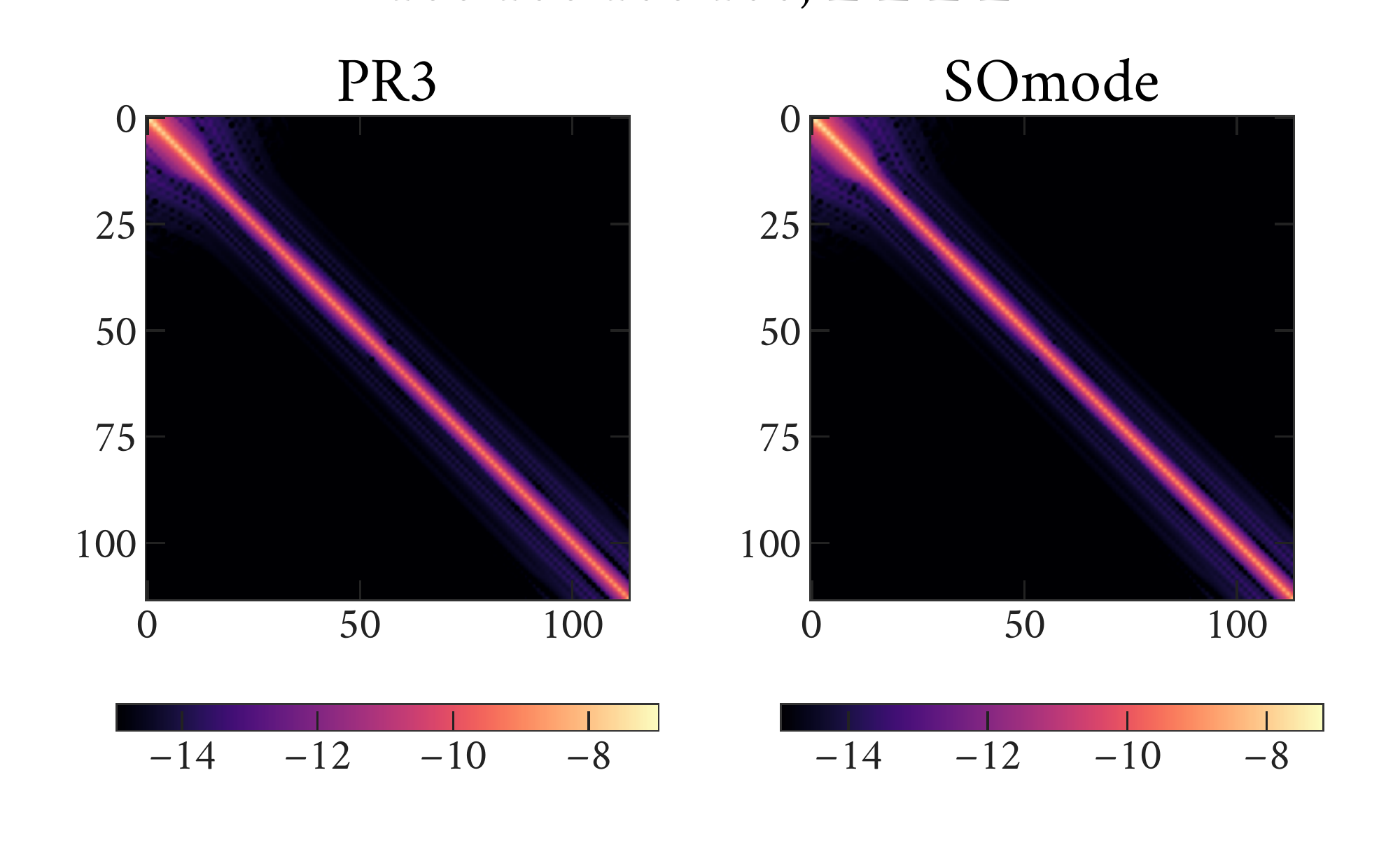}
\includegraphics[width=0.5\textwidth]{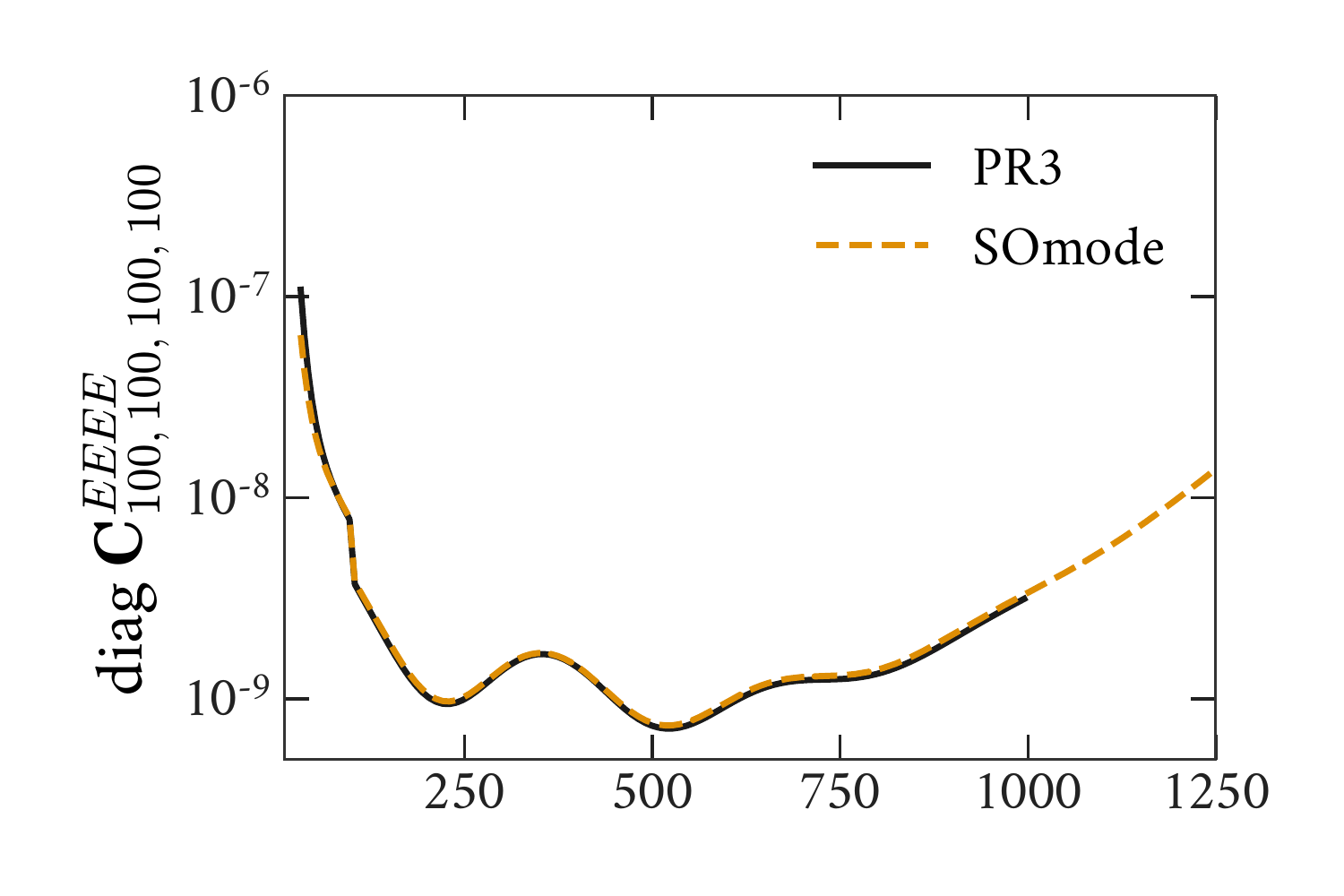}
\includegraphics[width=0.5\textwidth]{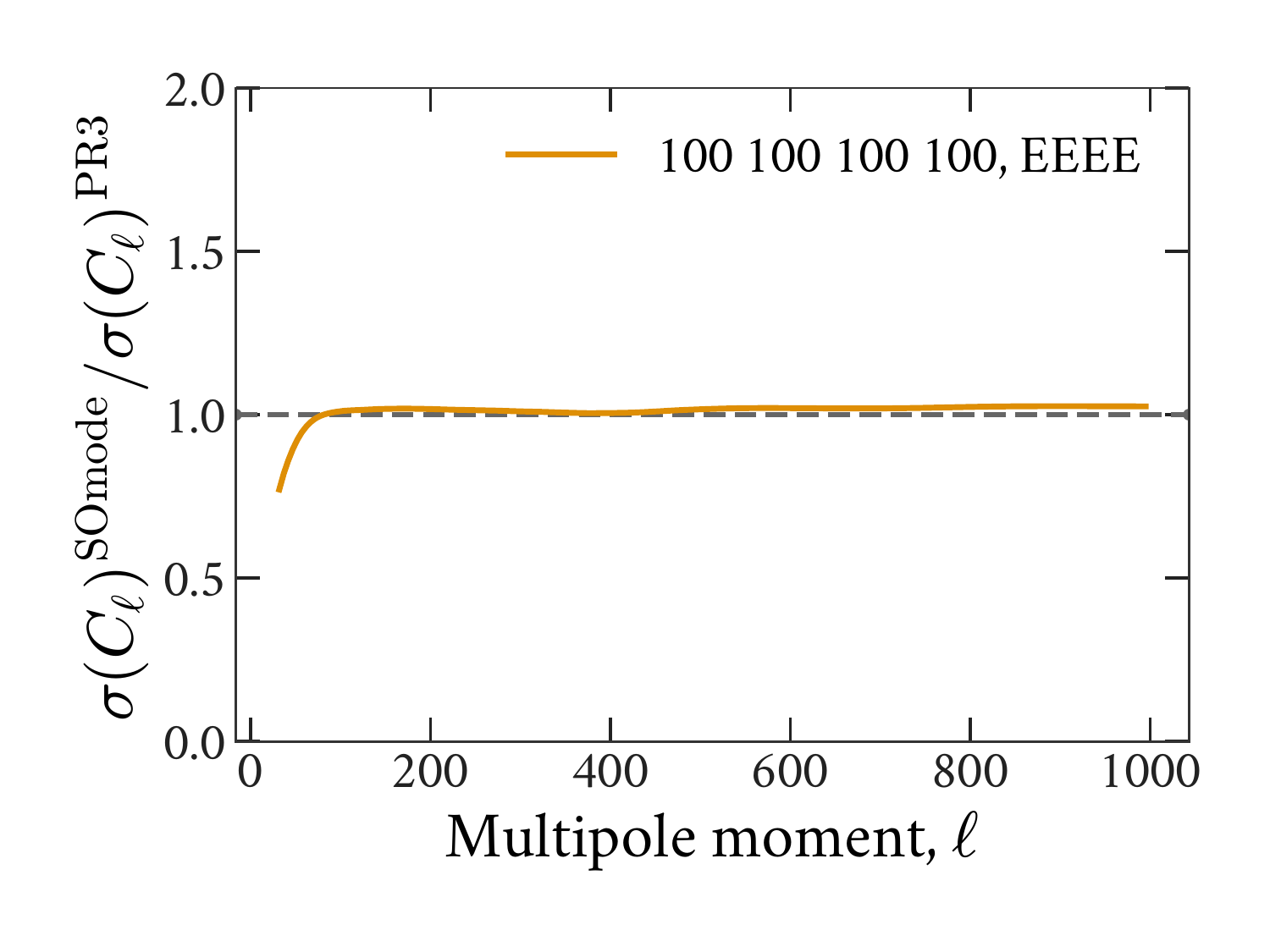}
\caption{A comparison of the E-mode square root of the covariance sub-block diagonals for cross-spectra between the half-missions. Our best candidate for the origin of this low-$\ell$ discrepancy is the  noise power spectrum estimate. We used the difference of the auto- and cross-spectra from the half-mission frequency maps, rather than the half-mission half-ring maps used in the PL20 analysis.}
\label{fig:ee_sub}
\end{figure}

\subsection{Comparison to PL20 covariances}

\begin{figure*}
\centering
\includegraphics[width=\textwidth]{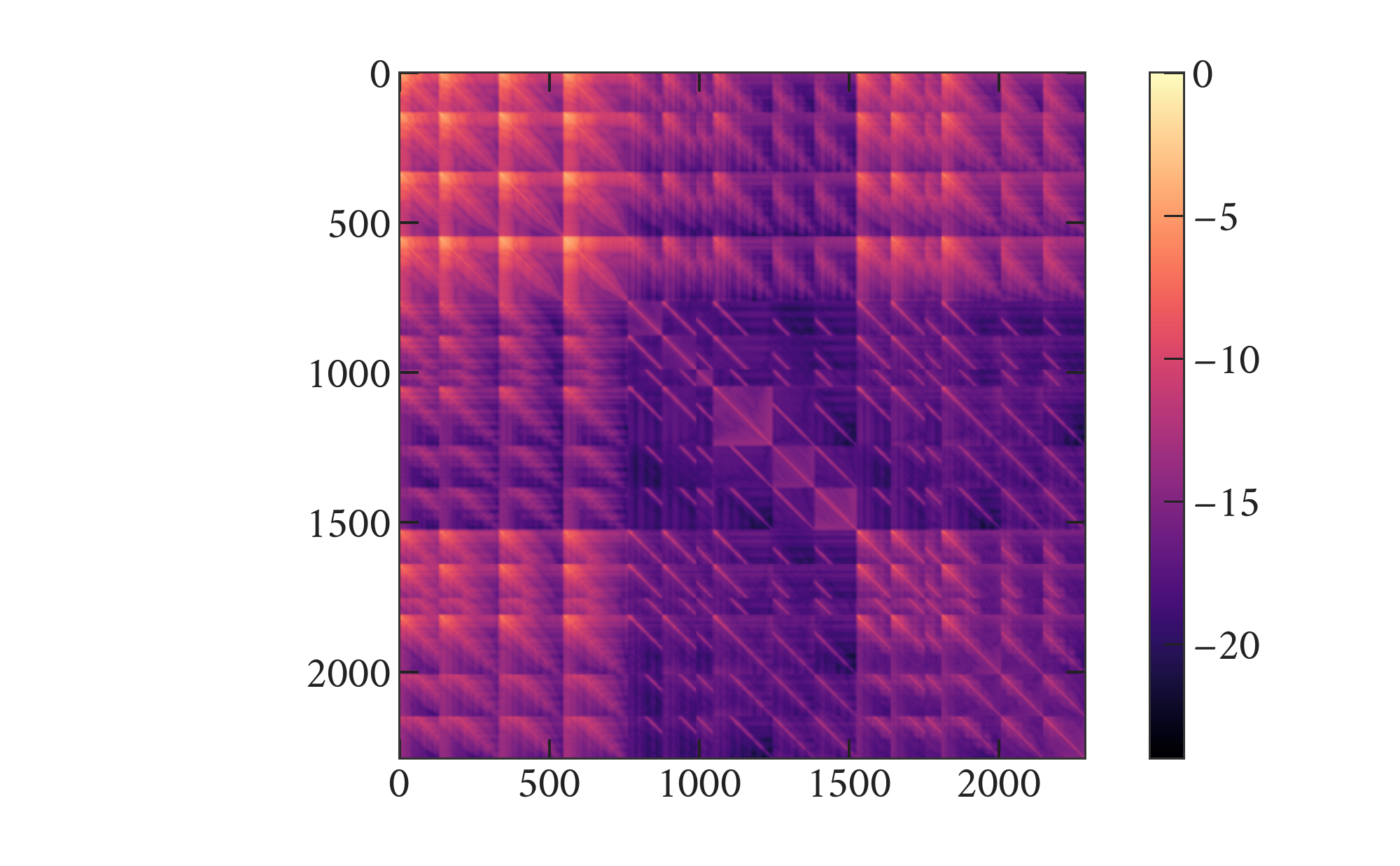}
\includegraphics[width=\textwidth]{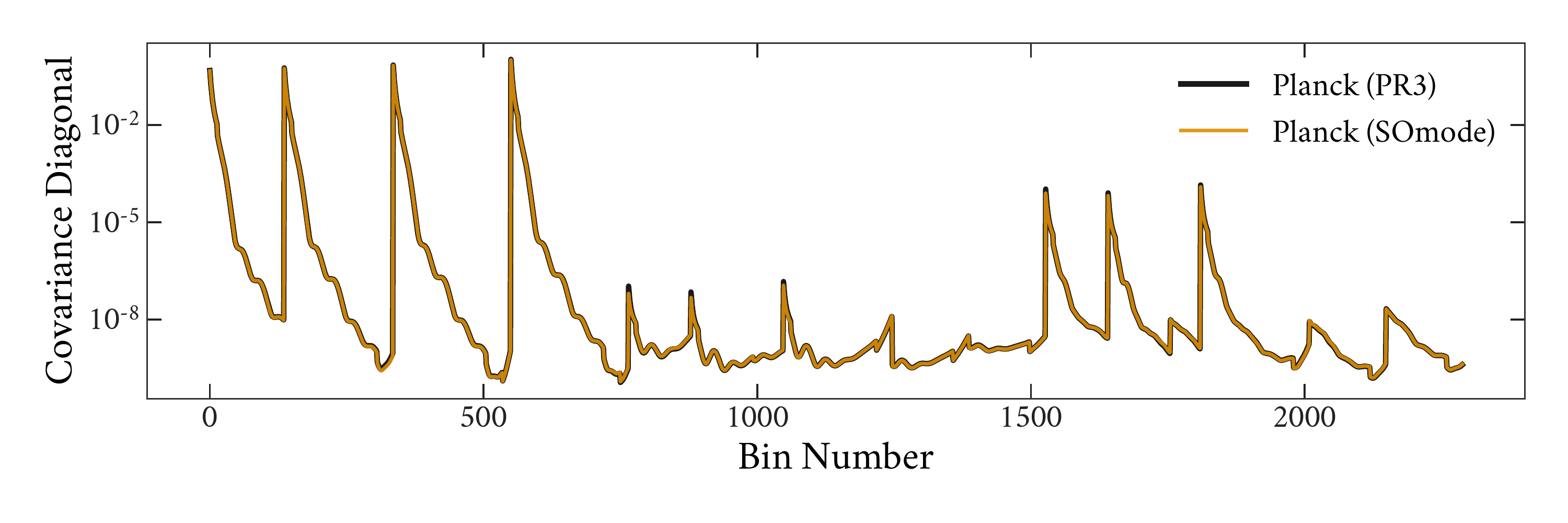}
\caption{(Top) The full covariance matrix for the \planck 2018 spectra, with (bottom) a comparison of the covariance matrix diagonals for \planck and our estimate. The data vector has 2289 elements and is ordered overall by $TT$, $EE$, $TE$. The ordering by frequency is given in Table~\ref{table:multipoleranges}.}
\vspace{1em}
\label{fig:covmat}
\end{figure*}

A comparison of the \planck PL20 and our covariance matrix for the 100~GHz $TT$ and E-mode covariance sub-blocks are shown in Figures~\ref{fig:tt_sub} and \ref{fig:ee_sub}. The length of the data vector in both cases is the number of multipole bins at this frequency, with $\ell_{\rm max}=1250$ at 100~GHz. The middle panels show the diagonal elements of the matrices as a function of multipole, with the ratio of the two estimates in the lower panels. The $TTTT$ covariance matrices agree on the diagonal to better than 6\% for all frequencies. The $EEEE$ covariance also shows good agreement, with a deviation at the largest scales; our method has a lower estimate at $\ell<100$ by $\sim$ 30\%. The off-diagonal structure also shows consistent behavior. The full covariance matrix, and comparison of the diagonals for the full data vector, is shown in Figure~\ref{fig:covmat}.


\section{Likelihood}
\label{sec:likelihood}
The Simons Observatory collaboration have also developed likelihood software to return the likelihood of the power spectra given some theoretical model, and we use it here to analyse these new inputs. 

The modeling of the \planck high-$\ell$ multi-frequency power spectra is described in detail in PL20 and ~\cite{planck_like2013, planck_like2015}, with PL20 presenting the modeling of the PR3 spectra and differences/improvements with respect to the previous releases. Here, our goal is to test the re-estimated spectra and covariance matrices, so we use the same modeling  assumptions used in the PL20 reference likelihood \texttt{Plik}, as summarized below. We have implemented this in \texttt{MFLikePlik}\footnote{\url{https://github.com/simonsobs/LAT_MFLike/tree/mflike-plik}}, a branch of the SO multi-frequency likelihood for the power spectrum analysis of the Large Aperture Telescope \texttt{MFLike}\footnote{\url{https://github.com/simonsobs/LAT_MFLike}}. 

The data vector of the spectra, $\mathbf{C_d}$, and its covariance matrix $\mathbf{\Sigma}$ from Sections~\ref{sec:spectra} and~\ref{sec:covmat} are used to form a Gaussian likelihood:
\begin{equation}
  \begin{split} 
 -2\ln{\cal L}(\mathbf{C_d} | \mathbf{C_{t}}(\theta)) = \left[\mathbf{C_d} -\mathbf{C_{t}}(\theta)\right]^{T} {\mathbf{\Sigma}}^{-1}
 \left[\mathbf{C_d} - \mathbf{C_{t}}(\theta)\right] \\  \quad + {\rm const.}\;, 
 \end{split}
\label{eqn:like}
\end{equation}
where $\mathbf{C_{t}}(\theta)$ is the theoretical prediction as function of cosmological, foreground and nuisance parameters $\theta$. For each cross-frequency binned spectrum (between frequencies $i$ and $j$) in temperature or polarization, this can be written as:
\begin{equation}
\mathbf{C_{t}^{ij}}(\theta)= \mathbf{C_t^{CMB}}(\theta_1) + \mathbf{C_t^{{sec},ij}}(\theta_2) + \mathbf{C_t^{{sys},ij}}, \label{eq:likemodel}
\end{equation}
where $\mathbf{C_t^{CMB}}$ is a binned CMB theory prediction for a given set of cosmological parameters $\theta_1$, and the other two terms described in more detail below account for frequency-dependent foregrounds and residual systematics, respectively. The spectra in PL20 are only binned at the last step of processing, resulting in a top-hat bandpower window function, followed by a $\ell (\ell+1)$ weighting within each bin. This binning operation is performed on the sum of the CMB theory and foreground components, to compute the likelihood.

\emph{Foregrounds:} $\mathbf{C_t^{sec,ij}}(\theta_2)$ is the secondary signal from Galactic and extragalactic emission and depends on 20 foreground parameters $\theta_2$. In temperature this term includes thermal and kinetic Sunyaev-Zel'dovich (SZ) effects, dusty star-forming and radio galaxies appearing as point sources, a clustering term for the cosmic infrared background (CIB), a correlation between SZ and CIB, and thermal dust emission from our Galaxy. In polarization the Galactic dust emission is modeled for both $TE$ and $EE$. We refer the reader to PL20 for the modeling of all these components in the likelihood, and in particular to Table 16 of PL20 for the definition of the $\theta_2$ parameters, their range of variation and the priors imposed on them. For our implementation of these components in \texttt{MFLikePlik} we adopt \texttt{fgspectra}\footnote{\url{https://github.com/simonsobs/fgspectra}}, an SO library that builds cross-spectra predictions for foreground components at different frequencies for a given set of model parameters. 

\emph{Systematics:} $\mathbf{C_t^{sys,ij}}$ is a term correcting the model for residual levels of systematic effects. A module in \texttt{MFLikePlik} includes the PL20 frequency-dependent templates for beam leakage, sub-pixel noise corrections, and correlated noise corrections to the model vector with fixed amplitudes. We do not re-estimate these templates in this analysis.

\emph{Calibration and polarization efficiencies:} The full theory vector is then calibrated and corrected for polarization efficiencies before comparing to the data.
As in PL20, the model is calibrated assuming a fixed temperature calibration equal to 1 at 143~GHz, and polarization efficiencies also fixed to 1.021, 0.966, and 1.04 for 100, 143, and 217\,GHz, respectively. The 100 and 217~GHz temperature calibration factors are added as nuisance parameters as a third component of the $\theta$ vector and varied with the same Gaussian priors used in PL20.

All the data products including the new spectra and covariance inputs presented here, the binning and weighting schemes, and the PL20 templates for systematic effects, are stored as plain text files in the \texttt{MFLikePlik} data folder. Additional templates needed to model foregrounds are stored in the \texttt{fgspectra} library.

\section{Tests of parameters}
\label{sec:params}
\begin{figure*}
\centering
\includegraphics[width=\textwidth]{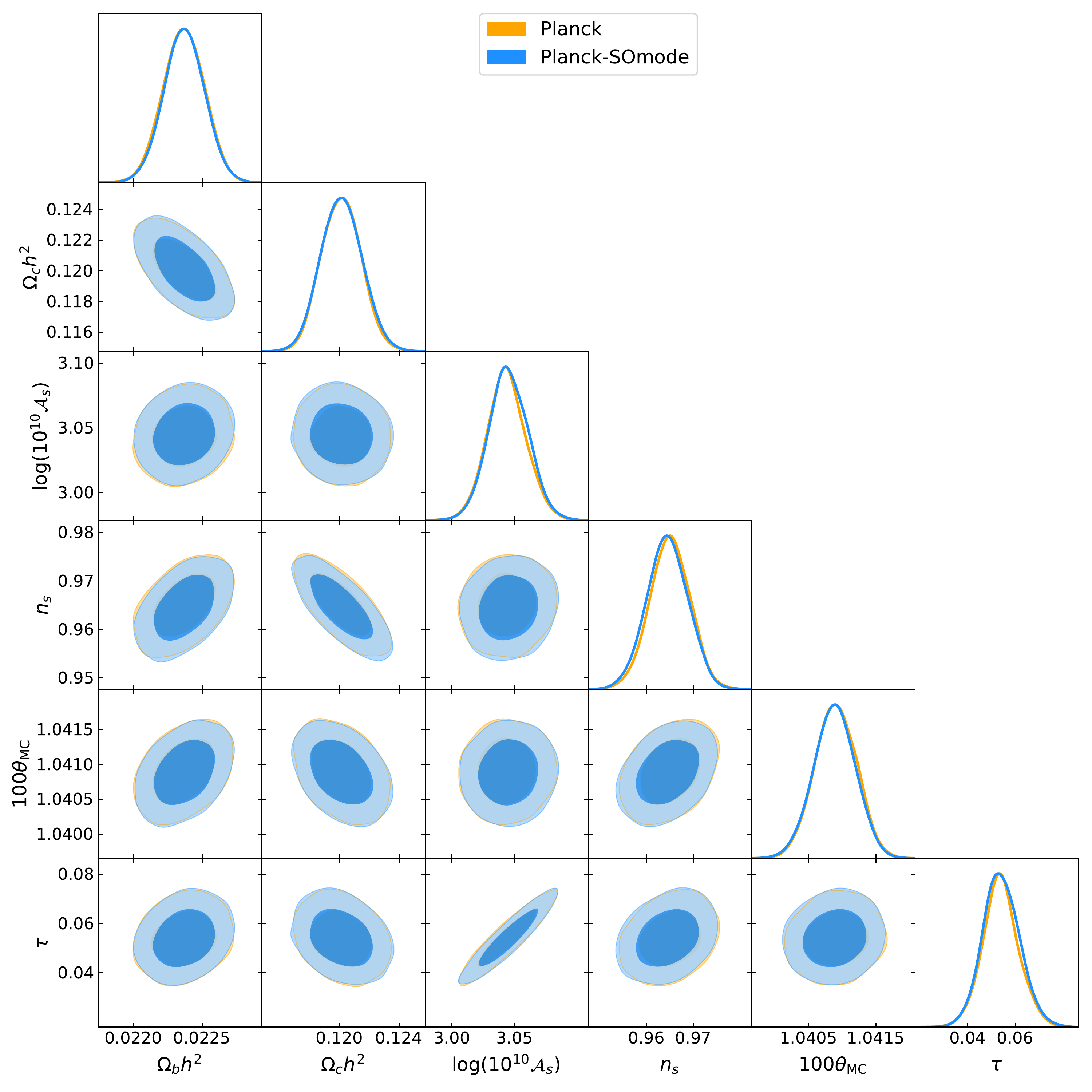}
\caption{Constraints on the $\Lambda$CDM cosmological parameters derived from the spectra and covariance matrix estimated in this work (blue), compared to the \planck\ legacy results (orange). The diagonal panels show the 1-dimensional posterior distributions; the contour plots show the 2-dimensional 68\% and 95\% confidence levels. The two pipelines give constraints that agree to within 0.1$\sigma$.}
\label{fig:lcdmpar}
\end{figure*}
Our re-estimated spectra and covariance products, which we will term \texttt{SOmode}, have been directly compared to the \planck PR3 products in previous sections. Here we look at the last step of the cosmological pipeline and test how cosmological parameters derived from our new inputs compare with those presented in PC20 for the basic $\Lambda$CDM model as well as a selection of extensions.

To do this we couple our likelihood to the publicly-available \texttt{Cobaya}\footnote{\url{https://cobaya.readthedocs.io/en/latest/}} code \citep{torrado/lewis:2021}. We extend the \texttt{Cobaya} \planck yaml files to call \texttt{MFLikePlik} for the high-$\ell$ spectra and we also include the PR3 likelihood implementations, wrapped in \texttt{Cobaya}, of the \planck `lowl' and `lowE' likelihoods modeling the large-scale temperature and $EE$ polarization data. 

\begin{figure*}
\centering
\includegraphics[width=12cm]{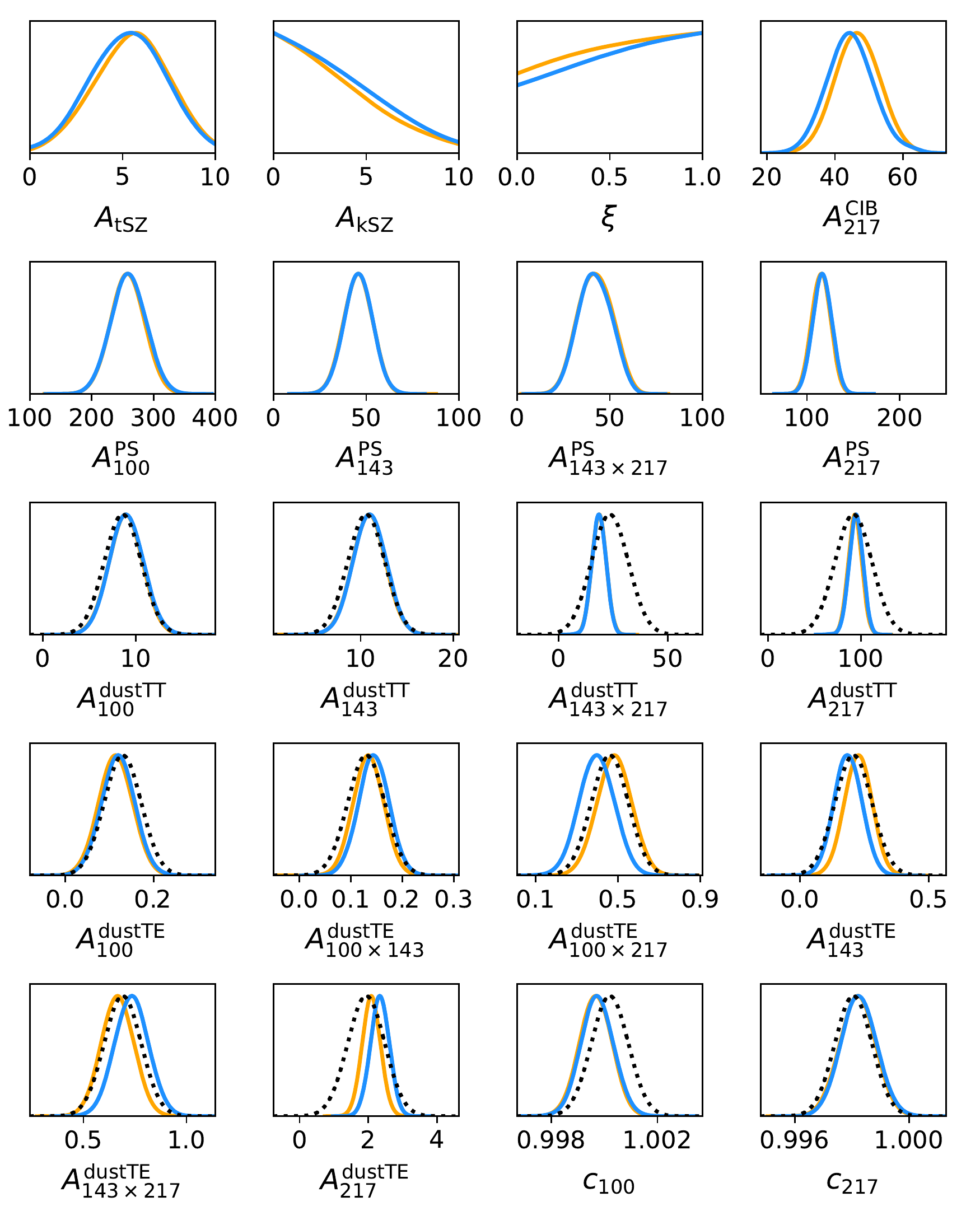}
\caption{Foreground and calibration parameters derived from the spectra and covariances estimated in this work (blue), compared the \planck\ legacy results (orange) -- from left to right and top to bottom these show: thermal and kinetic SZ amplitudes, the correlation between thermal SZ and CIB, the CIB amplitude, four point source amplitudes, four dust amplitudes in $TT$ and six in $TE$, and two calibrations (see PL20 for more details). The dotted black lines show the Gaussian priors imposed on individual parameters. An additional joint prior is used for the thermal and kinetic SZ amplitudes and not shown here. The two pipelines give consistent results, with small shifts only in some of the $TE$ dust amplitudes. This is attributed to small differences in the $EE$ spectra at low multipoles.}
\vspace{1em}
\label{fig:fgpar}
\end{figure*}

We find excellent agreement for the six basic \LCDM\ parameters ($\Omega_b h^2$, $\Omega_c h^2$, $A_s$, $n_s$, $\theta_{\rm MC}$, $\tau$) between our inputs and PC20, as shown in Figure~\ref{fig:lcdmpar}, with all parameters agreeing to within 0.1$\sigma$. The distributions of the foreground parameters are shown in Figure~\ref{fig:fgpar}. In this case, all the parameters are consistent with those reported in PC20 within 0.1$\sigma$, except for the amplitudes of thermal dust in $TE$ at frequencies higher than 100~GHz. These exhibit some small shifts, at the level of $0.2-0.9\sigma$. As in PC20, the dust amplitudes are varied in $TT$ and $TE$ with a Gaussian prior imposed, and are fixed in $EE$. 
We attribute the small shifts in $TE$ parameters to the small differences in the $EE$ spectrum at low multipoles (see Sec.~\ref{sec:specs}) where the dust contamination is significant, at 143 and 217 GHz. More precisely, since the dust amplitudes are fixed in $EE$, any difference at the $EE$ spectrum level will be leveraged by the free-to-vary $TE$ parameters. We tested that these shifts were indeed coming from spectrum differences, rather than differences in the covariance matrix, by estimating parameters using the PR3 spectra together with our new \texttt{SOmode} covariance matrix. This combination resulted in no parameter shifts in the foreground parameters. 

The small differences in the covariance matrix described in Sec.~\ref{sec:covmat} do manifest as a small difference in $\chi^2$. For \LCDM, the \planck\ best-fit $\chi^2$ (as per the .likestats Getdist file) for \texttt{plikTTTEEE+lowl+lowE} is $2768.9$, while for our combination of \texttt{MFLikePlik+lowl+lowE} we find $2698.7$. This corresponds to a $\Delta \chi^2=70.2$ which is dominated by the \texttt{plik} $\chi^2$ ($\Delta \chi^2_{\texttt{plik}}=71.9$). Finding a lower $\chi^2$ is consistent with the slightly increased noise and covariance elements we found with our products at $\ell>400$. We confirmed this computing $\chi^2$ for subsets of the data, the $\ell>800$ range, and TE in particular, have the larger impact on the difference in $\chi^2$. The difference in the TE spectra is dominated by the choice of computing the flat average of the various TE and ET cross-half-mission spectra that combine for the final result, rather than inverse-variance weighting.

\begin{figure}
\centering
\includegraphics[width=6.4cm]{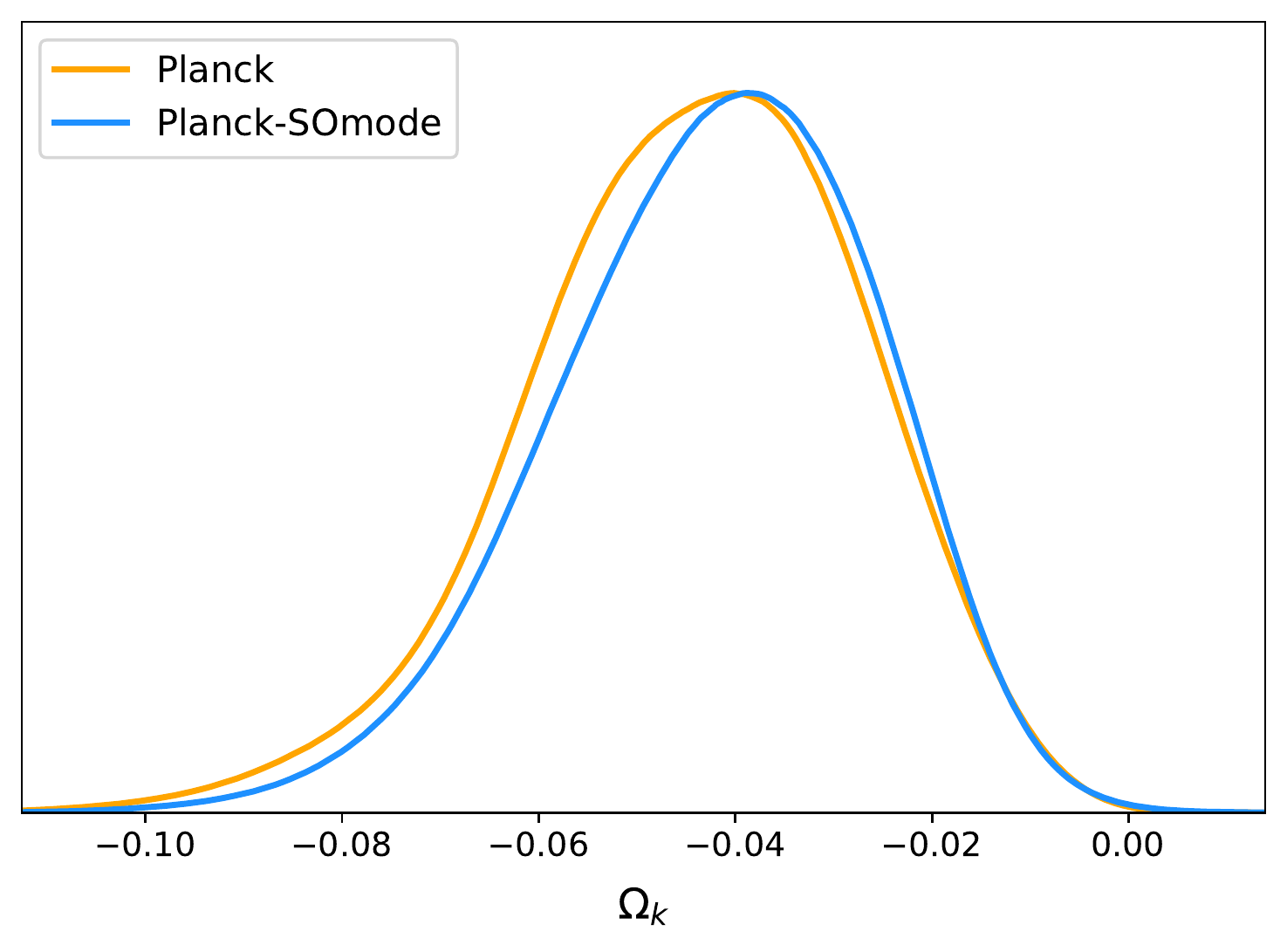}
\includegraphics[width=6.4cm]{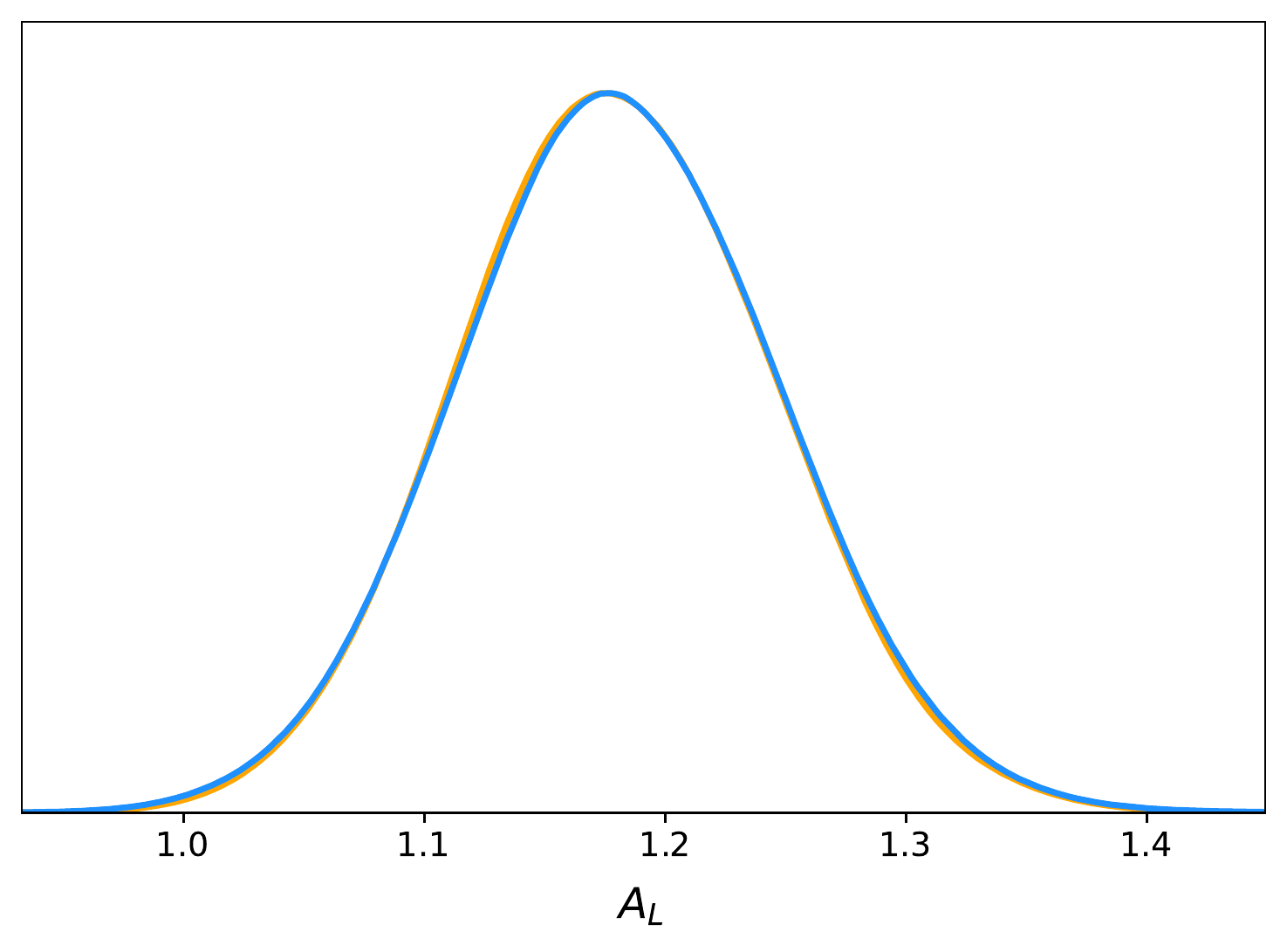}
\includegraphics[width=6.4cm]{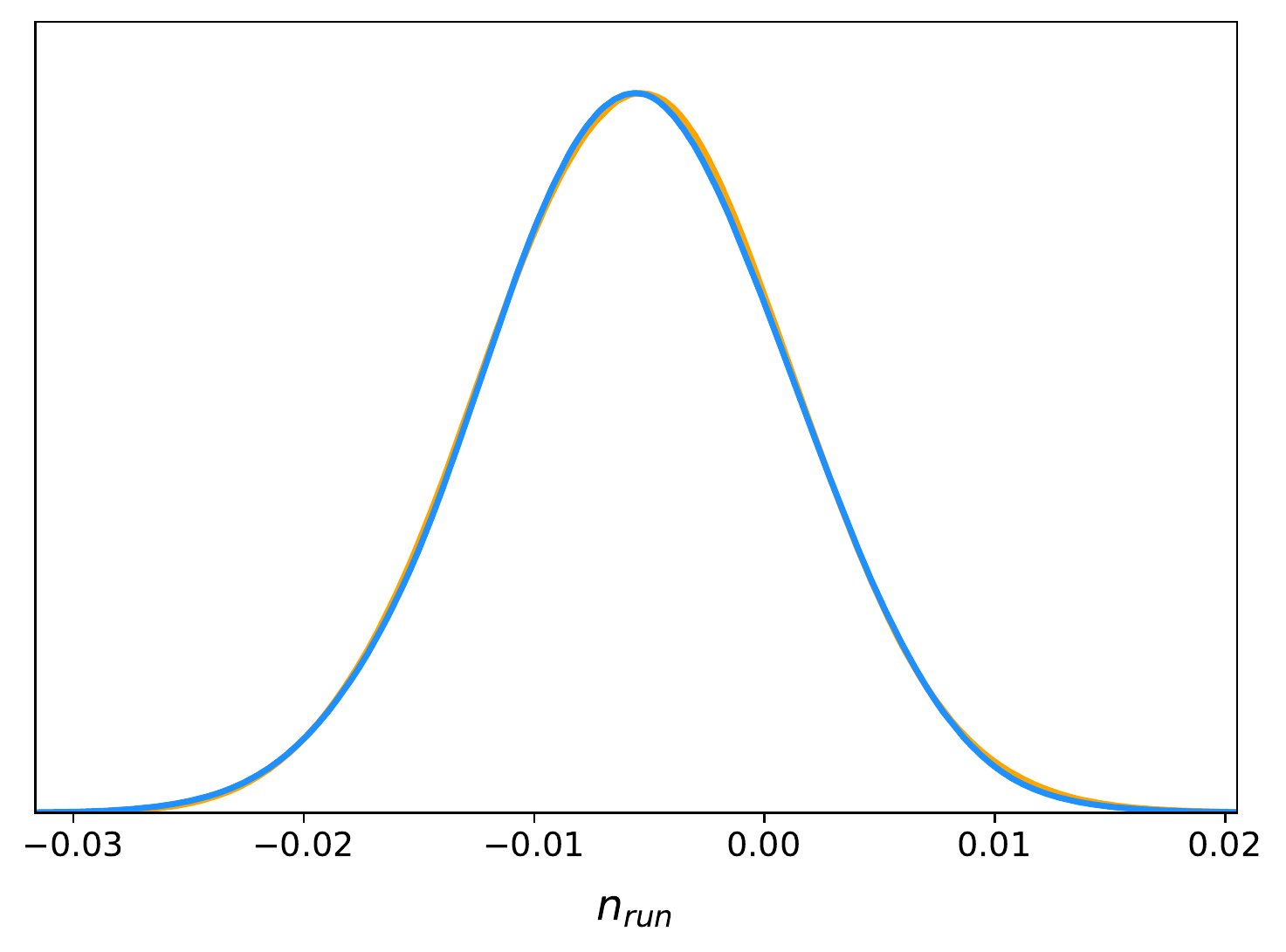}
\includegraphics[width=6.4cm]{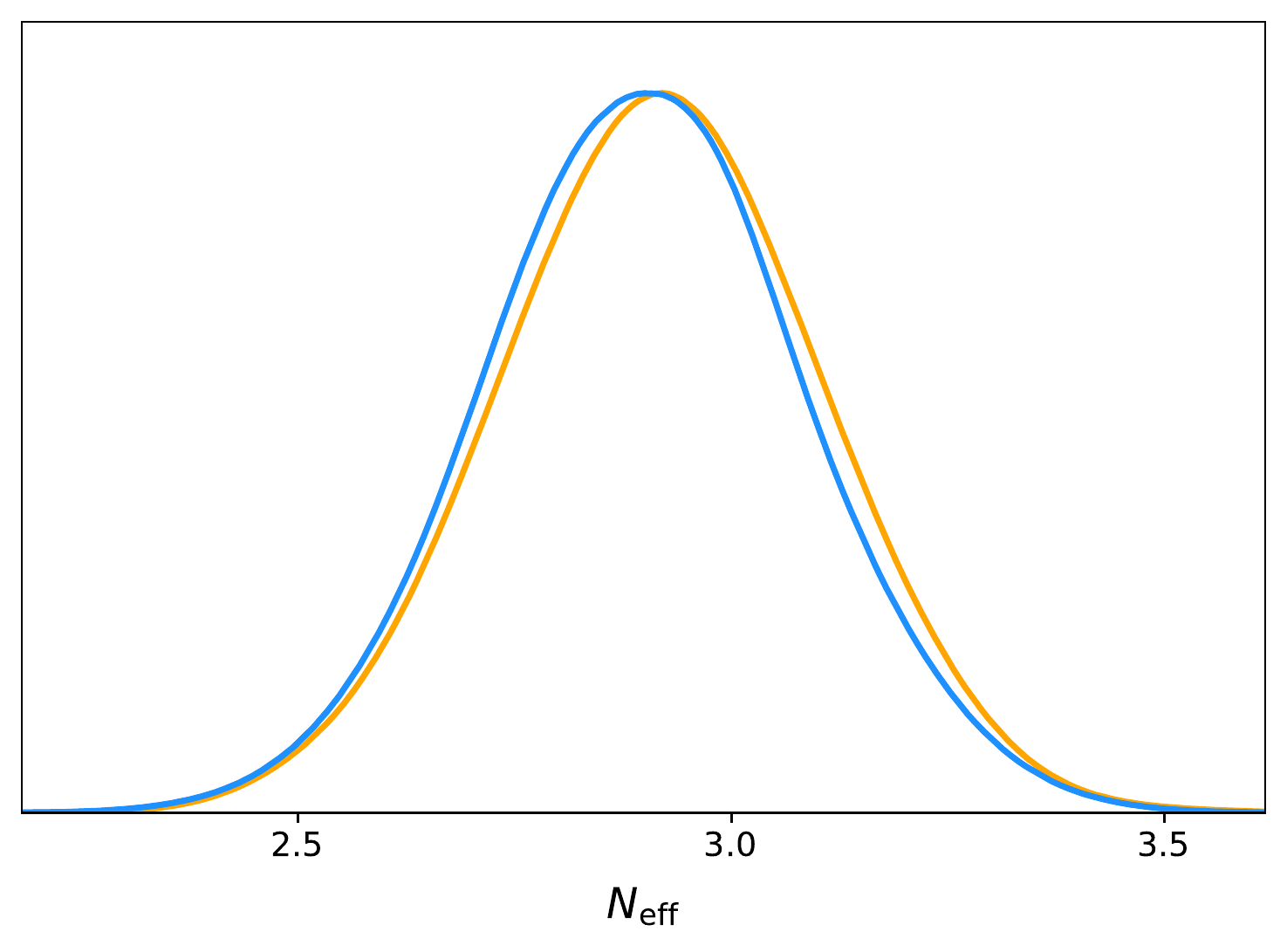}
\caption{Posterior distributions for four single-parameter extensions to the \LCDM\ model, derived using the spectra and covariance matrix estimated in this work (blue), compared to the \planck\ legacy results (orange). Only the extension parameter is shown for each model.}
\label{fig:extpar}
\end{figure}

We also explore a standard set of \LCDM\ extensions to validate our products on multiple theories/parametrizations. We estimate constraints on the effective number of relativistic species, $N_{\rm eff}$, the running of the spectral index, $n_{\rm run}$, the amplitude of lensing smoothing in the power spectra, $A_L$, and spatial curvature, $\Omega_k$. In all cases, we recover the PC20 results to within $0.1\sigma$ as shown in Figure~\ref{fig:extpar}.

\section{Conclusion}
\label{sec:conclude}

The \planck legacy dataset will be a critical part of cosmological constraints for at least the next decade. We expect exciting new data from existing and upcoming ground-based experiments 
to shed light on the unknown features of the cosmological model.
These data will have to be combined with \planck measurements of the large scale temperature modes in order to anchor parameters such as $A_s$, $n_s$, and $\tau$, in addition to {\it Planck}'s clean measurements of the first few acoustic peaks.

Drawing upon wisdom accrued from a decade of \planck analysis, we have 
reproduced
some of the main \planck cosmology products -- the CMB power spectrum and associated covariances, tested with the cosmological likelihood. This paper presents a complete, validated, and publicly available spectrum analysis, and we also provide all associated intermediary data products and software tools. We have reproduced the spectra to extremely high fidelity using the same tools we expect to apply to new ground-based data, and we construct covariance matrix estimates that lead to a remarkably consistent reproduction of the main \planck cosmology. We expect these to be broadly useful for making robust cosmological claims from the ground, and advancement past the era of {\it Planck}.

In this paper, we have provided a detailed and complete description of the scientific choices and approximations used in this pipeline, but we expect the code and scripts to provide substantial additional value for ongoing and future power spectrum analysis. However, future analysis will need to perform various necessary extensions to our work in order to adapt this pipeline to the particulars of their dataset.
\begin{itemize}
    \item We anticipate that spectra and covariances will require re-computation of the \planck spectra and covariances on new survey masks. We try to make substitution of masks in our pipeline straightforward.
    \item The azimuthally-averaged \planck beam varies based on sky patch due to the inhomoegenous scan pattern. We expect the future power spectrum analyst to also require new beams computed using \texttt{QuickPol} for \planck on their particular sky patch.
    \item The foreground model will change in future analysis, and that model is an input for the covariance matrix calculation as well as used in the likelihood.
\end{itemize}
We aim to make these tasks straightforward within the framework of \texttt{PSPipe} and \texttt{MFLike}, although re-estimating the Galactic contributions as inputs for the likelihood is beyond the scope of this current study.



A secondary aim of our analysis has been to provide an independent check of the \planck analysis. The excellent agreement we find with the primary \planck power spectrum analysis adds confidence in the implementation. 
Although the pipeline we present in this paper has hewed closely to the PR3 release, we did  experiment with choices such as methods for monopole/dipole subtraction, for masking missing pixels, using \texttt{PolSpice} versus other pseudo-$C_{\ell}$ codes, choices of noise modeling, and the choice of approximations in the estimation of the covariance matrix, all of which have order $\sim 1-10\%$ effect on the resulting products. The \planck dataset is complex, and details such as the computation of noise power spectra can have large effects on outputs like null tests. For example, the $\chi^2$ depends quadratically on the noise power spectrum through the $\mathcal{R}_{\ell}$ terms.

We note that \texttt{NPIPE}, an alternative analysis of \planck time-ordered data, is now publicly available. We are in the process of adapting our pipeline to these data, and leave a detailed analysis for future work. Future work will also incorporate adaptations for using the \planck data with current and future ground-based surveys.


\section{\label{sec:ack}Acknowledgments}

We are grateful to George Efstathiou and Steven Gratton for sharing useful information about the \planck analysis. We also thank Antony Lewis for helpful feedback. This work was performed using Princeton Research Computing resources, a consortium of groups including the Princeton Institute for Computational Science and Engineering (PICSciE) and the Office of Information Technology's High Performance Computing Center and Visualization Laboratory at Princeton University, as well as the Hawk high-performance computing cluster at the Advanced Research Computing at Cardiff (ARCCA). Research in Canada is supported by NSERC and CIFAR. JD gratefully acknowledges support from the Institute for Advanced Study and from NSF grant AST-2108126. DA is supported by the Science and Technology Facilities Council through an Ernest Rutherford Fellowship, grant reference ST/P004474. EC acknowledges support from the STFC Ernest Rutherford Fellowship ST/M004856/2 and STFC Consolidated Grant ST/S00033X/1; EC, HJ and UN acknowledge additional support from the Horizon 2020 ERC Starting Grant (Grant agreement No 849169).
GF acknowledges the support of the European Research Council under the Marie Sk\l{}odowska Curie actions through the Individual Global Fellowship No.~892401 PiCOGAMBAS.
SKC acknowledges support from NSF award AST-2001866.
Research at Perimeter Institute is supported in part by the Government of Canada through the Department of Innovation, Science and Industry Canada and by the Province of Ontario through the Ministry of Colleges and Universities.
P.D.M acknowledges support from the Netherlands organization for scientific research (NWO) VIDI grant (dossier 639.042.730).
Some of the results in this paper have been derived using the HEALPix~\citep{gorski/etal:2005}, healpy~\citep{healpy}, and Healpix.jl~\citep{tomasi2021} packages. We also acknowledge use of the matplotlib~ \citep{hunter:2007} package for producing plots in this paper. 

\bibliographystyle{aasjournal} 
\bibliography{main}



\appendix

\section{Spectrum residuals}
\begin{figure*}
\centering
\includegraphics[width=\textwidth]{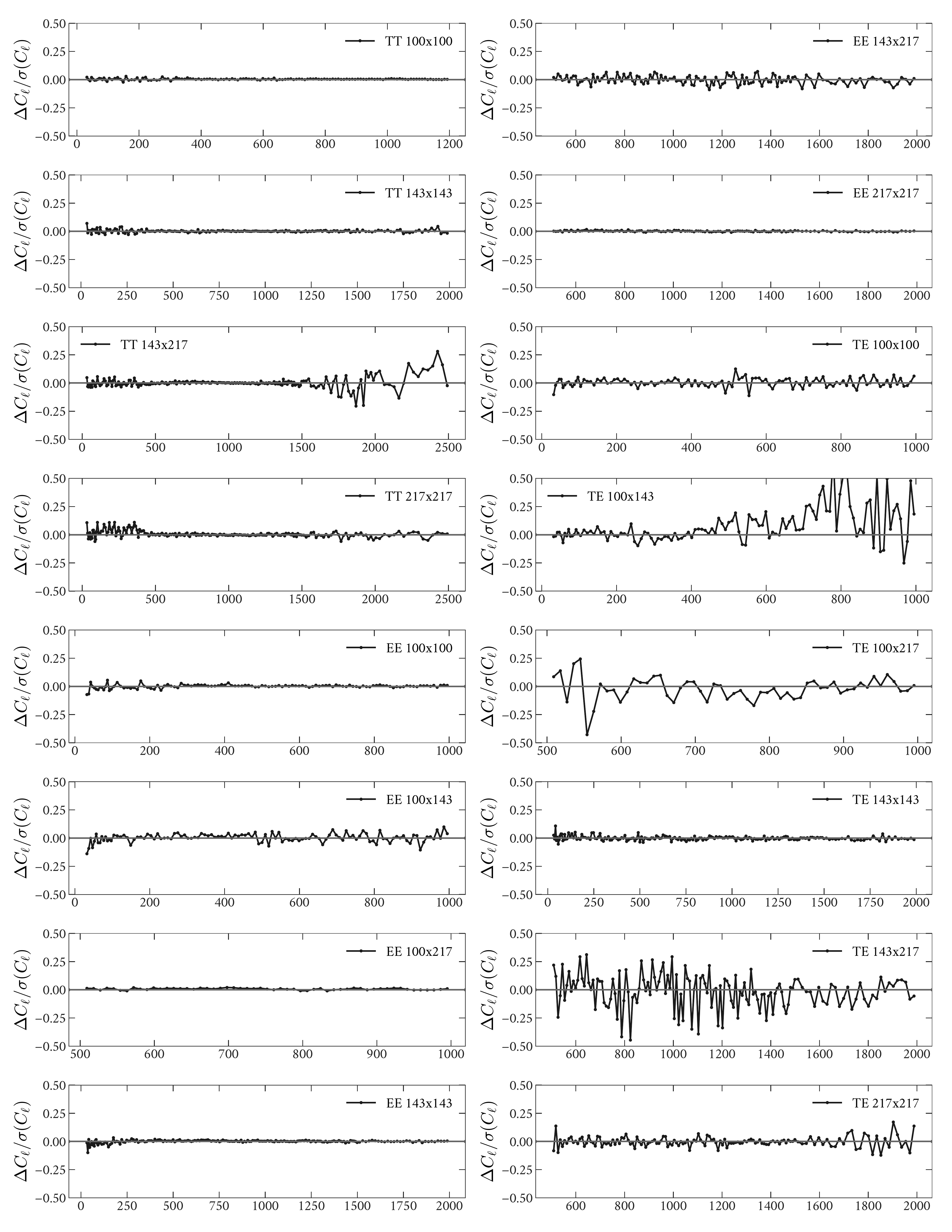}
\caption{All cross-spectra used in the likelihood, showing the residual of our Planck-SOmode estimates compared to the spectra from the official PR3 analysis.}
\label{fig:all_resid}
\end{figure*}

We show the full suite of residuals of our spectra compared to the PL20 public spectra in Figure~\ref{fig:all_resid}.

\section{Covariance Matrix Functions}

Following PL16 and PL20 we use index names
\begin{equation}
\begin{split}
    \alpha, \beta &\in \{ TT, TP, PT, PP\} \\
    X, Y &\in \{ \emptyset\emptyset, TT, PP \}.
\end{split}
\end{equation}
Here, indices $\alpha,\beta$ refer to the kind of cross-spectrum (i.e., cross-spectra between spin 0 and 2, spin 2 and 2, etc.). In contrast, $X,Y$ refer to the kind of noise-variance weighting applied to the mask. $X=\emptyset\emptyset$ refers to the mask alone, $X = TT$ refers to weighting by the $II$ variance, and $PP$ refers to weighting by the $QQ$ and $UU$ variance.

PL16 give the projector functions $\Xi^{X,Y}_{AB}\left[ (i,j)^{\alpha}, (p,q)^{\beta} \right]$, describing mode-coupling from the mask for maps $i,j,p,q$ on channel $AB \in \{TT, TE, EE\}$. They are given in Appendix C1 of PL16 and we repeat them here for completeness:
\begin{equation}
\begin{split}
    \Xi_{TT}^{X,Y} \left[ (i,j)^{\alpha}, (p,q)^{\beta} \right]_{\ell_1 \ell_2} &= \sum_{\ell_3} \frac{2 \ell_3 + 1}{4\pi} \begin{pmatrix}
\ell_1 & \ell_2 & \ell_3\\
0 & 0 & 0
\end{pmatrix} \\
&\times W^{X,Y} \left[ (i,j)^{\alpha}, (p,q)^{\beta} \right]_{\ell_3},
\end{split}
\end{equation}
\begin{equation}
\begin{split}
\Xi&_{TE}^{X,Y} \left[ (i,j)^{\alpha}, (p,q)^{\beta} \right]_{\ell_1 \ell_2} = \sum_{\ell_3} \frac{2 \ell_3 + 1}{8\pi} \left( 1 + (-1)^{\ell_1 + \ell_2 + \ell_3}\right)\\ &\times \begin{pmatrix}
\ell_1 & \ell_2 & \ell_3\\
0 & 0 & 0
\end{pmatrix} \begin{pmatrix}
\ell_1 & \ell_2 & \ell_3\\
-2 & 2 & 0
\end{pmatrix} 
W^{X,Y} \left[ (i,j)^{\alpha}, (p,q)^{\beta} \right]_{\ell_3},
\end{split}
\end{equation}
\begin{equation}
\begin{split}
\Xi&_{EE}^{X,Y} \left[ (i,j)^{\alpha}, (p,q)^{\beta} \right]_{\ell_1 \ell_2} = \sum_{\ell_3} \frac{2 \ell_3 + 1}{16\pi} \left( 1 + (-1)^{\ell_1 + \ell_2 + \ell_3}\right)^2\\ &\times \begin{pmatrix}
\ell_1 & \ell_2 & \ell_3\\
-2 & 2 & 0
\end{pmatrix}^2
W^{X,Y} \left[ (i,j)^{\alpha}, (p,q)^{\beta} \right]_{\ell_3}.
\end{split}
\end{equation}
The PL20 analysis ignores the effect of the BB spectrum in the covariance matrix estimates; we also state the corresponding BB projector here for completeness, which is also used in \cite{alonso/etal:2019}. Ignoring the conversion of $EE$ into $BB$ from coupling is only of order $5\%$ in $EE$ spectra at $\ell < 100$, and this is corrected from the diagonal through simulations, as we discuss later in this section.
\begin{equation}
\begin{split}
\Xi&_{BB}^{X,Y} \left[ (i,j)^{\alpha}, (p,q)^{\beta} \right]_{\ell_1 \ell_2} = \sum_{\ell_3} \frac{2 \ell_3 + 1}{16\pi} \left( 1 - (-1)^{\ell_1 + \ell_2 + \ell_3}\right)^2\\ &\times \begin{pmatrix}
\ell_1 & \ell_2 & \ell_3\\
-2 & 2 & 0
\end{pmatrix}^2
W^{X,Y} \left[ (i,j)^{\alpha}, (p,q)^{\beta} \right]_{\ell_3}.
\end{split}
\end{equation}
These projector functions depend on the cross-spectra $W$ between the masks, and require a Wigner 3j symbol that comes from integrals over the spherical harmonic basis. Note that the 3j symbols\footnote{We implement these 3j symbols in a fast and stable set of recurrence relations within the publicly available package \\\texttt{WignerFamilies.jl}.} are subject to a selection rule, such that the nonzero terms in these summations are limited to $|\ell_1 - \ell_2| \leq \ell_3 \leq \ell_1 + \ell_2$. 

The window functions $W^{X,Y} \left[ (i,j)^{\alpha}, (p,q)^{\beta} \right]$ in the projector functions, which are cross-spectra between the masks, are given in Appendix C2 of PL16; we repeat them here. 
\begin{equation*}
W^{\emptyset\emptyset,\emptyset\emptyset} \left[ (i,j)^{\alpha}, (p,q)^{\beta} \right]_{\ell} = \frac{1}{2\ell+1} \sum_m w_{\ell m}^{\emptyset \emptyset}(i,j)^{\alpha} w_{\ell m}^{*\emptyset \emptyset}(p,q)^{\beta}
\end{equation*}
\begin{equation*}
W^{\emptyset\emptyset,TT} \left[ (i,j)^{\alpha}, (p,q)^{TT} \right]_{\ell} = \frac{1}{2\ell+1} \sum_m w_{\ell m}^{\emptyset \emptyset}(i,j)^{\alpha} w_{\ell m}^{*II}(p,q)^{TT} 
\end{equation*}
\begin{equation*}
W^{TT,TT} \left[ (i,j)^{TT}, (p,q)^{TT} \right]_{\ell} = \frac{1}{2\ell+1} \sum_m w_{\ell m}^{II}(i,j)^{TT} w_{\ell m}^{*II}(p,q)^{TT} 
\end{equation*}
\begin{equation*}
\begin{split}
W^{\emptyset\emptyset,PP} \left[ (i,j)^{\alpha}, (p,q)^{PP} \right]_{\ell} = \frac{1}{2\ell+1} \sum_m \hspace{8em}  \\
\frac{1}{2} \left( w_{\ell m}^{\emptyset \emptyset}(i,j)^{\alpha} w_{\ell m}^{*QQ}(p,q)^{PP} + w_{\ell m}^{\emptyset \emptyset}(i,j)^{\alpha} w_{\ell m}^{*UU}(p,q)^{PP} \right)
\end{split}
\end{equation*}
\begin{equation*}
\begin{split}
W^{TT,PP} \left[ (i,j)^{TT}, (p,q)^{PP} \right]_{\ell} = \frac{1}{2\ell+1} \sum_m \hspace{8em}  \\
\frac{1}{2} \left( w_{\ell m}^{II}(i,j)^{TT} w_{\ell m}^{*QQ}(p,q)^{PP} + w_{\ell m}^{II}(i,j)^{TT} w_{\ell m}^{*UU}(p,q)^{PP} \right) 
\end{split}
\end{equation*}
\begin{equation}
\begin{split}
W^{PP,PP} \left[ (i,j)^{PP}, (p,q)^{PP} \right]_{\ell} = \frac{1}{2\ell+1} \sum_m \hspace{8em}  \\
\frac{1}{4} \big( w_{\ell m}^{QQ}(i,j)^{PP} w_{\ell m}^{*QQ}(p,q)^{PP} + w_{\ell m}^{UU}(i,j)^{PP} w_{\ell m}^{*UU}(p,q)^{PP} \\
+ w_{\ell m}^{QQ}(i,j)^{PP} w_{\ell m}^{*UU}(p,q)^{PP} + w_{\ell m}^{UU}(i,j)^{PP} w_{\ell m}^{*QQ}(p,q)^{PP} \big)
\end{split}
\end{equation}
These $w_{\ell m}$ are spherical harmonic transforms of the effective weight maps. The $w^{\emptyset\emptyset}_{\ell m}(i,j)$ are simply the coefficients of the spherical harmonic transform of the $i$ and $j$ masks. 
\begin{equation}
\begin{split}
    w_{\ell m}^{\emptyset\emptyset}(i,j)^{TT} &= \sum_{i}^{N_{\mathrm{pix}}} m^{i,T}_p m^{j,T}_p Y^*_{\ell m} ( \mathbf{\hat{n}}_p) \Omega_p,\\
    w_{\ell m}^{\emptyset\emptyset}(i,j)^{TP} &= \sum_{i}^{N_{\mathrm{pix}}} m^{i,T}_p m^{j,P}_p Y^*_{\ell m} ( \mathbf{\hat{n}}_p) \Omega_p,\\
    w_{\ell m}^{\emptyset\emptyset}(i,j)^{PT} &= \sum_{i}^{N_{\mathrm{pix}}} m^{i,P}_p m^{j,T}_p Y^*_{\ell m} ( \mathbf{\hat{n}}_p) \Omega_p,\\
    w_{\ell m}^{\emptyset\emptyset}(i,j)^{PP} &= \sum_{i}^{N_{\mathrm{pix}}} m^{i,P}_p m^{j,P}_p Y^*_{\ell m} ( \mathbf{\hat{n}}_p) \Omega_p,
\end{split}
\end{equation}
The $w^{II}_{\ell m}(i,j)$, $w^{QQ}_{\ell m}(i,j)$, and $w^{UU}_{\ell m}(i,j)$ are the coefficients of noise-variance weighted maps from II, QQ, and UU respectively.
\begin{equation}
\begin{split}
    w_{\ell m}^{II}(i,j)^{TT} &= \delta_{i,j} \sum_{i}^{N_{\mathrm{pix}}} \left(\sigma^{II}_p\right)^2 m^{i,T}_p m^{j,T}_p Y^*_{\ell m} ( \mathbf{\hat{n}}_p) \Omega_p,\\
    w_{\ell m}^{QQ}(i,j)^{TT} &= \delta_{i,j} \sum_{i}^{N_{\mathrm{pix}}} \left(\sigma^{QQ}_p\right)^2 m^{i,P}_p m^{j,P}_p Y^*_{\ell m} ( \mathbf{\hat{n}}_p) \Omega_p,\\
    w_{\ell m}^{UU}(i,j)^{TT} &= \delta_{i,j} \sum_{i}^{N_{\mathrm{pix}}} \left(\sigma^{UU}_p\right)^2 m^{i,P}_p m^{j,P}_p Y^*_{\ell m} ( \mathbf{\hat{n}}_p) \Omega_p,\\
\end{split}
\end{equation}
The $\sigma_p^2$ are pixel noise variances, and there is only a noise contribution if the two maps are the same.

The projector functions are linearly related to the mode-coupling matrices used in the pseudo-C$_{\ell}$ formalism to estimate unbiased spectra in the presence of a mask \citep{hivon2001}. For a pseudo-spectrum $\tilde{C}_{\ell}^{AB}$ and decoupled spectrum $\hat{C}_{\ell}^{AB}$, we have $\tilde{C}^{TT}_{\ell} = (2\ell_2 + 1)\Xi_{\ell, \ell^\prime}^{TT} \, \hat{C}^{TT}_{\ell}$ and $\tilde{C}^{TE}_{\ell} = (2\ell_2 + 1)\Xi_{\ell, \ell^\prime}^{TE} \, \hat{C}_{\ell}^{TE}$. However, the spin 2 fields EE, EB, BE, and BB requires the solution of a linear system,

\begin{equation}
\begin{split}
    \tilde{C}^{EE}_{\ell} &= (2\ell_2 + 1)\Xi_{\ell, \ell^\prime}^{EE} \, \hat{C}_{\ell}^{EE} + (2\ell_2 + 1)\Xi_{\ell, \ell^\prime}^{BB} \, \hat{C}_{\ell}^{BB} \\
    \tilde{C}^{BB}_{\ell} &= (2\ell_2 + 1)\Xi_{\ell, \ell^\prime}^{BB} \, \hat{C}_{\ell}^{EE} + (2\ell_2 + 1)\Xi_{\ell, \ell^\prime}^{EE} \, \hat{C}_{\ell}^{BB}
\end{split}
\end{equation}
Thus we compute the $C_{\ell}^{BB}$ pseudo-spectrum, in order to get decoupled spectra. Note that we follow PL20 and do not use the contribution of $C_{\ell}^{BB}$ to the covariance, as it is subdominant to the variance induced by cosmic variance and noise, as suggested by the signal-only simulations presented in Section~\ref{subsec:pstreat}.

\section{Covariance Matrix Blocks}


We repeat here the expressions analogous to Equation~\ref{eq:covTTTT} for $TT$, $TE$, and $EE$. The same $\mathcal{R}^{i,X}_{\ell}$ factors must be applied as in Section \ref{sec:covmat}.

\newcommand{\eq}[1]{Eq.~(\ref{#1})}

\newcommand{\ellp}{\ell^{\prime}}
\newcommand{\none}{\emptyset \emptyset}

\newcommand{\hiTTij}{^{TT \, i, j}}
\newcommand{\hiTTpq}{^{TT \, p, q}}
\newcommand{\hiTTip}{^{TT \, i, p}}
\newcommand{\hiTTiq}{^{TT \, i, q}}
\newcommand{\hiTTjp}{^{TT \, j, p}}
\newcommand{\hiTTjq}{^{TT \, j, q}}
\newcommand{\hiEE}{^{EE}}
\newcommand{\hiEEij}{^{EE \, i, j}}
\newcommand{\hiEEpq}{^{EE \, p, q}}
\newcommand{\hiEEip}{^{EE \, i, p}}
\newcommand{\hiEEiq}{^{EE \, i, q}}
\newcommand{\hiEEjp}{^{EE \, j, p}}
\newcommand{\hiEEjq}{^{EE \, j, q}}
\newcommand{\hiTE}{^{TE}}
\newcommand{\hiTEij}{^{TE \, i, j}}
\newcommand{\hiTEpq}{^{TE \, p, q}}
\newcommand{\hiTEip}{^{TE \, i, p}}
\newcommand{\hiTEiq}{^{TE \, i, q}}
\newcommand{\hiTEjp}{^{TE \, j, p}}
\newcommand{\hiTEjq}{^{TE \, j, q}}
\newcommand{\hiET}{^{ET}} 
\newcommand{\hiETij}{^{ET \, i, j}} 
\newcommand{\hiETpq}{^{ET \, p, q}}
\newcommand{\hiTP}{^{TP}} 
\newcommand{\hiPT}{^{PT}} 
\newcommand{\hiPP}{^{PP}}
\newcommand{\hiTT}{^{TT}}
\newcommand{\singleT}{T} 
\newcommand{\singleP}{P}
\newcommand{\TT}{{TT}} 
\newcommand{\TE}{{TE}} 
\newcommand{\ET}{{ET}} 
\newcommand{\EE}{{EE}} 
\newcommand{\BB}{{BB}}
\newcommand{\PP}{PP} 
\newcommand{\TP}{TP} 
\newcommand{\PT}{PT}
\newcommand{\II}{II} 
\newcommand{\QQ}{QQ} 
\newcommand{\UU}{UU}
\def\Var{\mathop{\rm Var}}

\newcommand{\wigner}[6]{\ensuremath{\begin{pmatrix} #1 & #2 & #3 \\
      #4 & #5 & #6 \end{pmatrix}}}
\newcommand{\argmin}{\operatornamewithlimits{argmin}}

\paragraph{$TTTT$ block:}
\begin{align}
\Var&(\hat{C}_{\ell}\hiTTij, \hat{C}_{\ellp}\hiTTpq) \nonumber\\
& \approx \sqrt{C_{\ell}\hiTTip C_{\ellp}\hiTTip C_{\ell}\hiTTjq
  C_{\ellp}\hiTTjq} \ \Xi_{\TT}^{\none, \none} \! \left[(i, p)\hiTT, (j,
  q)\hiTT \right]_{\ell \ellp} \nonumber\\
& + \sqrt{C_{\ell}\hiTTiq C_{\ellp}\hiTTiq C_{\ell}\hiTTjp
  C_{\ellp}\hiTTjp} \ \Xi_{\TT}^{\none, \none} \! \left[(i,q)\hiTT, (j,
  p)\hiTT\right]_{\ell \ellp} \nonumber\\
& + \sqrt{C_{\ell}\hiTTip C_{\ellp}\hiTTip} \ \Xi_{\TT}^{\none,
    } \! \left[(i, p)\hiTT, (j, q)\hiTT\right]_{\ell \ellp} \nonumber\\
& + \sqrt{C_{\ell}\hiTTjq C_{\ellp}\hiTTjq} \ \Xi_{\TT}^{\none,
  \TT} \! \left[(j, q)\hiTT, (i, p)\hiTT\right]_{\ell \ellp} \nonumber\\
& + \sqrt{C_{\ell}\hiTTiq C_{\ellp}\hiTTiq} \ \Xi_{\TT}^{\none,
  \TT} \! \left[(i, q)\hiTT, (j, p)\hiTT\right]_{\ell \ellp} \nonumber\\
& + \sqrt{C_{\ell}\hiTTjp C_{\ellp}\hiTTjp} \ \Xi_{\TT}^{\none, \TT} \! \left[(j,
   p)\hiTT, (i, q)\hiTT\right]_{\ell \ellp} \nonumber\\
& + \Xi_{\TT}^{\TT, \TT} \! \left[(i, p)\hiTT, (j, q)\hiTT\right]_{\ell \ellp} +
\Xi_{\TT}^{\TT, \TT} \! \left[(i, q)\hiTT, (j, p)\hiTT\right]_{\ell \ellp} \,.
\label{eq:hil_cov_mat_tttt_block}
\end{align}

\paragraph{$TTTE$ block:}
\begin{align}
\Var&(\hat{C}_{\ell}\hiTTij, \hat{C}_{\ellp}\hiTEpq) \nonumber\\
& \approx \frac{1}{2} \sqrt{C_{\ell}\hiTTip C_{\ellp}\hiTTip}
\left( C_{\ell}\hiTEjq + C_{\ellp}\hiTEjq \right) \ \Xi_{\TT}^{\none,
  \none} \! \left[(i, p)\hiTT, (j, q)\hiTP\right]_{\ell \ellp} \nonumber\\
& + \frac{1}{2} \sqrt{C_{\ell}\hiTTjp C_{\ellp}\hiTTjp}
\left( C_{\ell}\hiTEiq + C_{\ellp}\hiTEiq \right) \ \Xi_{\TT}^{\none,
  \none} \! \left[(i, q)\hiTP, (j, p)\hiTT\right]_{\ell \ellp} \nonumber\\
& + \frac{1}{2} \left( C_{\ell}\hiTEjq + C_{\ellp}\hiTEjq \right)
\ \Xi_{\TT}^{\none, \TT} \! \left[(j, q)\hiTP, (i, p)\hiTT\right]_{\ell \ellp}
\nonumber\\
& + \frac{1}{2} \left( C_{\ell}\hiTEiq + C_{\ellp}\hiTEiq \right)
\ \Xi_{\TT}^{\none, \TT} \! \left[(i, q)\hiTP, (j, p)\hiTT\right]_{\ell \ellp} \,.
\label{eq:hil_cov_mat_ttte_block}
\end{align}

\paragraph{$TETE$ block}
\begin{align}
\Var&(\hat{C}_{\ell}\hiTEij, \hat{C}_{\ellp}\hiTEpq) \nonumber\\
& \approx \sqrt{C_{\ell}\hiTTip C_{\ellp}\hiTTip C_{\ell}\hiEEjq
  C_{\ellp}\hiEEjq} \ \Xi_{\TE}^{\none, \none} \! \left[(i, p)\hiTT, (j,
  q)\hiPP\right]_{\ell \ellp} \nonumber\\
& + \frac{1}{2} \left( C_{\ell}\hiTEiq C_{\ellp}\hiTEjp +
C_{\ell}\hiTEjp C_{\ellp}\hiTEiq \right) \ \Xi_{\TT}^{\none,
  \none} \! \left[(i, q)\hiTP, (j, p)\hiPT\right]_{\ell \ellp} \nonumber\\
& + \sqrt{C_{\ell}\hiTTip C_{\ellp}\hiTTip} \ \Xi_{\TE}^{\none,
  \PP} \! \left[(i, p)\hiTT, (j, q)\hiPP\right]_{\ell \ellp} \nonumber\\
& + \sqrt{C_{\ell}\hiEEjq C_{\ellp}\hiEEjq} \ \Xi_{\TE}^{\none,
  \TT} \! \left[(j, q)\hiPP, (i, p)\hiTT\right]_{\ell \ellp} \nonumber\\
& + \Xi_{\TE}^{\TT, \PP} \! \left[(i, p)\hiTT, (j, q)\hiPP\right]_{\ell \ellp} \,.
\label{eq:hil_cov_mat_tete_block}
\end{align}

\paragraph{$TTEE$ block:}
\begin{align}
\Var&(\hat{C}_{\ell}\hiTTij, \hat{C}_{\ellp}\hiEEpq) \nonumber\\
& \approx \frac{1}{2} \left( C_{\ell}\hiTEip C_{\ellp}\hiTEjq +
C_{\ell}\hiTEjq C_{\ellp}\hiTEip \right) \ \Xi_{\TT}^{\none,
  \none} \! \left[(i, p)\hiTP, (j, q)\hiTP\right]_{\ell \ellp} \nonumber\\
& + \frac{1}{2} \left( C_{\ell}\hiTEiq C_{\ellp}\hiTEjp +
C_{\ell}\hiTEjp C_{\ellp}\hiTEiq \right) \ \Xi_{\TT}^{\none,
  \none} \! \left[(i,q)\hiTP, (j, p)\hiTP\right]_{\ell \ellp}
\label{eq:hil_cov_mat_ttee_block}
\end{align}

\paragraph{$TEEE$ block:}
\begin{align}
\Var&(\hat{C}_{\ell}\hiTEij, \hat{C}_{\ellp}\hiEEpq) \nonumber\\
& \approx \frac{1}{2} \sqrt{C_{\ell}\hiEEjq C_{\ellp}\hiEEjq}
\left(C_{\ell}\hiTEip + C_{\ellp}\hiTEip \right) \ \Xi_{\EE}^{\none,
  \none} \! \left[(i, p)\hiTP, (j, q)\hiPP\right]_{\ell \ellp} \nonumber\\
& + \frac{1}{2} \sqrt{C_{\ell}\hiEEjp C_{\ellp}\hiEEjp}
\left(C_{\ell}\hiTEiq + C_{\ellp}\hiTEiq \right) \ \Xi_{\EE}^{\none,
  \none} \! \left[(i,q)\hiTP, (j, p)\hiPP\right]_{\ell \ellp} \nonumber\\
& + \frac{1}{2} \left( C_{\ell}\hiTEip + C_{\ellp}\hiTEip \right)
\ \Xi_{\EE}^{\none, \PP} \! \left[(i, p)\hiTP, (j, q)\hiPP\right]_{\ell \ellp} \nonumber\\
& + \frac{1}{2} \left( C_{\ell}\hiTEiq + C_{\ellp}\hiTEiq \right)
\ \Xi_{\EE}^{\none, \PP} \! \left[(i, q)\hiTP, (j, p)\hiPP\right]_{\ell \ellp} \,.
\label{eq:hil_cov_mat_teee_block}
\end{align}

\paragraph{$EEEE$ block:}
\begin{align}
\Var&(\hat{C}_{\ell}\hiEEij, \hat{C}_{\ellp}\hiEEpq) \nonumber\\
& \approx \sqrt{C_{\ell}\hiEEip C_{\ellp}\hiEEip C_{\ell}\hiEEjq
  C_{\ellp}\hiEEjq} \ \Xi_{\EE}^{\none, \none} \! \left[(i, p)\hiPP, (j,
  q)\hiPP\right]_{\ell \ellp} \nonumber\\
& + \sqrt{C_{\ell}\hiEEiq C_{\ellp}\hiEEiq C_{\ell}\hiEEjp
  C_{\ellp}\hiEEjp} \ \Xi_{\EE}^{\none, \none} \! \left[(i, q)\hiPP, (j,
  p)\hiPP\right]_{\ell \ellp} \nonumber\\
& + \sqrt{C_{\ell}\hiEEip C_{\ellp}\hiEEip} \ \Xi_{\EE}^{\none, \PP} \! \left[(i,
  p)\hiPP, (j, q)\hiPP\right]_{\ell \ellp} \nonumber\\
& + \sqrt{C_{\ell}\hiEEjq C_{\ellp}\hiEEjq} \ \Xi_{\EE}^{\none, \PP} \! \left[(j,
  q)\hiPP, (i, p)\hiPP\right]_{\ell \ellp} \nonumber\\
& + \sqrt{C_{\ell}\hiEEiq C_{\ellp}\hiEEiq} \ \Xi_{\EE}^{\none, \PP} \! \left[(i,
  q)\hiPP, (j, p)\hiPP\right]_{\ell \ellp} \nonumber\\
& + \sqrt{C_{\ell}\hiEEjp C_{\ellp}\hiEEjp} \ \Xi_{\EE}^{\none, \PP} \! \left[(j,
  p)\hiPP, (i, q)\hiPP\right]_{\ell \ellp} \nonumber\\
& + \Xi_{\EE}^{\PP, \PP} \! \left[(i, p)\hiPP, (j, q)\hiPP\right]_{\ell \ellp} +
\Xi_{\EE}^{\PP, \PP} \! \left[(i, q)\hiPP, (j, p)\hiPP\right]_{\ell \ellp} \,.
\label{eq:hil_cov_mat_eeee_block}
\end{align}

\section{Noise spectrum fits}
\begin{figure}
\centering
\includegraphics[width=0.35\textwidth]{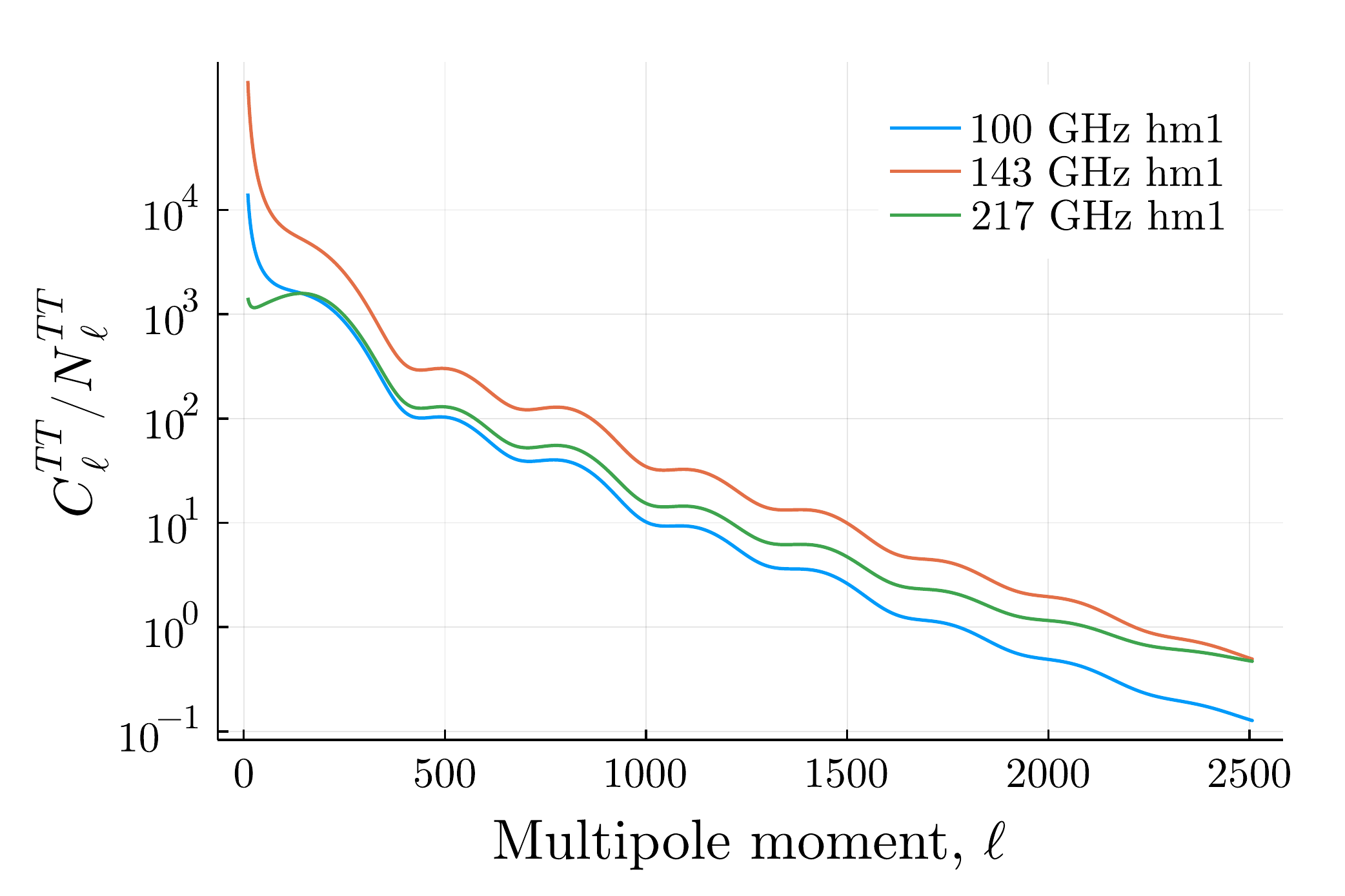}
\includegraphics[width=0.35\textwidth]{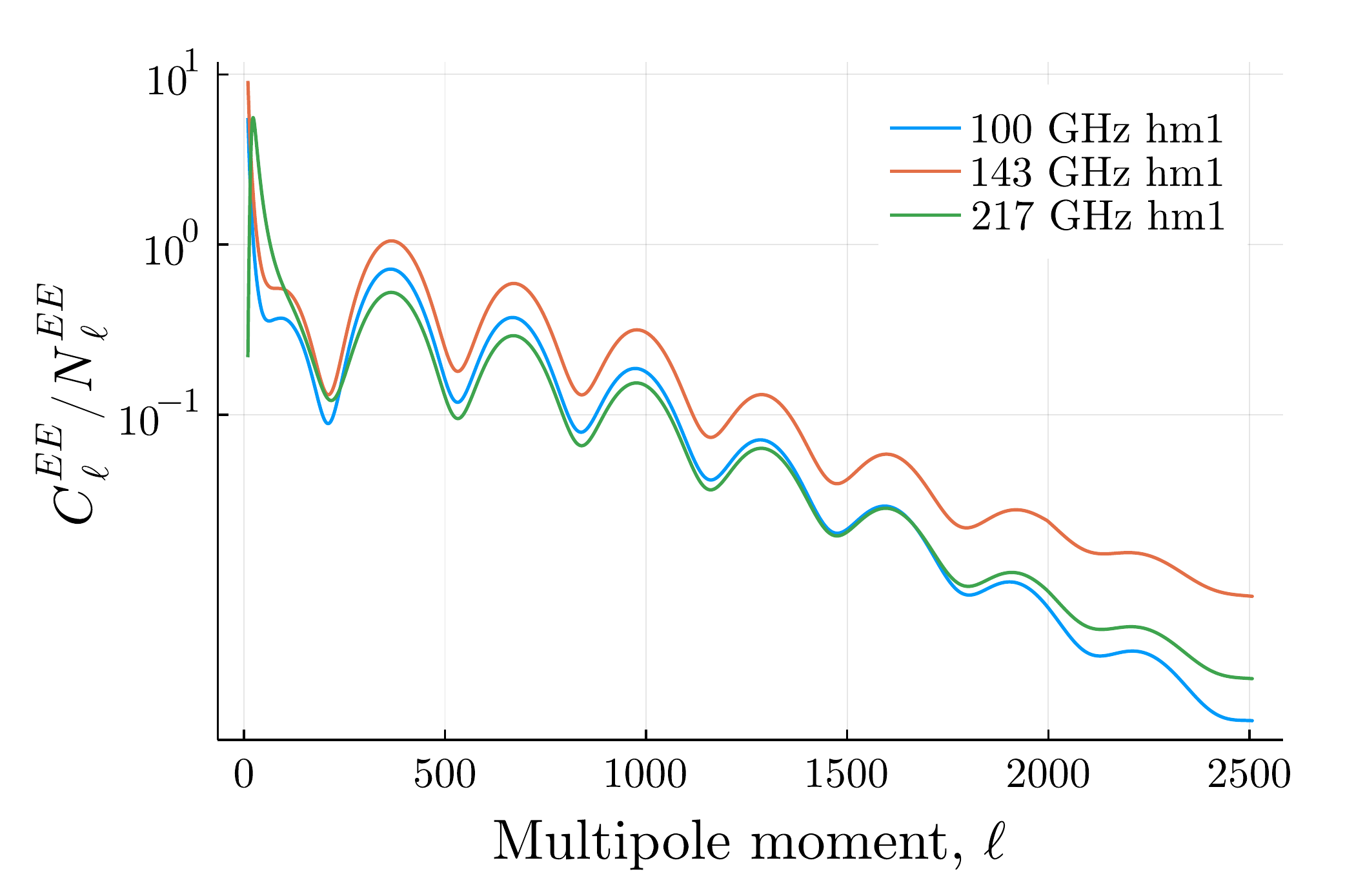}
\caption{The ratio of signal power spectrum (including foregrounds) and noise power spectrum for the half-mission 1 frequency maps in temperature (above) and polarization (below).}
\label{fig:signal_to_noise}
\end{figure}

\begin{figure*}
    \includegraphics[width=0.32\linewidth]{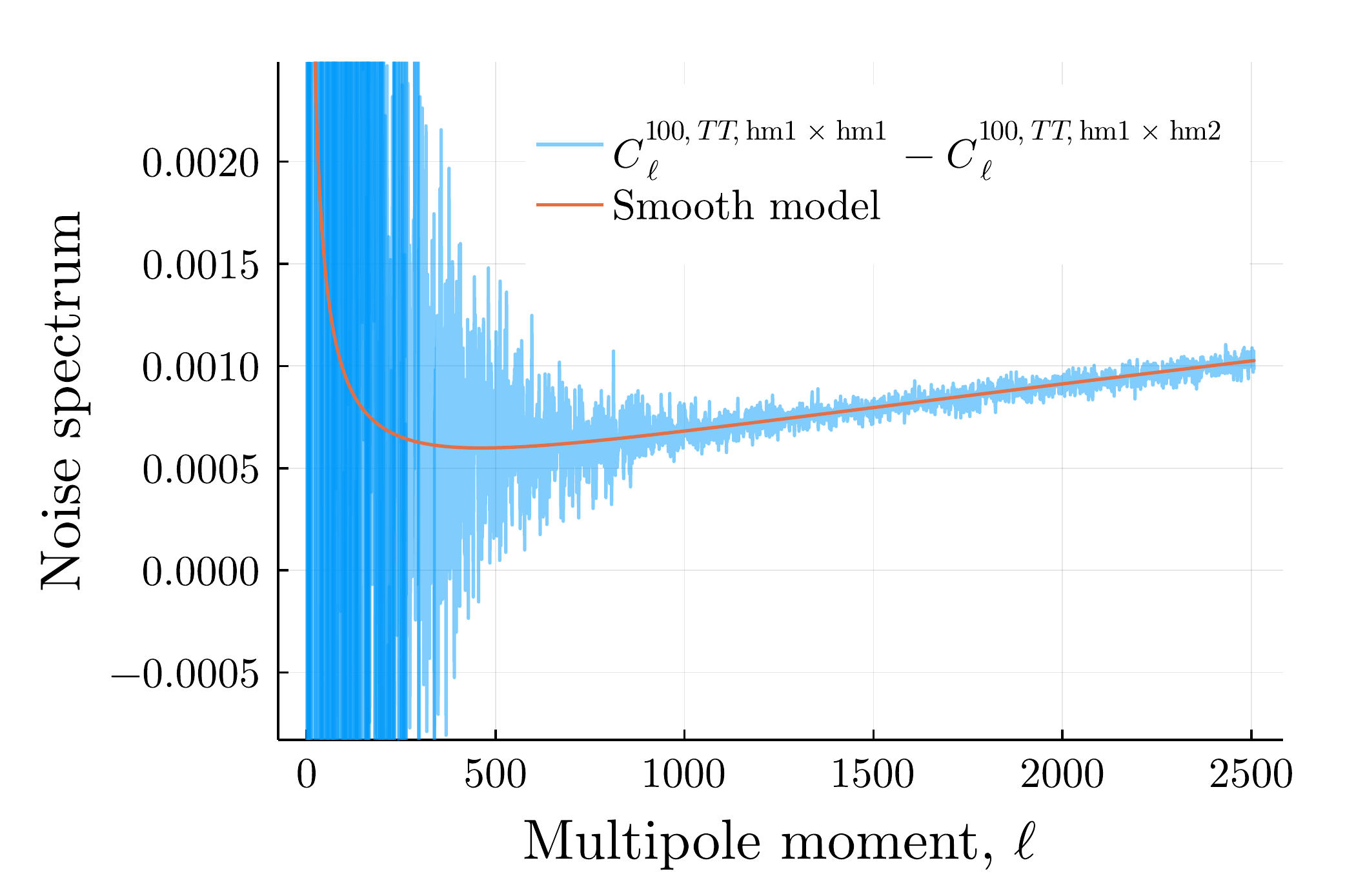}
    \includegraphics[width=0.33\linewidth]{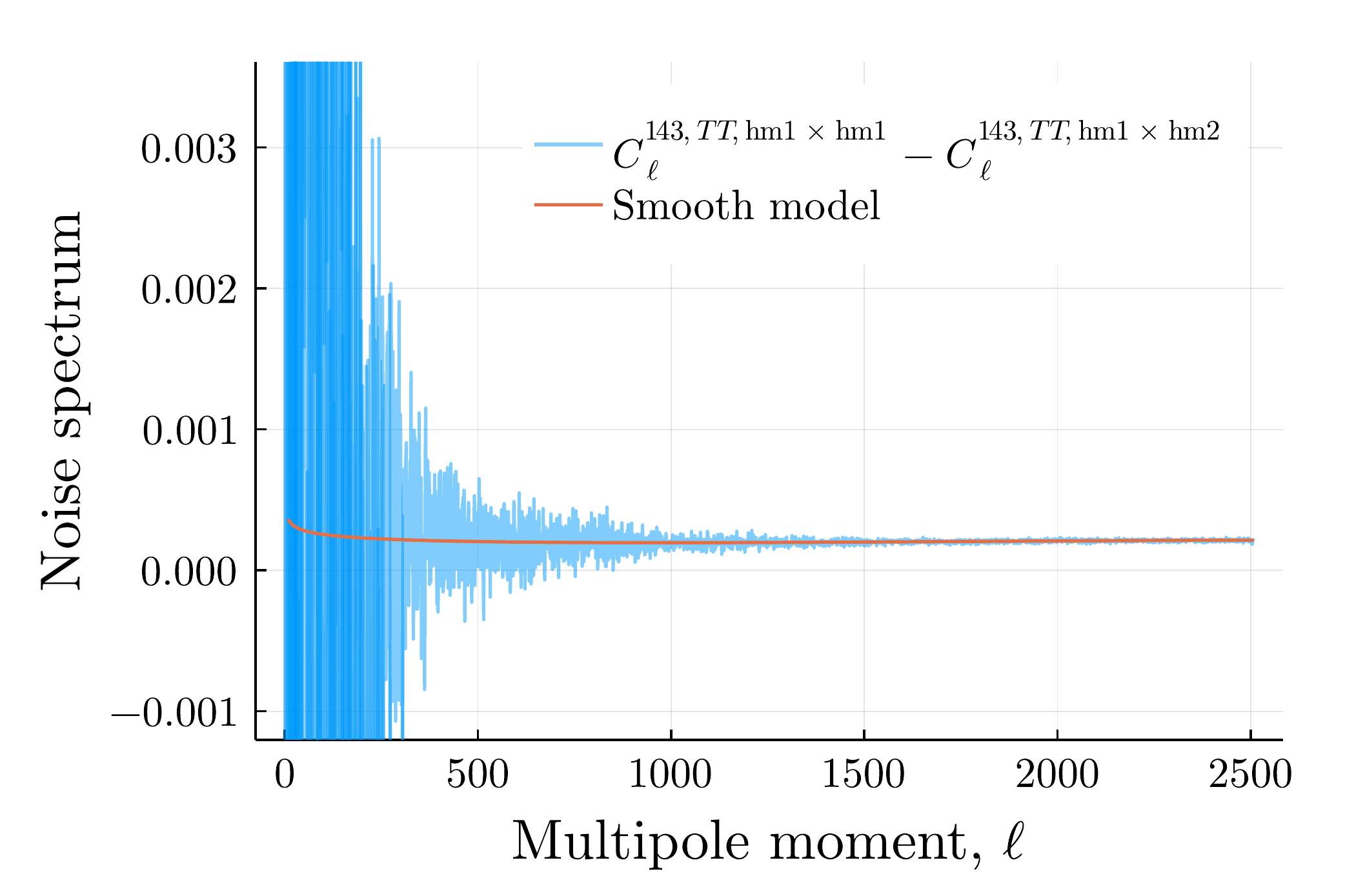}
    \includegraphics[width=0.33\linewidth]{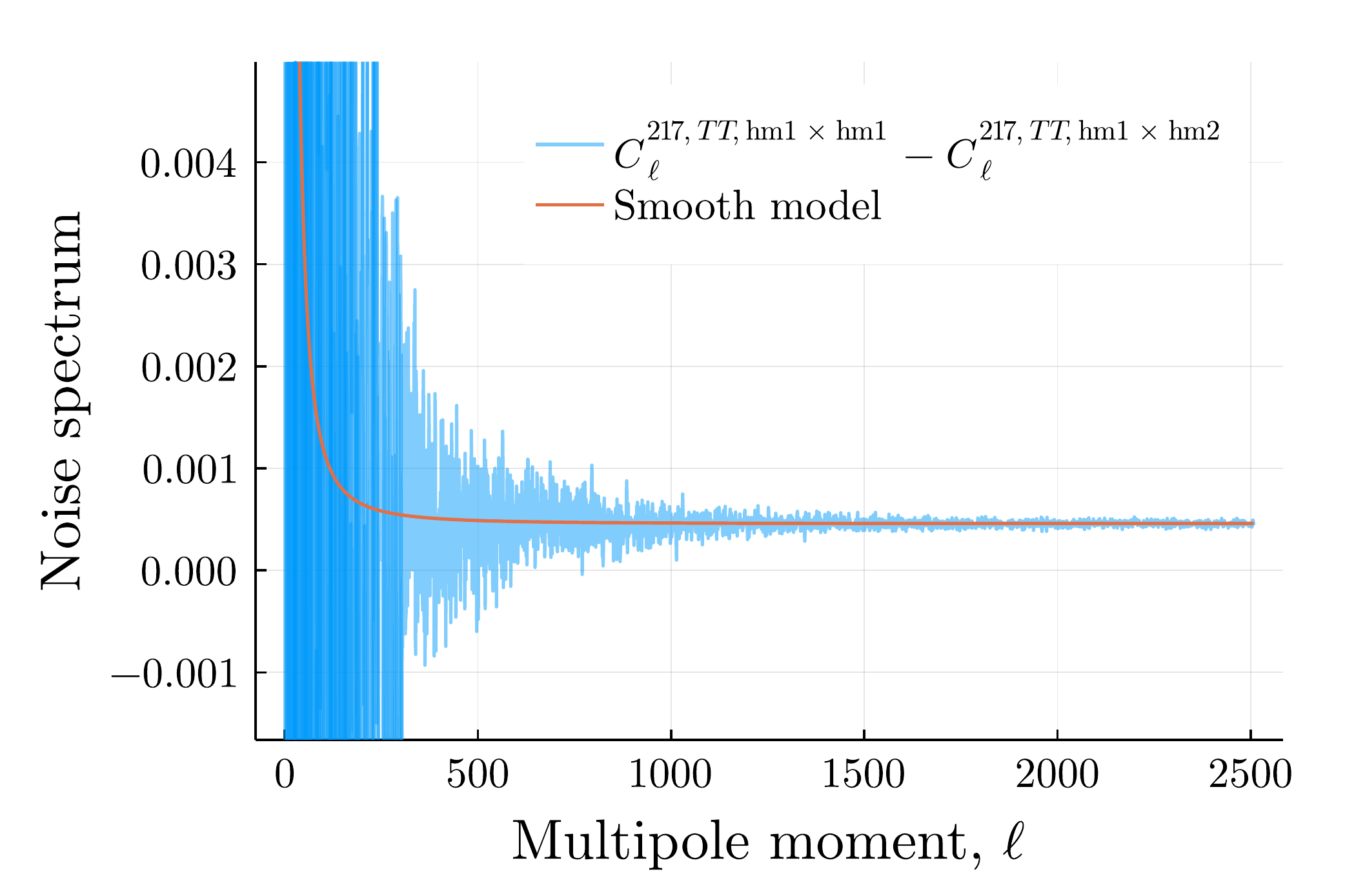}
    \includegraphics[width=0.32\linewidth]{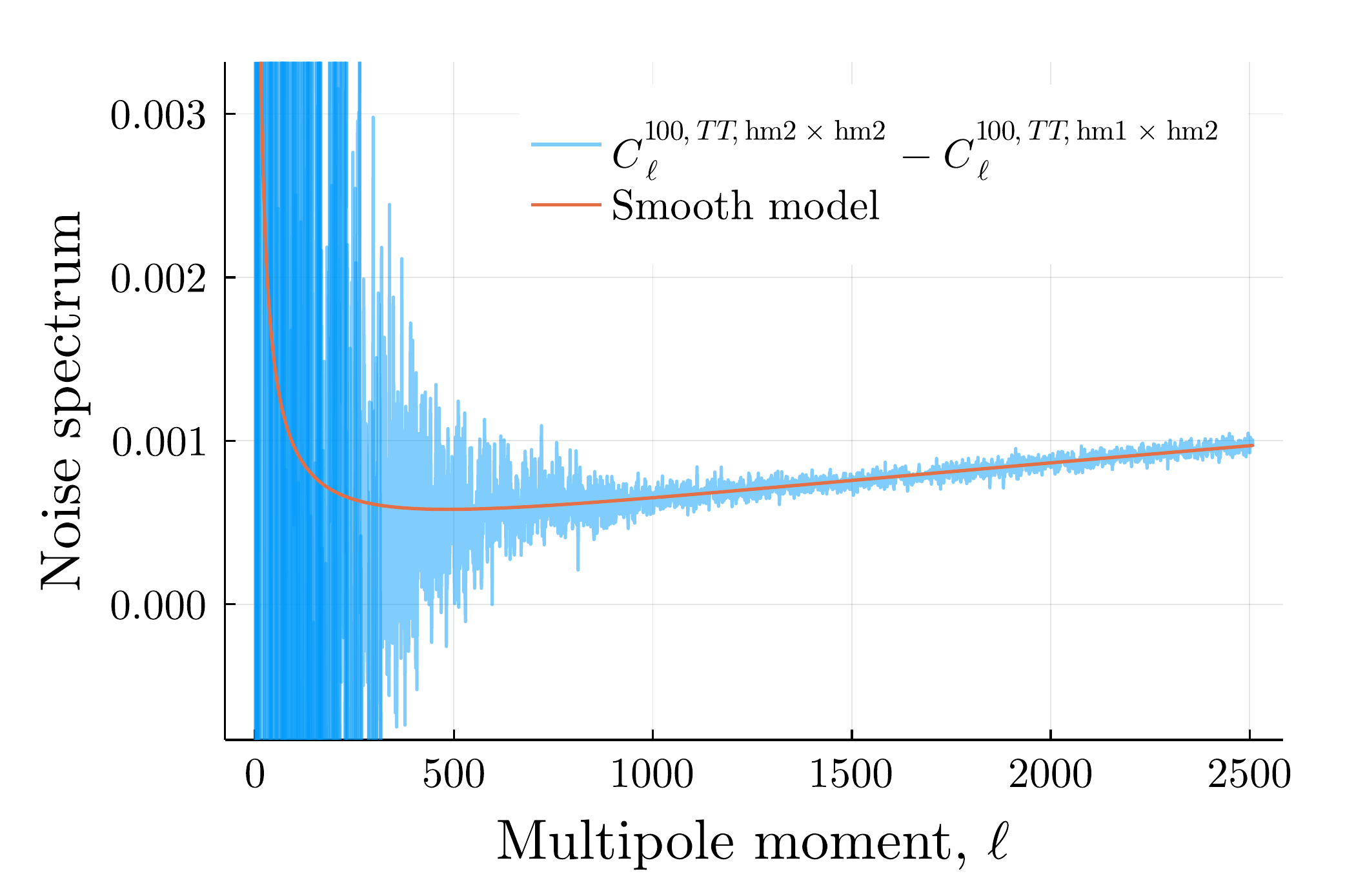}
     \includegraphics[width=0.33\linewidth]{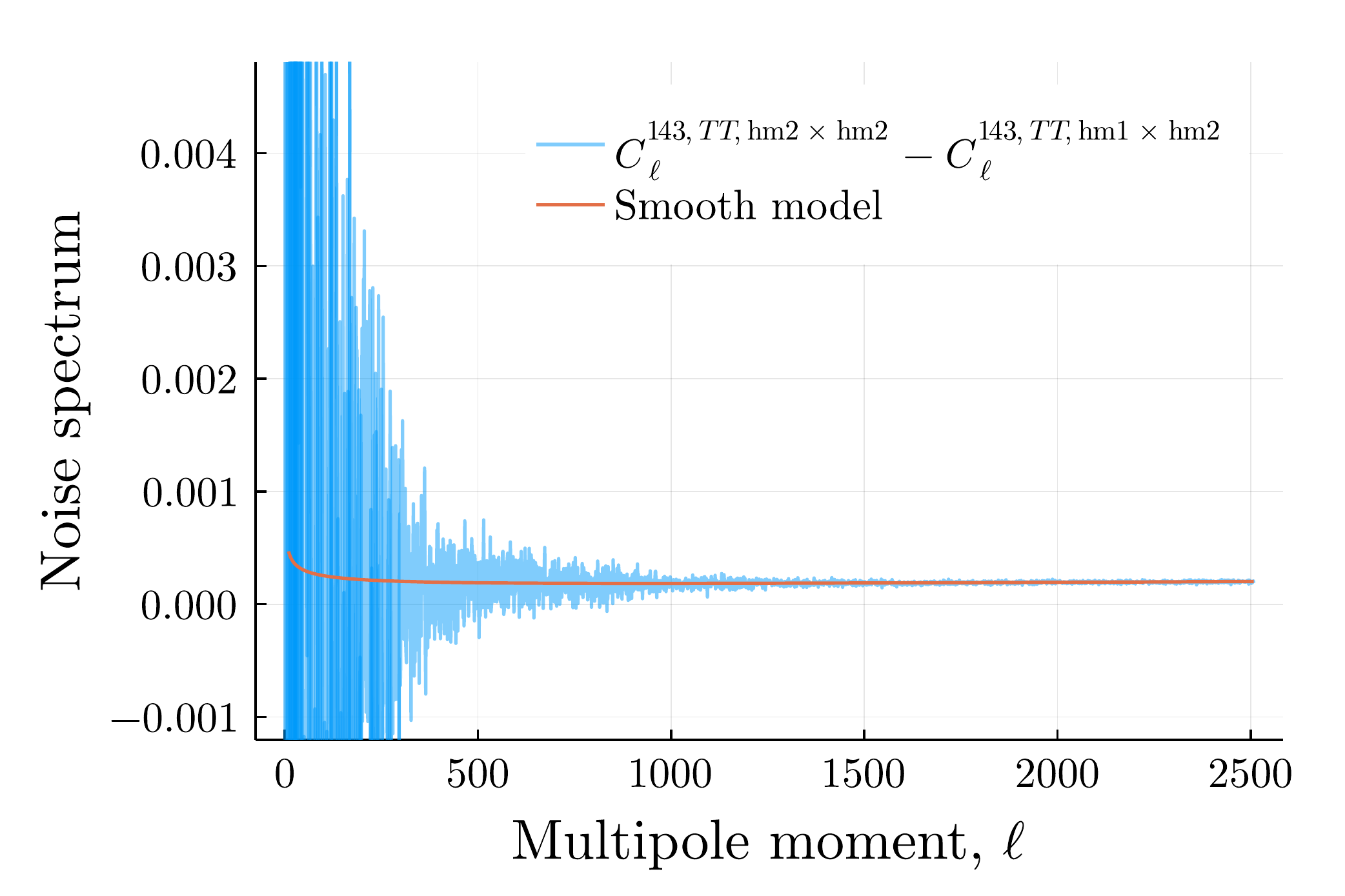}
     \includegraphics[width=0.33\linewidth]{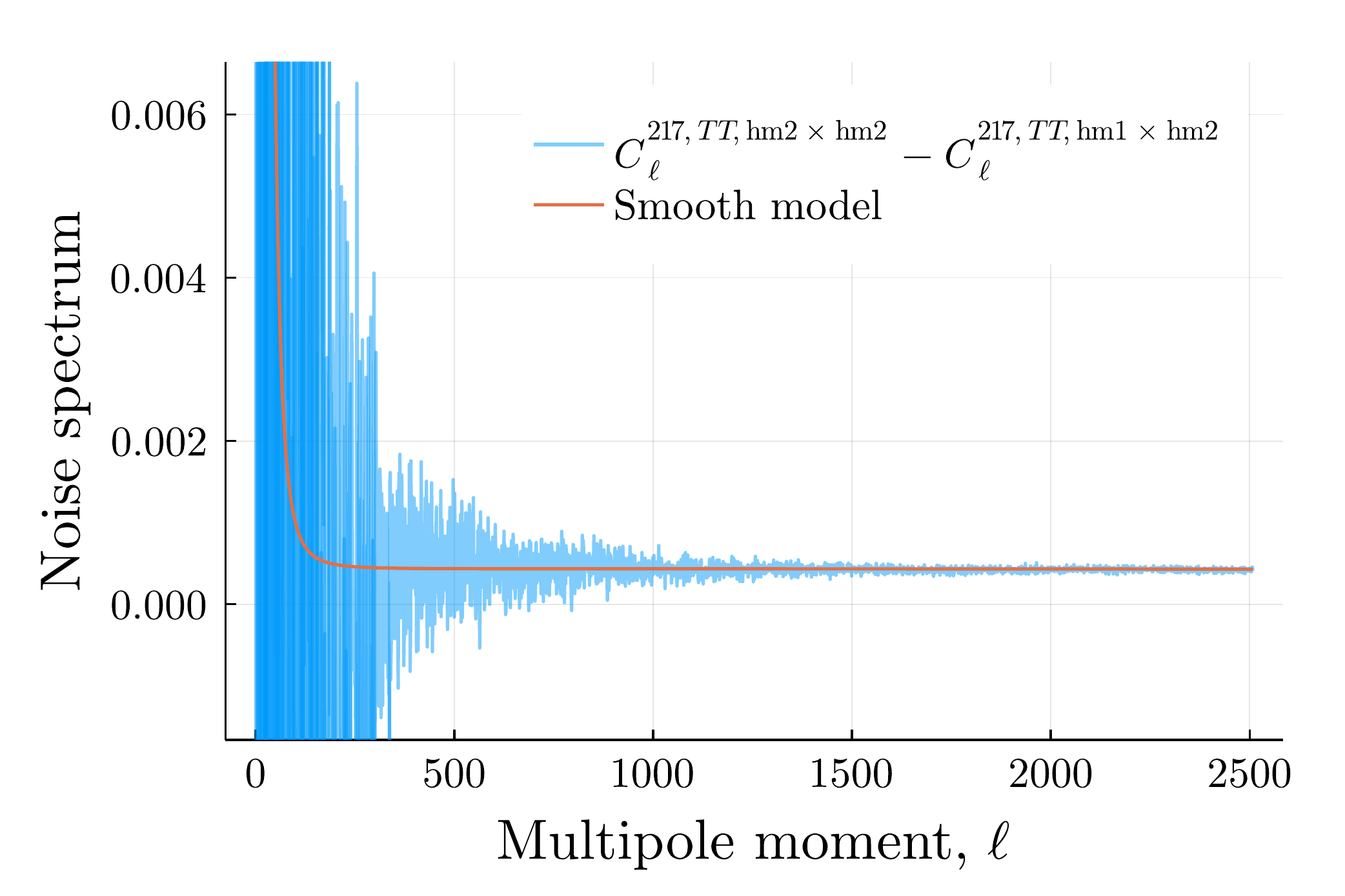}
\caption{The estimated noise power spectrum (blue) and our smooth model fits (orange) for the \planck temperature maps. At large scales ($\ell < 500$), the spectra are signal-dominated. The estimate from the difference of auto- and cross-spectra from the half-mission splits is thus contaminated by noise $\times$ signal terms. We note that the covariance matrix is not sensitive to the noise power spectrum at such scales, since those scales are signal-dominated.}
\label{fig:TT_noise_models}
\end{figure*}

\begin{figure*}
   \includegraphics[width=0.32\linewidth]{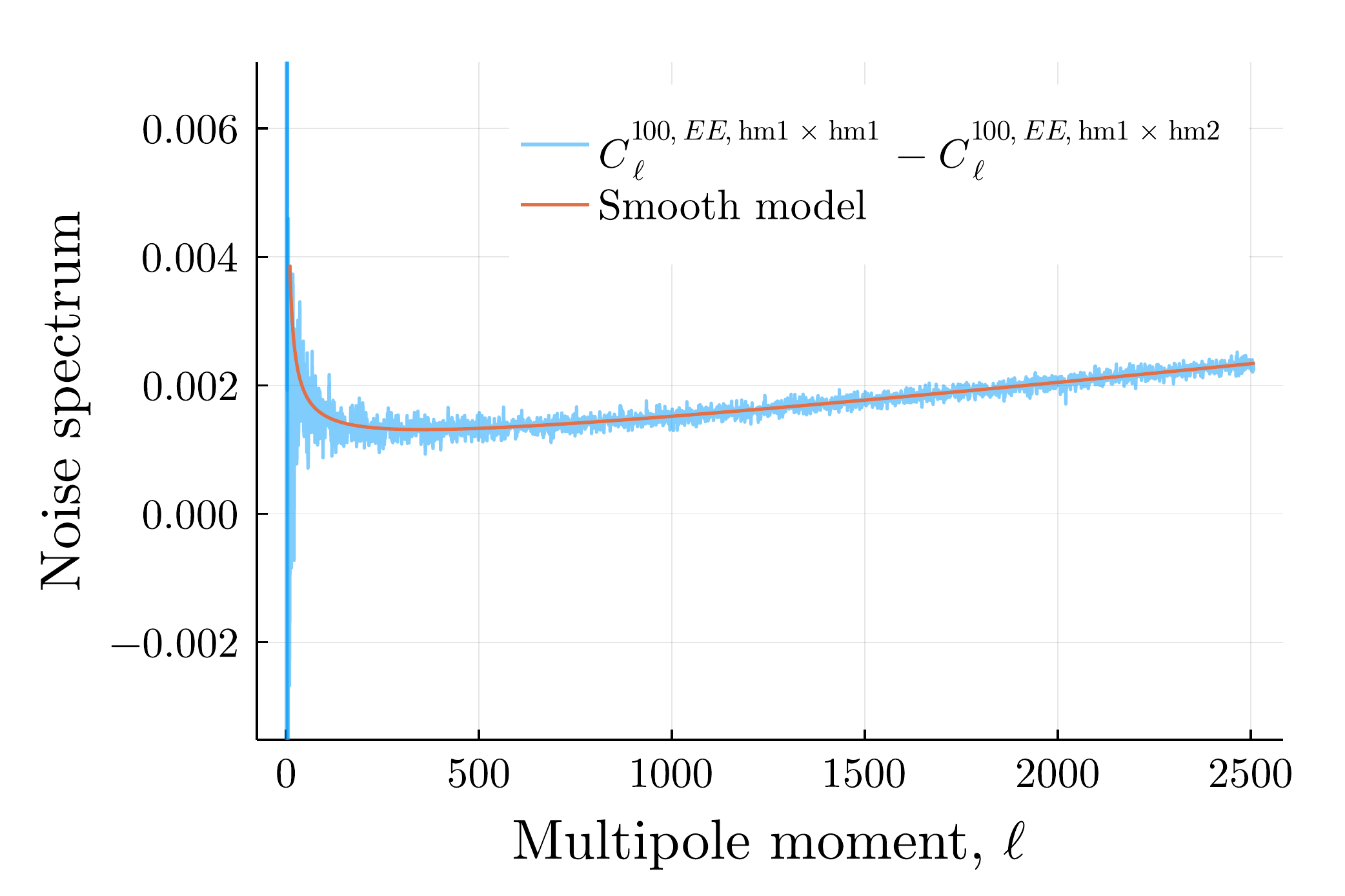}
    \includegraphics[width=0.33\linewidth]{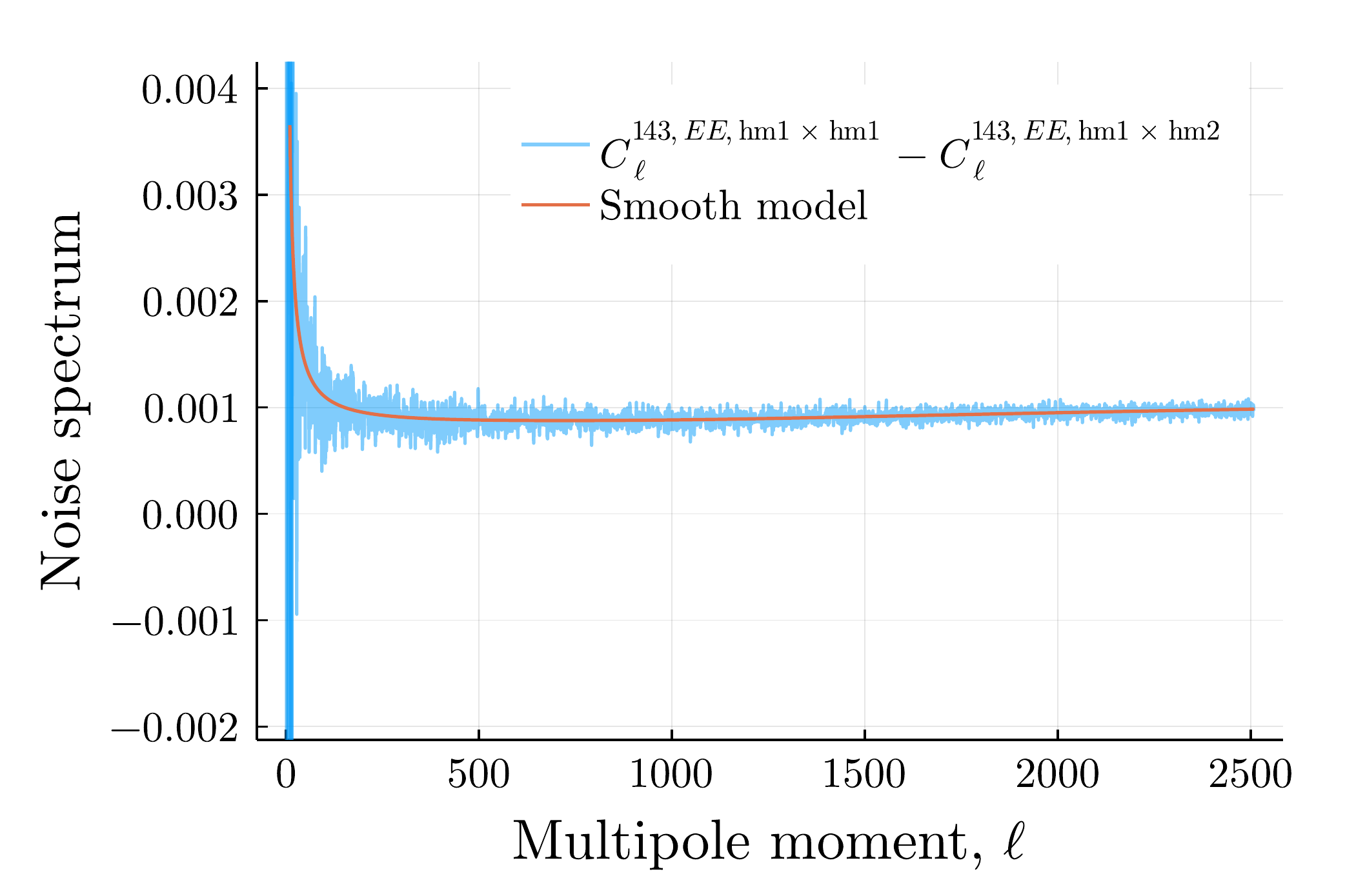}
    \includegraphics[width=0.33\linewidth]{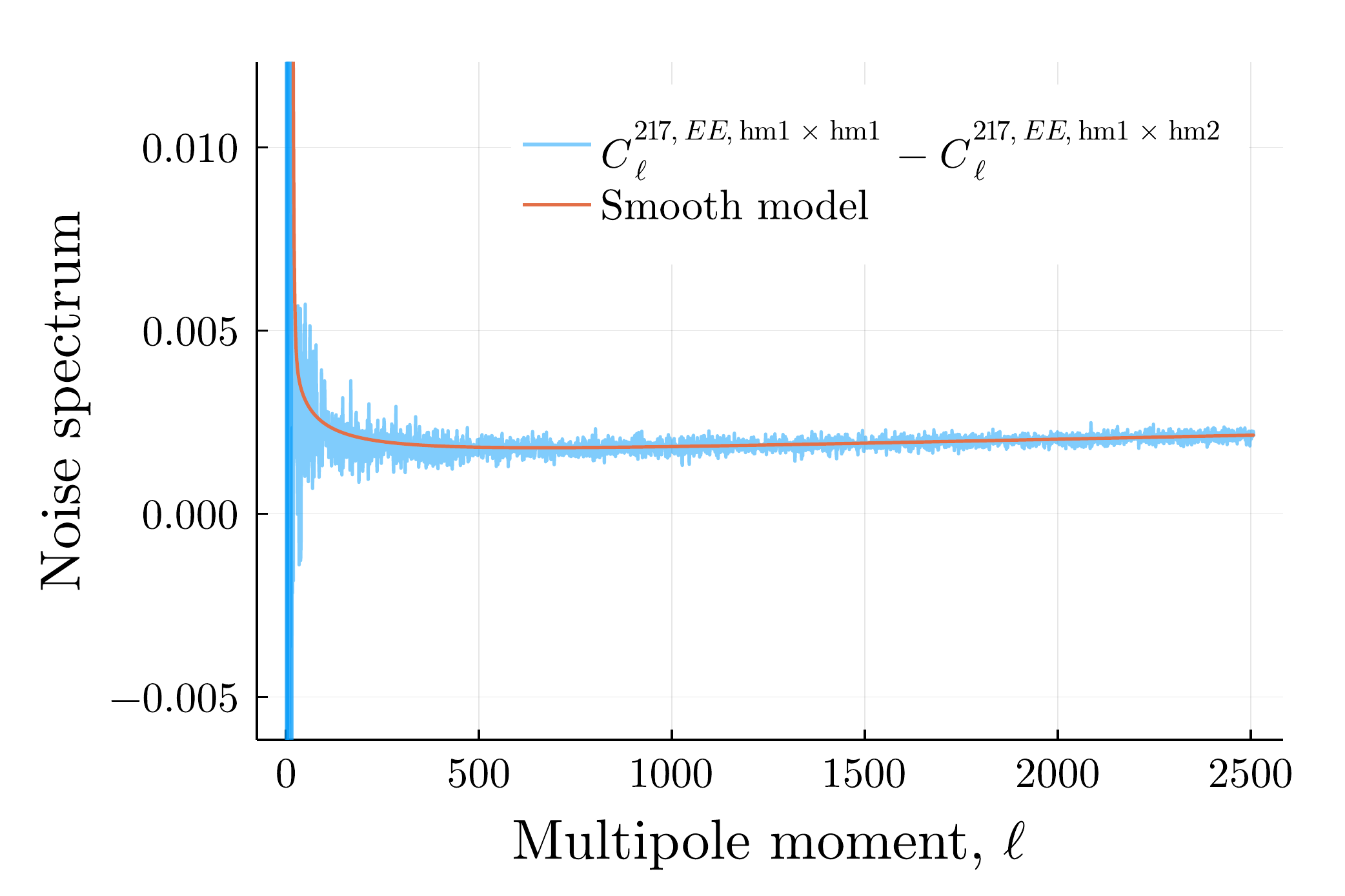}
    \includegraphics[width=0.32\linewidth]{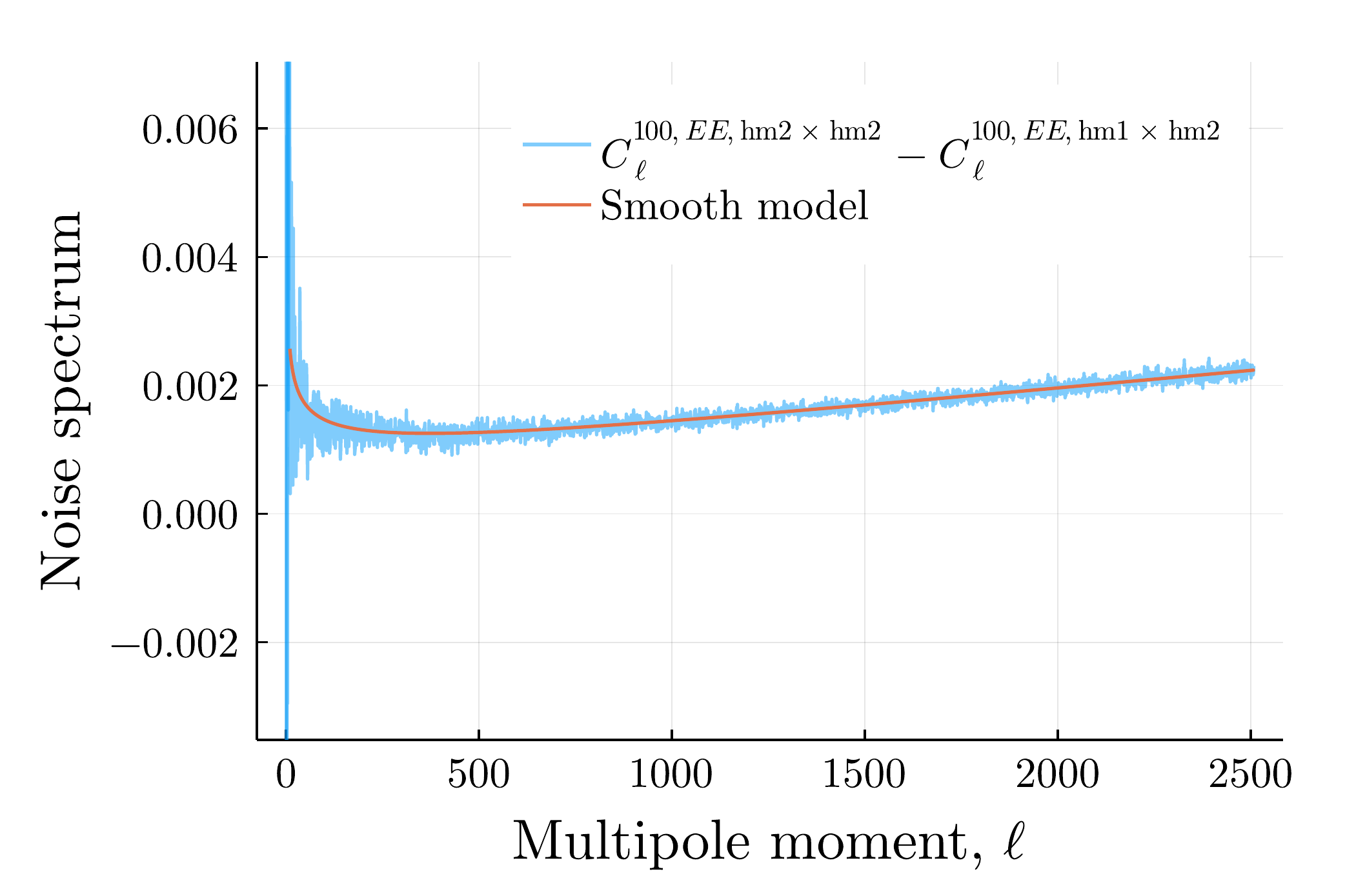}
     \includegraphics[width=0.33\linewidth]{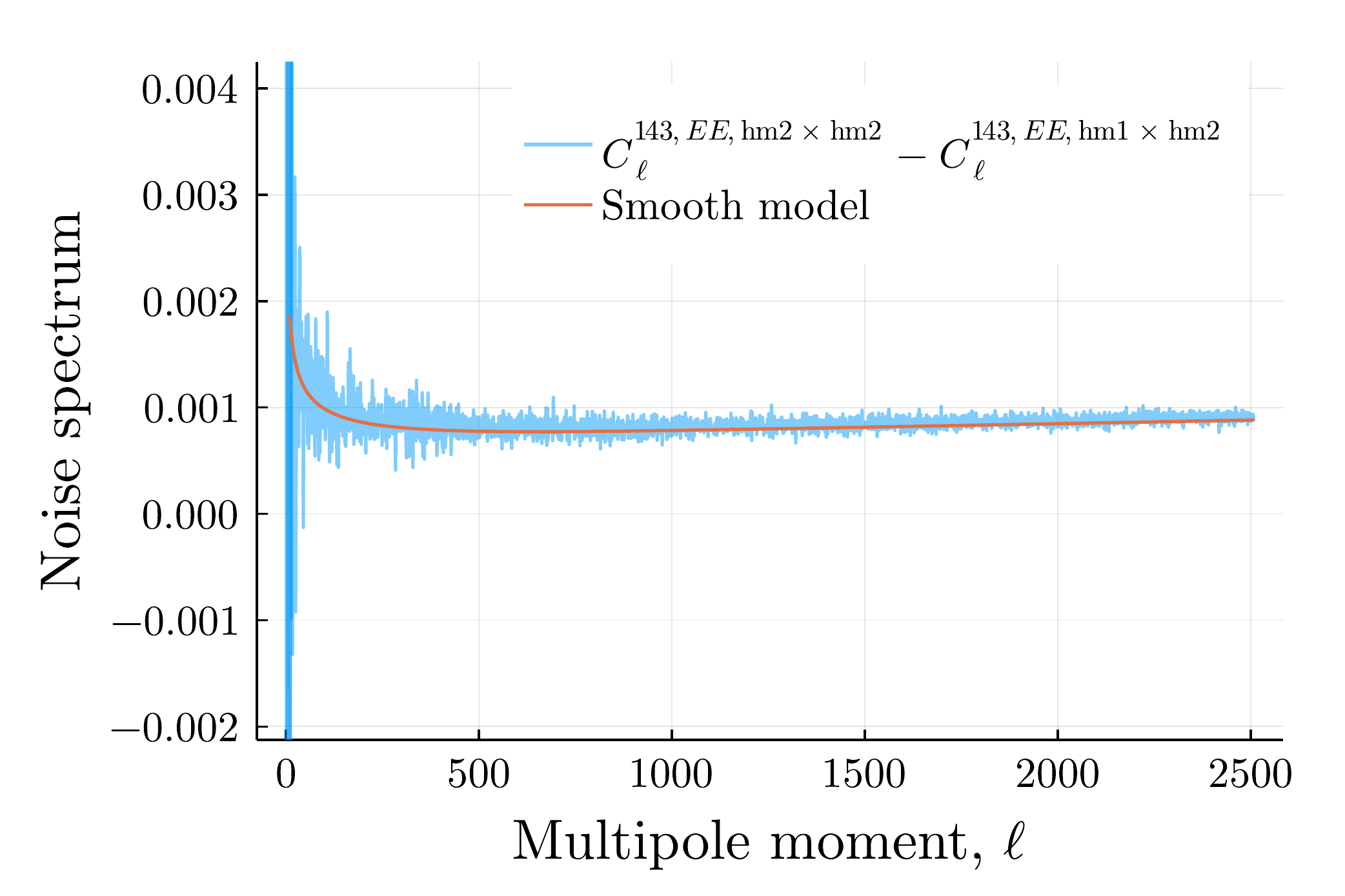}
     \includegraphics[width=0.33\linewidth]{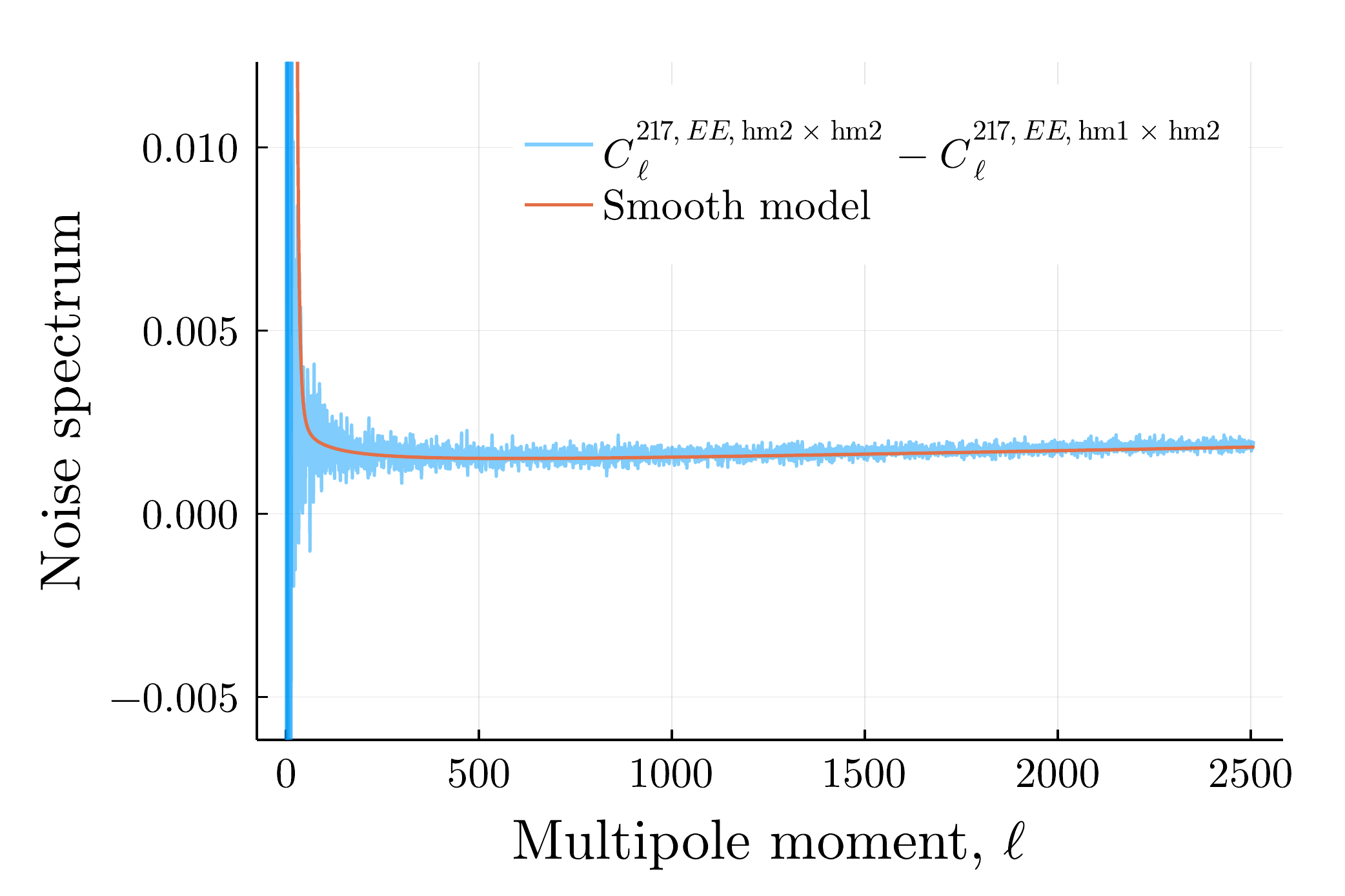}
\caption{The estimated noise power spectrum (blue) and our smooth model fits (orange) for the \planck polarization maps. Unlike temperature, the polarization power spectra are always noise-dominated in every bin, leading to a well-determined noise power spectrum and resulting model fit.}
\label{fig:EE_noise_models}
\end{figure*}

In Figure \ref{fig:signal_to_noise} we show the ratio of signal power spectrum to noise power spectrum for the half-mission 1 maps, to illustrate how at large angular scale in temperature the noise spectrum is significantly smaller than the signal spectrum, and thus the covariance is dominated by cosmic variance. In polarization the noise spectrum is not sub-dominant at any scale. Figures \ref{fig:TT_noise_models} and \ref{fig:EE_noise_models} show the smooth noise models fit to the estimated noise power spectra, for temperature and polarization respectively.

\end{document}